\documentclass[twocolumn,english,aps,prb,twocolum,superscriptaddress,bibnotes,amsmath,amssymb,floatfix]{revtex4-2}

\pdfoutput=1
\usepackage{amsfonts}
\usepackage{amsmath}
\usepackage{amssymb}
\usepackage[colorlinks=true,citecolor=blue,linkcolor=blue,breaklinks=true]{hyperref}
\usepackage{graphicx}

\usepackage{soul}
\usepackage{url}
\usepackage{dcolumn}
\usepackage{bm}
\usepackage{stmaryrd}
\usepackage[final]{changes}
\usepackage{multirow}
\usepackage{rotating}
\usepackage{titlesec}
\usepackage{tabularx}
\usepackage{textcomp}
\usepackage[flushleft]{threeparttable}

\usepackage{titlesec}
\renewcommand{\thesection}{\arabic{section}}
\titleformat{\section}
{\normalfont\large\bfseries}{\thesection.}{1em}{}

\begin{document}

\title{A squeezed mechanical oscillator with milli-second quantum decoherence}

\author{Amir Youssefi}\thanks{These authors contributed equally.}
\author{Shingo Kono} \thanks{These authors contributed equally.}
\author{Mahdi Chegnizadeh} \thanks{These authors contributed equally.}
\author{Tobias~J.~Kippenberg}
\email[]{tobias.kippenberg@epfl.ch}
\affiliation{Laboratory of Photonics and Quantum Measurement, Swiss Federal Institute of Technology Lausanne (EPFL), Lausanne, Switzerland}
\affiliation{Center for Quantum Science and Engineering, EPFL, Lausanne, Switzerland}
\maketitle

\textbf{Quantum control and measurement of mechanical oscillators has applications ranging from quantum metrology~\cite{aasi2013enhanced,mason2019continuous,whittle2021approaching} and quantum computing~\cite{pechal2018superconducting, wallucks2020quantum}, to fundamental test of quantum mechanics itself~\cite{fiaschi2021optomechanical,marinkovic2018optomechanical} 
	or searches for dark matter~\cite{carney2021mechanical,manley2021searching}. This has been achieved by coupling mechanical oscillators to auxiliary degrees of freedom in the form of optical or microwave cavities~\cite{RMP_optomechanics}, or superconducting qubits~\cite{clerk2020hybrid,chu2020perspective}, allowing numerous advances such as 
	mechanical squeezing~\cite{wollman2015quantum,pirkkalainen2015squeezing,lecocq2015quantum}, quantum state transfer~\cite{reed2017faithful,chu2018creation}, quantum transduction~\cite{mirhosseini2020superconducting,andrews2014bidirectional}, or teleportation~\cite{fiaschi2021optomechanical}.
	An enduring challenge in constructing such hybrid systems is the dichotomy of engineered coupling to an auxiliary degree of freedom, while being mechanically well isolated from the environment, that is, low quantum decoherence -- which consists of both thermal decoherence and dephasing.
	Although Hertz-level thermal decoherence has been achieved in optomechanical crystals at mK temperature~\cite{maccabe2020nano}, such systems suffer from large dephasing. 
	Currently employed opto- and electro-mechanical~\cite{rossi2018measurement,palomaki2013coherent,magrini2021real} as well as qubit-coupled mechanical systems~\cite{chu2018creation,wollack2022quantum,satzinger2018quantum} have significantly higher thermal decoherence. 
	Here we overcome this challenge by introducing a superconducting circuit optomechanical platform which exhibits an ultra-low quantum decoherence while having a large optomechanical coupling to prepare with high fidelity the quantum ground and squeezed states of motion. We directly measure a thermal decoherence rate of only 20.5~Hz (corresponding to $T_1 = 7.7$~ms) as well as a pure dephasing rate of 0.09 Hz, on par with and better than, respectively, the motional degree of freedom of trapped ion systems in high vacuum~\cite{gaebler2016high, leibfried2003quantum}, and 100-fold improvement of quantum-state lifetime compared to the prior optomechanical systems~\cite{rossi2018measurement,maccabe2020nano,reed2017faithful}.
	This enables us to reach to 0.07 quanta motional ground state occupation (93\% fidelity) and realize mechanical squeezing of -2.7 dB below zero-point-fluctuation. To directly measure the quantum-state lifetime, we observe the free evolution of the phase-sensitive squeezed state for the first time, preserving its non-classical nature over milli-second timescales.
	Such ultra-low quantum decoherence not only increases the fidelity of quantum control and measurement of macroscopic mechanical systems, but may equally benefit interfacing with qubits~\cite{gely2021phonon,pechal2018superconducting}, 
	and places the system in a parameter regime suitable for tests of quantum gravity~\cite{gely2021superconducting,liu2021gravitational}.}

The decoherence of a mechanical oscillator induced by the interaction with its environment conceals macroscopic quantum phenomena and limits the realization of mechanical oscillator-based quantum protocols~\cite{marinkovic2018optomechanical,fiaschi2021optomechanical,wallucks2020quantum,kotler2021direct,rossi2018measurement, delaney2019measurement,mirhosseini2020superconducting}.
The quantum decoherence can be characterized with two independent rates: the thermal decoherence rate ($\Gamma_\mathrm{th} = (n_\mathrm{m}^\mathrm{th}+1) \Gamma_\mathrm{m}$, where $n_\mathrm{m}^\mathrm{th}$ is the thermal bath occupation and $\Gamma_\mathrm{m}$ is the bare damping rate), which describes the rate at which phonons are exchanged with the thermal bath, and the pure dephasing rate ($\Gamma_\varphi$), caused by the mechanical frequency fluctuations, i.e.\ phonon-number conserving interactions with the environment~\cite{gardiner2004quantum}.
Even though the lowest thermal decoherence has been achieved in optomechanical crystals at mK temperature ($\Gamma_\mathrm{th}/2\pi \simeq 0.1$~Hz), such systems experience a large dephasing ($\Gamma_\varphi/2\pi \simeq 4$~kHz), limiting their quantum coherence~\cite{maccabe2020nano}. Soft clamped dissipation diluted $\mathrm{Si_3N_4}$ membranes~\cite{rossi2018measurement} and levitated particles~\cite{magrini2021real,tebbenjohanns2021quantum} are other examples of optomechanical platforms interfacing with light, which achieved thermal decoherence rates of $\mathcal{O}(1\text{ kHz})$, but support limited optomechanical protocols as they operate in the non-resolved-sideband or cavity-free regimes, or suffer from optical heating~\cite{delic2020cooling,piotrowski2022simultaneous}.
One of the most widely and successfully used optomechanical platforms is microwave superconducting circuit optomechanics~\cite{teufel2011sideband}, which exhibits large optomechanical coupling in the resolved-sideband regime, and can be integrated with superconducting qubits~\cite{reed2017faithful,clerk2020hybrid}. These circuits have been used for numerous advances including mechanical squeezing~\cite{wollman2015quantum,pirkkalainen2015squeezing,lecocq2015quantum}, entanglement~\cite{kotler2021direct,ockeloen2018stabilized,palomaki2013entangling}, non-classical state storage~\cite{reed2017faithful,palomaki2013coherent}, and non-reciprocal circuits~\cite{Bernier2017}. However, it has been a challenge to achieve a high quantum coherence in this platform - state of the art decoherence rates are $\mathcal{O}$(1 kHz)~\cite{palomaki2013coherent}. 
Enhancing the quantum coherence in such systems improves the fidelity of quantum optomechanical protocols~\cite{wallucks2020quantum,pechal2018superconducting,gely2021phonon} and may equally benefit future tests of quantum mechanics~\cite{fiaschi2021optomechanical,marinkovic2018optomechanical,gely2021superconducting,liu2021gravitational}.
While some attempts~\cite{seis2022ground,liu2021optomechanical} have been made to integrate ultra-coherent soft clamped $\mathrm{Si_3N_4}$ membranes~\cite{tsaturyan2017ultracoherent} with superconducting circuits, these hybrid systems have been compounded by insufficient optomechanical coupling, resulting in unwanted cavity heating.
\begin{figure*}[t!]
	\centering
	\includegraphics[scale=1.2]{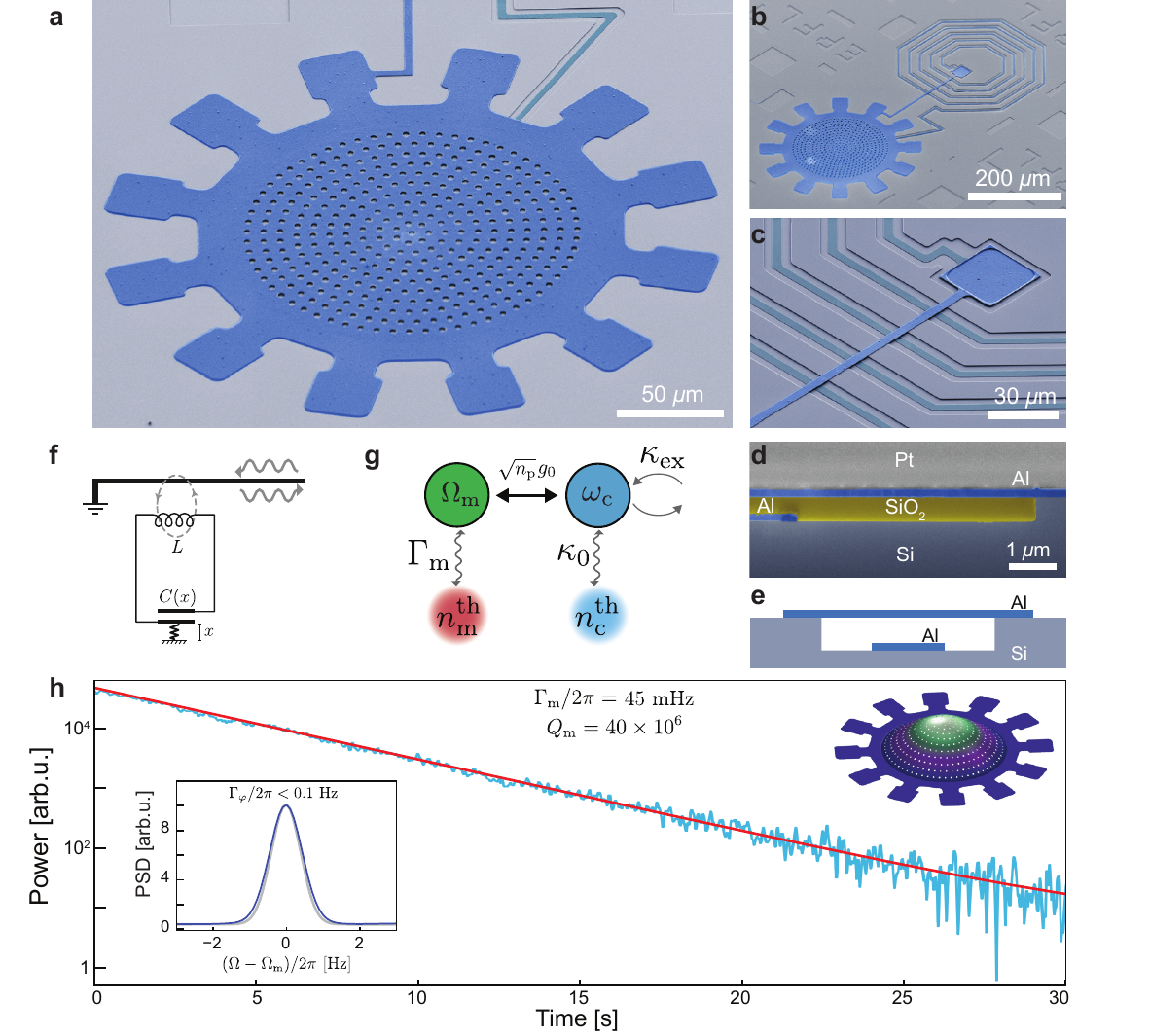}
	\caption{\footnotesize \linespread{1}
		\textbf{Ultra-coherent circuit optomechanics}
		\textbf{a},~False colored SEM image of a mechanically compliant parallel plate capacitor. 
		\textbf{b},~A microwave superconducting LC resonator consisting of the capacitor shunted by a spiral inductor. 
		\textbf{c}, Magnified image showing a silicon-etched trench, inductor air-bridges, and the galvanic connection. 
		\textbf{d},~Focused ion beam cross-section of a test capacitor --with a higher gap size than the main device-- before removing a SiO$_2$ sacrificial layer, where Pt is used as the FIB protective layer.
		\textbf{e},~Schematic cross-section of the suspended capacitor over the trench. 
		\textbf{f},~Mode diagram of an optomechanical system. $\kappa_0$ and $\kappa_\mathrm{ex}$ are internal loss and external coupling rates of the cavity respectively.  
		(\textbf{g},~Equivalent circuit diagram of the system. 
		\textbf{h},~Ring-down trace, showing the energy decay of the mechanical oscillator with a rate of $\Gamma_\mathrm{m}/2\pi = 45$ mHz. The red line is the exponential fit. 
		The top inset shows the FEM simulation of the fundamental mechanical mode of the drum. The blue line in the bottom inset shows the averaged PSD of the mechanics with 1~Hz measurement resolution bandwidth (gray line), indicating the frequency instability of less than 0.1~Hz.
		\label{fig:1}}
\end{figure*}
Here we demonstrate a circuit optomechanical platform which simultaneously realizes an ultra-low quantum decoherence, i.e. both thermal decoherence and dephasing, while exhibiting an efficient optomechanical coupling for quantum control and measurement.
By observing free evolution of the prepared ground state as well as squeezed state we report a thermal decoherence rate of 20.5~Hz (corresponding to 130 quanta/sec motional heating rate) and a pure dephasing rate of 0.09~Hz, showing the quantum decoherence is dominated by the thermal decoherence, comparable with motional decoherences achieved in trapped ion systems in high vacuum~\cite{leibfried2003quantum}, where the thermal decoherence is typically $\mathcal{O}(10~\mathrm{Hz})$, while the dephasing rate reaches $\mathcal{O}(100~\mathrm{Hz})$~\cite{gaebler2016high}.

\begin{figure*}[t!]
	\centering
	\includegraphics[scale=1.2]{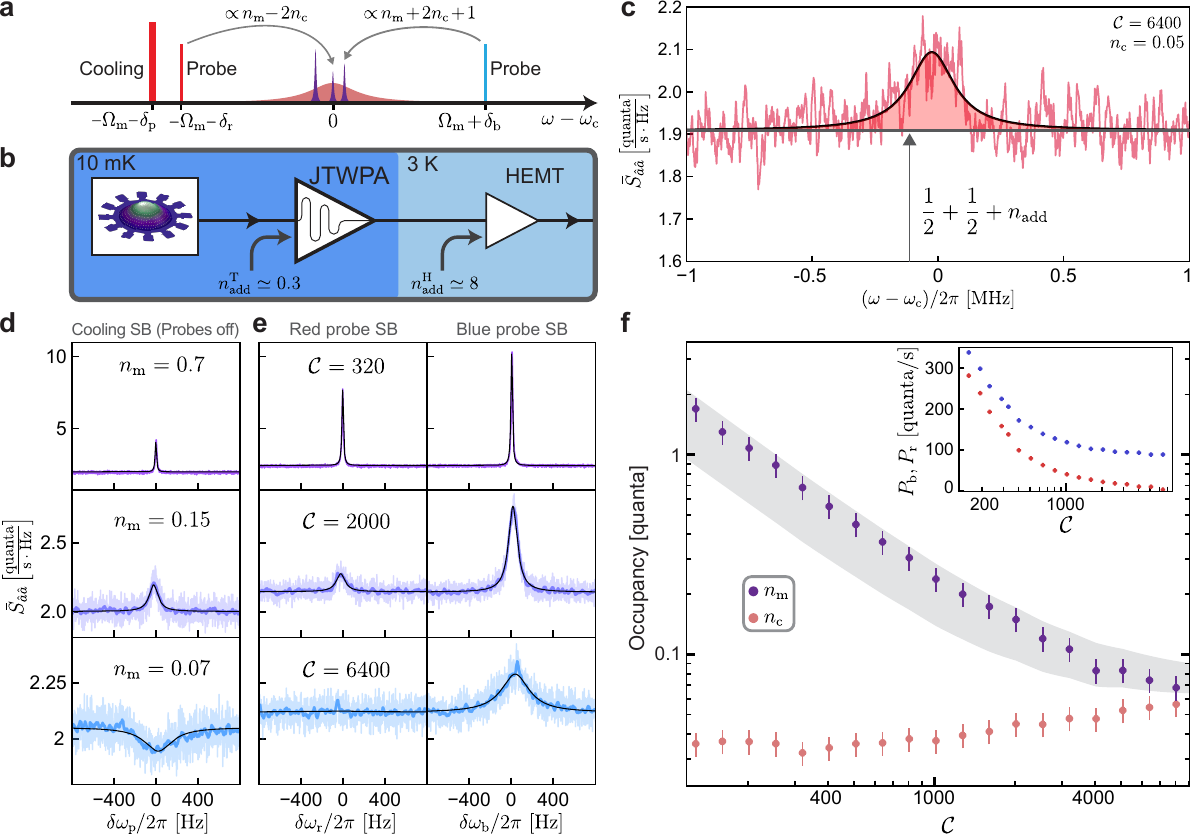}
	\caption{\footnotesize \linespread{1}
		\textbf{High-fidelity optomechanical ground-state cooling.} 
		\textbf{a},~Frequency landscape for optomechanical cooling and sideband asymmetry measurement. The Stokes and anti-Stokes sideband powers are proportional to $n_\mathrm{m}+2n_\mathrm{c}+1$ and $n_\mathrm{m}-2n_\mathrm{c}$ respectively. 
		\textbf{b},~Simplified experimental setup for optomechanical cooling. 
		\textbf{c},~An example of the PSD of the microwave cavity thermal emission measured at $\mathcal{C}=6400$ on top of the noise floor of $1+n_\mathrm{add}$. 
		\textbf{d},~PSD of optomechanical sidebands scattered from the cooling pump when two probes are off. The black line shows the Lorentzian fit. \textbf{e}~PSD of optomechanical sidebands scattered from red and blue probes. 
		\textbf{f},~Measured occupation of the mechanical (violet dots) and microwave (red dots) modes as a function of the cooperativity when the two probes are off. 
		The shaded area shows theoretically expected phonon occupation in the range of system parameters errors.  
		The inset shows the measured powers of the Stokes (blue) and anti-Stokes (red) sidebands generated, $P_\mathrm{b}$ and $P_\mathrm{r}$, by the probes, exhibiting high quantum efficiency of the readout chain. 
		\label{fig:2}}
\end{figure*}
\subsection*{Ultra-coherent circuit optomechanical platform}
We develop a nanofabrication process based on a silicon-etched trench, which enables us to significantly enhance the mechanical quality factor, $Q_\mathrm{m}$. 
Figure~\ref{fig:1}a shows a vacuum gap capacitor with a top plate suspended on a circular trench with a gap size of 180 nm. The capacitor is shunted by a spiral inductor (Figs.~\ref{fig:1}b and c), forming a microwave LC resonator with a frequency of $\omega_\mathrm{c}/2\pi = 5.55$~GHz and a total decay rate of $\kappa/2\pi = 250$~kHz which is inductively coupled to a waveguide (Fig.~\ref{fig:1}f). This superconducting circuit is operated in a dilution fridge with $\sim11$~mK base temperature. The flat geometry of the top plate (Figs.~\ref{fig:1}d and e) ensures minimal clamp and radiative mechanical losses, as well as stress relaxation in the aluminum thin-film. A mechanical ring-down measurement (Fig.~\ref{fig:1}h) clearly exhibits the extremely low dissipation rate of $\Gamma_\mathrm{m} / 2\pi = 45$~mHz for the fundamental drumhead mode with a frequency of $\Omega_\mathrm{m}/2\pi = 1.8$~MHz, corresponding to $Q_\mathrm{m} = 40\times10^6$. 
This can be explained by the loss dilution factor~\cite{schmid2011damping} estimated to be $D_Q \simeq 100$ from finite element method (FEM) simulation for such a flat drumhead (Fig.~\ref{fig:1}h top inset). The single-photon optomechanical coupling rate is measured to be $g_0/2\pi = 13.4\pm0.5$~Hz (see SI). We note that lower gap sizes lead to higher $g_0$ values, but not implemented in this work. Furthermore, the frequency fluctuation is observed below 0.1 Hz - inferred as an upper bound for dephasing- by measuring the power spectral density (PSD) of a thermomechanical sideband averaged over more than an hour and subtracting a measurement resolution bandwidth of 1 Hz (Fig.~\ref{fig:1}h bottom inset, see SI for more information).

\subsection*{High-fidelity optomechanical ground state cooling}
\begin{figure*}[t!]
	\includegraphics[width=\textwidth]{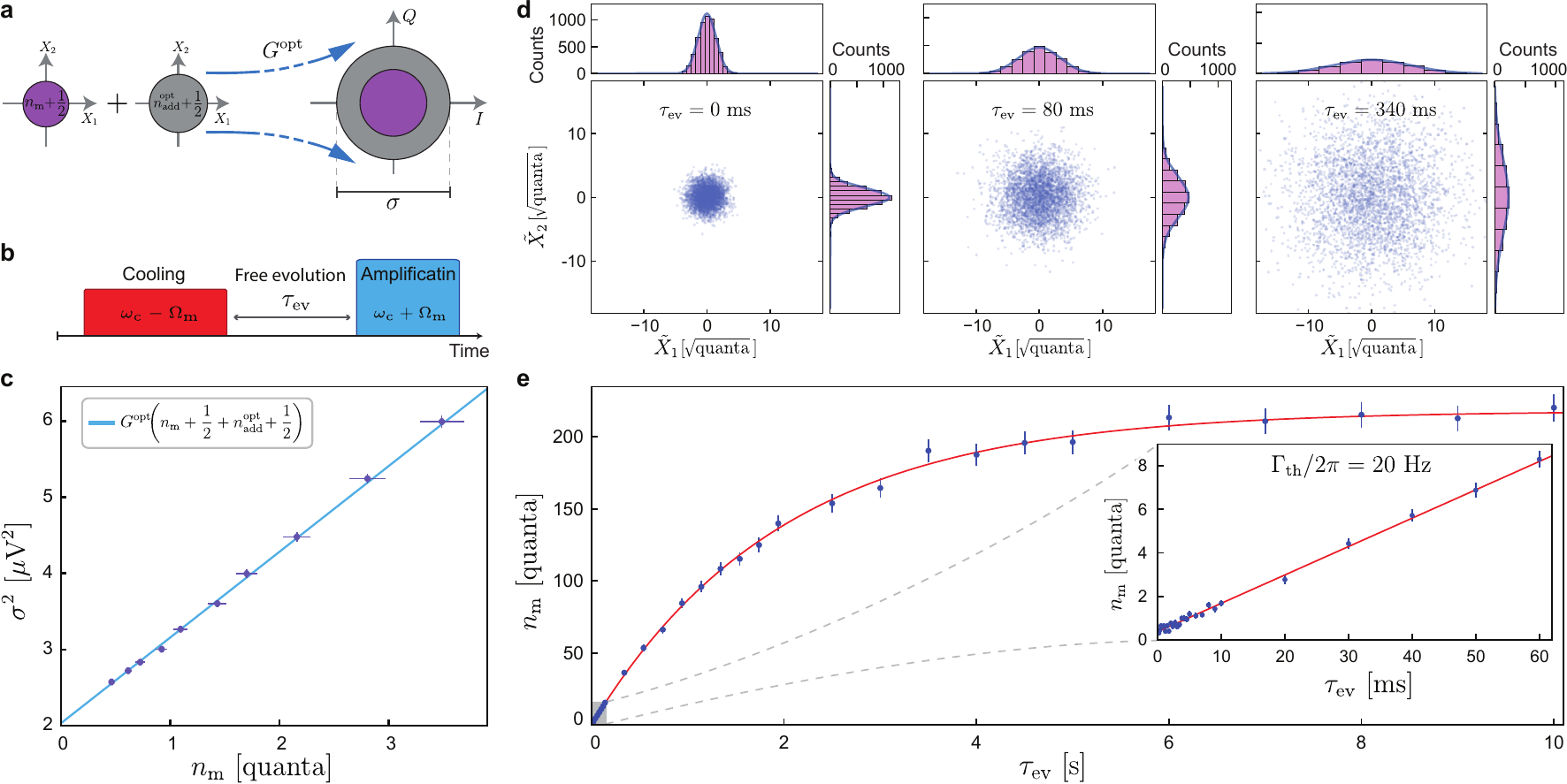}
	\caption{\footnotesize \linespread{1}
		\textbf{Recording the motional heating rate out of the quantum ground state.} 
		\textbf{a},~Schematic diagram showing optomechanical amplification process, where an initial mechanical state is amplified by $G^\mathrm{opt}$ with an added noise of $\frac{1}{2}+n_\mathrm{add}^\mathrm{opt}$ referred to input. 
		\textbf{b},~Pulse sequence for observing the thermalization of the ground state. 
		\textbf{c},~Quadrature variance of the optomechanical sideband signal, $\sigma^2$, as a function of the prepared phonon occupation that is calibrated by the sideband asymmetry measurement, enabling us to reliably determine $G^\mathrm{opt}$ and $n_\mathrm{add}^\mathrm{opt}$. 
		\textbf{d},~Scatter plots of the measured quadratures of motion normalized by $\sqrt{G^\mathrm{opt}}$ for different evolution times. The marginal histogram for each quadrature is shown by purple bars.
		A Gaussian curve with the calculated variance is shown by thick blue lines, reproducing each histogram.
		\textbf{e},~Extracted mechanical occupations as a function of the free evolution time. The blue dots are experimental data with error bars while the red line is the exponential fit. The right inset shows the same experiment for a short evolution time range, exhibiting a thermal decoherence rate of $\Gamma_\mathrm{th}/2\pi=20.5\pm0.6$ Hz obtained from the linear fit.}
	\label{fig:3}
\end{figure*}
The extremely high mechanical quality factor, together with the sufficient optomechanical coupling, enables us to perform an effective optomechanical sideband cooling~\cite{teufel2011sideband} to prepare the mechanical oscillators in its quantum ground state with high fidelity. 
As schematically shown in Fig.~\ref{fig:1}g, in the resolved-sideband regime, where the quantum back-action does not influence the final phonon occupation~\cite{RMP_optomechanics}, i.e., $(\kappa/4\Omega_\mathrm{m})^2=0.001\ll1$ in our case, the phonon occupation of the mechanical oscillator in the presence of a cooling pump red-detuned by $\Omega_\mathrm{m}$ from the cavity frequency is given by
\begin{equation}
	n_\mathrm{m} = \frac{n_\mathrm{m}^\mathrm{th}}{1+\mathcal{C}} + \frac{\mathcal{C}}{1+\mathcal{C}}  n_\mathrm{c},
	\label{eq:cooling}
\end{equation}
where $\mathcal{C} = 4n_\mathrm{p} g_0^2/(\kappa\Gamma_m)$ is the optomechanical cooperativity with the intracavity pump photon number $n_\mathrm{p}$ and $n_\mathrm{c} = \kappa_0n_\mathrm{c}^\mathrm{th}/\kappa$ is the cavity thermal photon number induced by a finite loss rate of $\kappa_0$ to an intrinsic photon bath with $n_\mathrm{c}^\mathrm{th}$.
The strong cooling pump may heat up the intrinsic photon bath occupation and consequently $n_\mathrm{c}$, which normally imposes the minimal achievable phonon occupancy in the large cooperativity limit, i.e., $\mathcal{C}\gg1$~\cite{teufel2011sideband}. We discovered that the thin native oxide layer in the galvanic connection between top and bottom layers (shown in Fig.~\ref{fig:1}c) is the dominant source of such cavity heating, which has been ubiquitous in all microwave optomechanical experiments. We significantly reduced the heating by removing the oxide to achieve $n_\mathrm{c}\sim 0.05$~quanta at high cooperativities (see SI for details).
\begin{figure*}[t!]
	\centering
	\includegraphics[width=\textwidth]{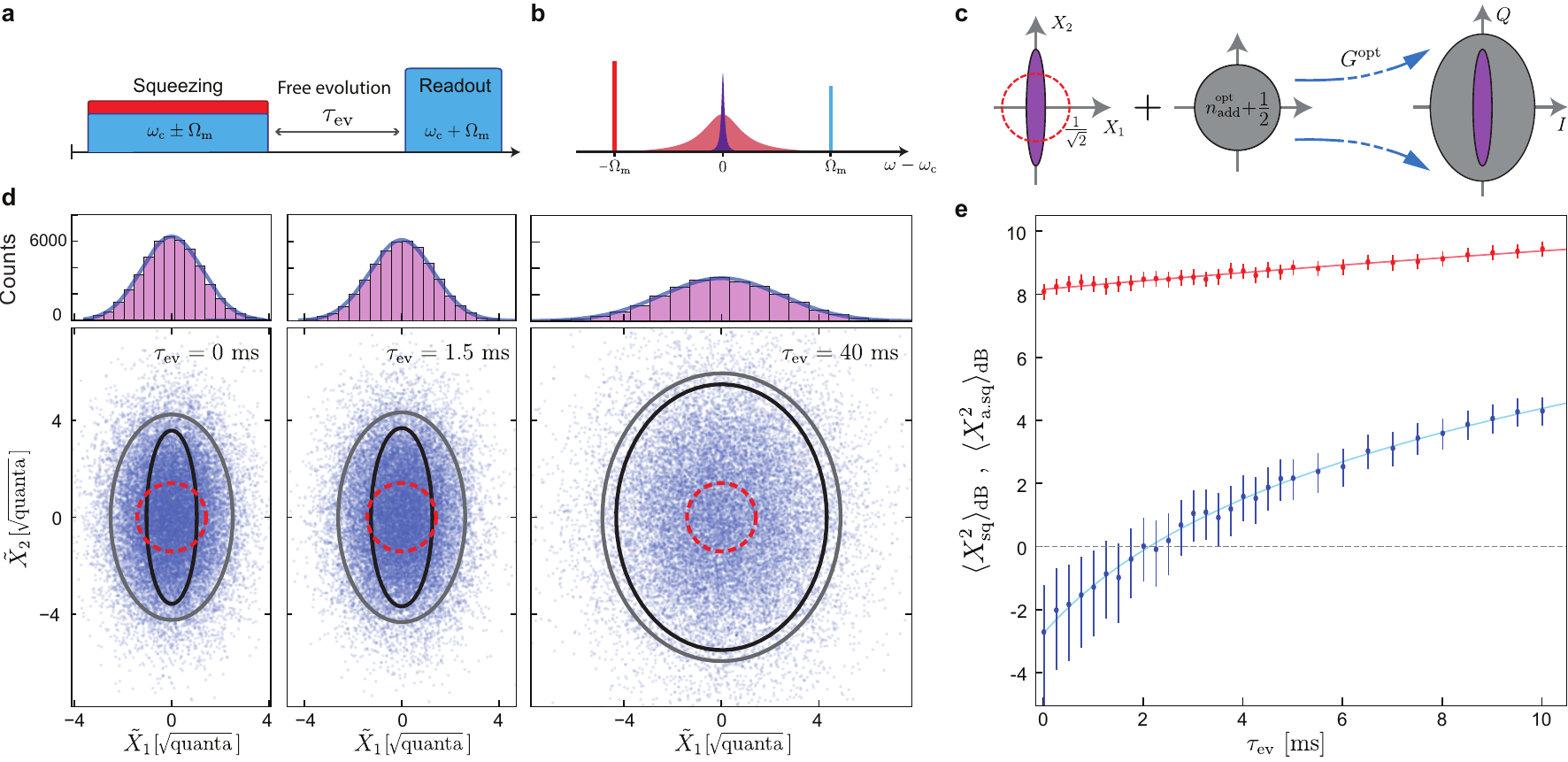}
	\caption{\footnotesize \linespread{1}
		\textbf{Tracking the free evolution of a mechanical squeezed state.}
		\textbf{a},~Pulse sequence for recording the free evolution of a squeezed state.
		\textbf{b},~Frequency landscape of optomechanical dissipative squeezing.
		\textbf{c},~Schematic diagram showing optomechanical amplification process of an initially prepared squeezed state.
		\textbf{d},~The measured quadrature scatter plots for initially prepared squeezed state and its time evolution. The gray and black ellipses represent standard deviation contours scaled by factor of 2 (for better visualization) of the Gaussian density function before (measured state) and after (inferred squeezed state) noise subtraction. Dashed red circles indicates the same for the zero-point-fluctuation (ideal ground-state). 
		\textbf{e},~Variance of the squeezed (blue) and anti-squeezed (red) quadratures, relative to the zero-point-fluctuation, as a function of the free evolution time. The initial state shows $-2.7^{+1.4}_{-2.3}$~dB squeezing and $8.1^{+0.3}_{-0.3}$ dB anti-squeezing. The blue and red solid lines are the linear fits, revealing a slight difference in decoherence rates of two quadratures, corresponding to a pure dephasing rate of $\Gamma_\varphi/2\pi =0.09 (\pm0.05)$~Hz.}
	\label{fig:4}
\end{figure*}
To reliably characterize the phonon occupation close to the ground state, we use optomechanical sideband asymmetry~\cite{weinstein2014observation} as an out-of-loop calibration. 
As shown in Fig.~\ref{fig:2}a, we apply a strong cooling pump, and two weak, blue- and red-detuned, probes with balanced powers to generate Stokes and anti-Stokes optomechanical sidebands, respectively, on the cavity resonance with a few kHz spacing to individually measure them. 
Figure~\ref{fig:2}b shows the simplified experimental setup, where a Josephson traveling wave parametric amplifier (JTWPA)~\cite{macklin2015near} is used to amplify microwave signals with an added noise of $\frac{1}{2}+n_\mathrm{add}^\mathrm{T}\simeq\frac{1}{2}+0.3$~quanta and sufficient gain of $G_\mathrm{T} = 25$~dB to suppress the classical noise dominated by the HEMT amplifier ($n_\mathrm{add}^\mathrm{H}\simeq8$), enabling a nearly quantum-limited measurement of thermomechanical noise spectrum with an effective added noise of $\frac{1}{2}+n_\mathrm{add}\simeq\frac{1}{2}+0.9$ (see the calibrated noise floor in Fig.~\ref{fig:2}c).
Figures~\ref{fig:2}c and e show the measured thermomechanical noise spectrum of the cavity thermal emission, as well as the Stokes and anti-Stokes sidebands, which used for obtaining their powers, expressed by $P_\mathrm{c}$, $P_\mathrm{b}$, and $P_\mathrm{r}$, respectively, by fitting a Lorentzian to the PSD of the sidebands. 
While the sideband asymmetry may allow us to perform the calibration-free measurement of $n_\mathrm{m}$, a finite cavity heating distorts the asymmetry, i.e., $P_\mathrm{b} \not \propto n_\mathrm{m} + 1$ and $P_\mathrm{r} \not \propto n_\mathrm{m}$, preventing us from extracting $n_\mathrm{m}$ without the prior knowledge of $n_\mathrm{c}$~\cite{weinstein2014observation}.
Nevertheless, we are able to simultaneously extract both $n_\mathrm{c}$ and $n_\mathrm{m}$ without any calibration of the measurement chain by analytically obtaining them from the two sideband powers normalized by the cavity thermal emission power, expressed by
\begin{equation}
	\frac{P_\mathrm{b}}{P_\mathrm{c}} = \frac{\Gamma_\mathrm{b}}{\kappa} \frac{n_\mathrm{m} +1+ 2n_\mathrm{c}}{n_\mathrm{c}}  \ \ \text{and}\ \ \frac{P_\mathrm{r}}{P_\mathrm{c}} = \frac{\Gamma_\mathrm{r}}{\kappa} \frac{n_\mathrm{m} - 2n_\mathrm{c}}{n_\mathrm{c}},
	\label{eq:nm_asym}
\end{equation}
where $\Gamma_\mathrm{r} (\Gamma_\mathrm{b})$ is the optomechanical (anti-)damping rate of the red (blue) probe.
Importantly, this analysis enables us to calibrate the scaling factor between the actual occupations and the measured powers, which can be used to directly extract $n_\mathrm{c}$ and $n_\mathrm{m}$  independently from the cavity thermal emission and the sideband induced by the cooling pump, even when the two probes are off -- therefore avoiding the quantum back-action induced by the blue probe and an additional cavity heating (see SI for more information). 
Using the PSDs of the thermomechanical sideband from the cooling pump (Fig.~\ref{fig:2}d) and the cavity thermal emission when two probes are off, we thus extract $n_\mathrm{c}$ and $n_\mathrm{m}$ as a function of the cooling pump cooperativity, as shown in Fig.~\ref{fig:2}f.
The result shows a high-fidelity ground state cooling down to $n_\mathrm{m}^{(\mathrm{min})}=6.8(\pm0.9)\times10^{-2}$~quanta (93\% ground state occupation which is $-8.7$~dB of the zero-point energy), mainly limited by the cavity heating.

\subsection*{Measurement of motional heating rate}

Next, we directly measure the thermal decoherence by recording the thermalization of the mechanical oscillator out of the ground state using a time-domain protocol~(Fig.~\ref{fig:3}b), where we first prepare the ground state, and leave the system to freely evolve for a certain time of $\tau_\mathrm{ev}$.
Using optomechanical amplification technique with a blue-detuned pump~\cite{palomaki2013entangling,delaney2019measurement}, we intrinsically amplify the mechanical motion by $\sim50$~dB with a minimal added noise, and measure both quadratures of motion encoded in a generated optomechanical sideband signal~(Fig.~\ref{fig:3}a). 
Repeating this pulse sequence allows us to capture the quadrature distribution of the mechanical state, realizing quantum-state tomography (Fig.~\ref{fig:3}d).
As shown in Fig.~\ref{fig:3}c, we are able to precisely calibrate the amplification process using different phonon occupations as an input state, which are well-calibrated by the sideband asymmetry measurement, resulting in $n^\mathrm{opt}_\mathrm{add} = 0.80\pm0.09$~quanta.

Figure \ref{fig:3}d shows examples of the measured quadrature distributions at different evolution times in units of $\left[\sqrt{\text{quanta}}\right]$.
Figure \ref{fig:3}e shows the free evolution from the ground state to the thermal equilibrium. 
The exponential fit results in a bare dissipation rate of $\Gamma_\mathrm{m}/2\pi= 80$~mHz in the low phonon occupation regime, close to the value measured from the ring-down experiment with $n_\mathrm{m} > 10^7$ (Fig.~\ref{fig:1}h). 
The right inset of the Fig.~\ref{fig:3}~e shows the thermalization in shorter evolution times, where the thermal decoherence rate is directly measured as $\Gamma_\mathrm{th} /2\pi= 20.5\pm0.6$~Hz, corresponding to a phonon lifetime of $T_1 = 7.7$~ms. 

\subsection*{Recording thermalization of squeezed mechanical state}
Finally, we generate a quantum squeezed state of our oscillator. Since squeezed state is a phase-sensitive quantum state, its free evolution is subject to the dephasing in the system. Tracking its time evolution enables us to directly measure the quantum-state lifetime and verify minimal dephasing in our mechanical oscillator.
The ability to squeeze the	mechanical oscillator critically relies on the residual thermal occupation upon cooling, which in our case is below 0.1~quanta, implying that strong squeezing below zero-point-fluctuation is possible.	
We use optomechanical dissipative squeezing technique~\cite{kronwald2013arbitrarily,wollman2015quantum,pirkkalainen2015squeezing,lecocq2015quantum} by simultaneously applying two red- and blue-detuned pumps symmetrically with respect to the cavity frequency (Fig.~\ref{fig:4}b), and achieve $\langle X_\mathrm{sq}^2\rangle / \frac{1}{2} = -2.7^{+1.4}_{-2.3}$~dB squeezing in one quadrature of motion below the vacuum fluctuation and $\langle X_\mathrm{a.sq}^2\rangle/ \frac{1}{2} = 8.1^{+0.3}_{-0.3}$~dB anti-squeezing in the other quadrature. These are obtained by subtracting the accurately calibrated $n^\mathrm{opt}_\mathrm{add}$ in the optomechanical amplification (Figs.~\ref{fig:4} a and c).

Figure~\ref{fig:4}d shows measured quadrature scatter plots of the prepared squeezed state and its time evolution. 
We are able to record the free evolution of a prepared squeezed state (Fig.~\ref{fig:4}d) and observe the decoherence of both the quadratures to the thermal equilibrium (Fig.~\ref{fig:4}e). A slight difference is observed in the decoherence rates of the two quadratures, $(\Gamma^\mathrm{sq}-\Gamma^\mathrm{a.sq}) /2\pi= 1.1(\pm0.6)$~Hz. Comparing it with a numerical simulation allows us to characterize the dephasing rate of $\Gamma_\varphi/2\pi =0.09 (\pm0.05)$~Hz in our platform (see SI), in agreement with the measured frequency fluctuation discussed earlier.
We observe that the variance of the squeezed quadrature remains below the zero-point-fluctuation up to 2 ms, demonstrating a significantly long quantum state storage time in a macroscopic mechanical oscillator.

\subsection*{Conclusion and outlook}
The high-fidelity quantum control and measurement of mechanical oscillators with such extremely low thermal decoherence and pure dephasing rates may benefit the implementation of qubit-mechanics interfaces~\cite{reed2017faithful}, generation of mechanical non-classical states~\cite{gely2021phonon}, and can realize long life-time memories for quantum computation and communication~\cite{wallucks2020quantum,pechal2018superconducting}. Furthermore, such a low quantum decoherence sets the stage to perform fundamental tests of quantum mechanics in macroscopic scales such as quantum gravity tests~\cite{liu2021gravitational,gely2021superconducting}, as well as high fidelity Bell tests~\cite{marinkovic2018optomechanical}, and quantum teleportation~\cite{fiaschi2021optomechanical}.  

\subsection*{Acknowledgment}
We thank MIT Lincoln Laboratory and Prof. William D. Oliver for providing the JTWPA. We thank A. Arabmoheghi for helpful discussions on the theory of mechanical dissipation.
This work was supported by the EU H2020 research and innovation programme under grant No. 101033361 (QuPhon), and from the European Research Council (ERC) grant No. 835329 (ExCOM-cCEO). This work was also supported by the Swiss National Science Foundation (SNSF) under grant No. NCCR-QSIT: 51NF40\_185902 and No. 204927.
All devices were fabricated in the Center of MicroNanoTechnology (CMi) at EPFL.
\subsection*{Author contributions}
A.Y. conceived the experiment. S.K., M.C., and A.Y. developed the theory. A.Y. designed and simulated devices. A.Y. developed the fabrication process with the assistance of M.C. M.C. and A.Y. fabricated the samples. A.Y. and M.C. developed the experimental setup. The measurement was performed by A.Y. and M.C., with the assistance of S.K. The data analysis was performed by A.Y with the assistance of S.K. S.K. introduced the phonon number calibration based on sideband asymmetry and conducted the numerical simulation for extracting the mechanical dephasing. The manuscript was written by A.Y., S.K., M.C., and T.J.K. T.J.K. supervised the project.
\subsection*{Data and codes availability}
The data and codes used to produce the plots within this paper are available on Zenodo\\ (\href{https://doi.org/10.5281/zenodo.7833893}{https://doi.org/10.5281/zenodo.7833893}). All other data used in this study are available from the corresponding author on reasonable request.

\let\oldaddcontentsline\addcontentsline
\renewcommand{\addcontentsline}[3]{}

\let\addcontentsline\oldaddcontentsline

\clearpage
\onecolumngrid

\begin{center}
	\large{\textbf{Supplementary Information for: A squeezed mechanical oscillator with milli-second quantum decoherence}}
\end{center}
\begin{center}
	Amir Youssefi$^{*}$, Shingo Kono$^{*}$, Mahdi Chegnizadeh$^{*}$, and Tobias J. Kippenberg$^{\dagger}$\\[.1cm]
	{\itshape
		Laboratory of Photonics and Quantum Measurement, Swiss Federal Institute of Technology Lausanne (EPFL), Lausanne, Switzerland\\
	}
	$^\dagger$Electronic address: tobias.kippenberg@epfl.ch\\
\end{center}

\setcounter{equation}{0}
\renewcommand{\thefigure}{S\arabic{figure}}
\renewcommand{\theHfigure}{S\arabic{figure}}
\setcounter{figure}{0}
\setcounter{table}{0}

\setcounter{subsection}{0}
\setcounter{section}{0}

\tableofcontents

\pagebreak

\section{System parameters and variables}

\begin{table}[h]
	\caption{System parameters}
	\label{SI:glassory}
	\begin{center}
		\begin{tabular}{| l  | c | l |} \hline
			\multicolumn{1}{|c|}{\textbf{Parameter}}  & 
			\textbf{Symbol} & 
			\multicolumn{1}{|c|}{\textbf{Value}} \\ \hline\hline
			
			Microwave cavity frequency  &
			$\omega_\mathrm{c}$ & $2\pi \; \cdot$ 5.5 GHz \\ \hline
			
			Microwave cavity linewidth $\left(\kappa_\mathrm{ex}+\kappa_\mathrm{0} \right)$ &
			$\kappa$ & $2\pi \; \cdot$ 250 kHz \\ \hline
			
			Microwave cavity external coupling rate &
			$\kappa_\mathrm{ex}$ & $\sim \ 2\pi \; \cdot$ 200 kHz \\ \hline
			
			Microwave cavity  internal loss rate &
			$\kappa_\mathrm{0}$ & $\sim \ 2\pi \; \cdot$ 50 kHz \\ \hline

			Mechanical frequency &
			$\Omega_\mathrm{m}$ & $2\pi  \; \cdot$ 1.8 MHz \\ \hline
			
			Mechanical bare damping rate&
			$\Gamma_\mathrm{m}$ & $2\pi  \; \cdot$ 45 mHz \\ \hline
			
			Mechanical quality factor&
			$Q_\mathrm{m}$ & 40 $\times 10^6$ \\ \hline
			
			Single-photon optomechanical coupling rate  &
			$g_0$ & $2\pi  \; \cdot$ 13.4 Hz \\ \hline			
			
			Thermal decoherence rate of  mechanical oscillator&
			$\Gamma_\mathrm{th}$ & $2\pi  \; \cdot 20.5$~Hz  \\ \hline
			
			Pure dephasing rate of mechanical oscillator&
			$\Gamma_{\varphi}$ & $2\pi  \; \cdot 0.09$~Hz \\ \hline
			
			%
			
		\end{tabular}
	\end{center}
\end{table}

\begin{table}[h]
	\caption{Variables}
	\label{SI:glassory}
	\begin{center}
		\begin{tabular}{| l  | c |} \hline
			\multicolumn{1}{|c|}{\textbf{Variables}}  & 
			\textbf{Symbol} \\ \hline\hline
			
			Annihilation operator for microwave cavity &
			$\hat a$  \\ \hline					
			
			Annihilation operator for mechanical oscillator&
			$\hat b$ \\ \hline					
			
			Noise operator for microwave intrinsic bath &
			$\hat a_0^{\rm{in}} $ \\ \hline									
			
			Noise operator for microwave external bath&
			$\hat a_{\rm{ex}}^{\rm{in}} $  \\ \hline									
			
			Noise operator for mechanical intrinsic bath&
			$\hat b_0^{\rm{in}} $  \\ \hline			
			
			Microwave cavity thermal bath occupation  &
			$n_\mathrm{c}^\mathrm{th}$\\ \hline	
			
			Microwave cavity thermal occupation  &
			$n_\mathrm{c}$ \\ \hline	
			
			Mechanical thermal bath occupation  &
			$n_\mathrm{m}^\mathrm{th}$ \\ \hline			
			
			Mechanical occupation  &
			$n_\mathrm{m}$  \\ \hline	
			
			Intracavity photon number induced by cooling pump&
			$n_\mathrm{p}$ \\ \hline			
			
			
			
			Optomechanical coupling rate induced by cooling pump, red probe, or blue probe&
			$g_{\mathrm{p,r,b}}$\\ \hline

			Optomechanical (anti-)damping rate induced by the (blue probe,) cooling pump, red probe $\left(4 g_{\mathrm{b,p,r}}^2/\kappa\right)$&
			$\Gamma_\mathrm{opt}^\mathrm{b,p,r}$ \\ \hline
			
			Total damping rate of the mechanical oscillator $\left({\Gamma _{\rm{m}}} + \Gamma _{{\rm{opt}}}^{\rm{p}} + \Gamma _{{\rm{opt}}}^{\rm{r}} - \Gamma _{{\rm{opt}}}^{\rm{b}}\right)$&
			$\Gamma_\mathrm{tot}$ \\ \hline

			Optomechanical cooperativity $\left(\frac{4g_0^2 }{\kappa\Gamma_\mathrm{m}}n_\mathrm{p}\right)$  &
			$\mathcal{C}$ \\ \hline
			
			Quadratures of filtered microwave field &
			$I,Q$ \\ \hline			
			
			Inferred quadratures of mechanical oscillator &
			$X_{1,2}$  \\ \hline
			
			Measured quadratures of mechanical oscillator, including added noise $n_\mathrm{add}^\mathrm{opt}$ &
			${{\tilde X}_{1,2}}$  \\ \hline			
			
			Total gain of microwave measurement chain&
			$G$ \\ \hline	
			
			Total added noise of microwave measurement chain &
			$n_{\rm{add}}$ \\ \hline	
			
			Conversion factor of optomechanical amplification $\left[\frac{\mu V^2}{\mathrm{quanta}}\right]$ &
			$G^\mathrm{opt}$ \\ \hline	
			
			Total added noise of optomechanical amplification &
			$n^\mathrm{opt}_{\rm{add}}$ \\ \hline
			HEMT effective added noise &
			$n_{\rm{add}}^{\rm{H}}$  \\ \hline	
			
			JTWPA added noise &
			$n_{\rm{add}}^{\rm{T}}$  \\ \hline

		\end{tabular}
	\end{center}
\end{table}
\newpage


\section{Theory}
In this section, we provide the detailed theory for the continuous-wave measurement (sideband asymmetry) and time-domain measurement (optomechanical amplification). 

\subsection{Sideband asymmetry}
Here we discuss the theory of optomechanical sideband cooling in the presence of two additional probes used in sideband asymmetry experiment.
As schematically shown in Fig.~\ref{fig:SI_SB_asym}, to extract the mechanical occupation in our experiment, we pump our device with three microwave tones simultaneously: cooling pump, red probe, and blue probe. The cooling pump is red-detuned from the cavity and has relatively higher power compared to those of the red and blue probes. The red and blue probes have a balanced power in a way that their effective dynamical back-actions are canceled out each other.

The Langevin equations of an optomechanical system in the presence of the three microwave tones  are given by
\begin{equation}
	\begin{aligned}
		\frac{{d\hat a}}{{dt}} &=  - \frac{\kappa }{2}\hat a - i{g_0}(\hat b + {{\hat b}^\dag })\hat a + \sqrt {{\kappa _{{\rm{ex}}}}} \left({\alpha _{\mathrm{b}}}{e^{ - i(\Omega_{\rm{m}} + {\delta _{\mathrm{b}}})t}} + {\alpha _{\mathrm{r}}}{e^{ + i(\Omega_{\rm{m}} + {\delta _{\mathrm{r}}})t}} + {\alpha _{\mathrm{p}}}{e^{ + i(\Omega_{\rm{m}} + {\delta _{\mathrm{p}}})t}} + {{\hat a}_{\rm{ex}}^{{\rm{in}}}}(t) \right) + \sqrt {{\kappa _{\rm{0}}}} {{\hat a}_{\rm{0}}^{\rm{in}}(t)},\\
		\frac{{d\hat b}}{{dt}} &=  - i{\Omega_{\rm{m}}}\hat b - \frac{{{\Gamma _{\rm{m}}}}}{2}\hat b - i{g_0}{{\hat a}^\dag }\hat a + \sqrt \Gamma_{\rm{m}} {{\hat b}_0^{{\rm{in}}}(t)},
	\end{aligned}
	\label{eq:lang}
\end{equation}
where, $\hat a (\hat b)$ is the annihilation operator for the microwave cavity (mechanical oscillator), $g_0$ is the single photon optomechanical coupling rate, $\kappa_{\rm{ex}(0)}$ is the microwave external coupling (internal loss) rate, $\Gamma_{\rm{m}}$ is the mechanical intrinsic damping rate, ${{\hat a}_0^{\rm{in}}}({{\hat b}_0^{{\rm{in}}}})$ is the noise operator for the intrinsic loss of the microwave cavity (mechanical oscillator), and $\hat a_{\rm{ex}}^{\rm{in}}$ is the noise operator for the external field. Furthermore, $\alpha_{\rm{p,r,b}}$ denotes the coherent amplitude of the cooling pump with a detuning of $-\Omega_\mathrm{m}-\delta_{\mathrm{p}}$, the red probe with a detuning of  $-\Omega_\mathrm{m}-\delta_{\mathrm{r}}$, and the blue probe with a detuning of $\Omega_\mathrm{m}+\delta_{\mathrm{b}}$, respectively. 
Note that the Langevin equation for the cavity is described in the rotating frame of the cavity frequency.
The noise correlators satisfy
\begin{equation}
	\begin{aligned}
		\left\langle {{{{\hat b}_0^{{\rm{in}}\dag}(t)}} {{{\hat b}_0^{{\rm{in}}}(t')}}} \right\rangle  &= {n_{{\rm{m}}}^{\rm{th}}}\delta (t - t'),\\
		\left\langle {{{\hat b}_{\rm{0}}^{\rm{in}}}(t)\hat b_0^{{\rm{in}}\dag} (t')} \right\rangle  &= ({n_{{\rm{m}}}^{\rm{th}}} + 1)\delta (t - t'),\\
		\left\langle {\hat a_{{\rm{0}}}^{\rm{in}\dagger} (t){{\hat a}_0^{\rm{in}}}(t')} \right\rangle  &= {{ n}_{\rm{c}}^{\rm{th}}}\delta (t - t'),\\
		\left\langle {{{\hat a}_0^{\rm{in}}}(t)\hat a_{{\rm{0}}}^{\rm{in}\dag} (t')} \right\rangle  &= ({n_{\rm{c}}^{\rm{th}}} + 1)\delta (t - t'),\\
		\left\langle {{\hat a_{\rm{ex}}}^{\rm{in}\dag}(t){{\hat a}_{\rm{ex}}^{\rm{in}}}(t')} \right\rangle  &= 0,\\
		\left\langle {{{\hat a}_{\rm{ex}}^{\rm{in}}}(t){{\hat a}_{\rm{ex}}^{\rm{in}\dag}}(t')} \right\rangle  &= \delta (t - t'),
	\end{aligned}
	\label{eq:noise}
\end{equation}
where $n_{\rm{m}}^{\rm{th}} (n_{\rm{c}}^{\rm{th}})$ is the mechanical (cavity) thermal bath occupation. 
Due to the sufficient attenuation in the input line, we can assume that the input noise from the waveguide corresponds to the vacuum noise, regardless of the amount of input pump power. 
This can be experimentally confirmed by the fact that there is no significant dip on the noise floor of the power spectral density at the cavity frequency, showing that the effective temperature of the external bath is equal to the intrinsic bath temperature that can be safely assumed to be zero when a strong pump field is not applied. Furthermore, there is no significant increase in the noise floor even when stronger pump fields are applied.

To solve the equations, we first divide the cavity field into coherent amplitudes and fluctuation, i.e., $\hat a(t) \to {{\bar a}_{\mathrm{b}}}{e^{ - i(\Omega_{\rm{m}} + {\delta _{\mathrm{b}}})t}} + {{\bar a}_{\mathrm{r}}}{e^{ + i(\Omega_{\rm{m}} + {\delta _{\mathrm{r}}})t}} + {{\bar a}_{\mathrm{p}}}{e^{ + i(\Omega_{\rm{m}} + {\delta _{\mathrm{p}}})t}} + \hat a(t)$, where $\bar a_{\mathrm{p,r,b}}$ is the coherent amplitude induced by the cooling pump, red probe, and blue probe, respectively, and $\hat a(t)$ is the fluctuation. Considering sufficiently small $g_0$ compared to the other cavity parameters, we can find approximated solutions for the coherent amplitudes:
\begin{equation}
	\begin{aligned}
		{{\bar a}_{\mathrm{b}}} &\simeq \frac{{\sqrt {{\kappa _{{\rm{ext}}}}} }}{{ - i(\Omega_{\rm{m}}  + {\delta _{\mathrm{b}}}) + \kappa /2}}{\alpha _{\mathrm{b}}},\\
		{{\bar a}_{\mathrm{r,p}}} &\simeq \frac{{\sqrt {{\kappa _{{\rm{ext}}}}} }}{{ + i(\Omega_{\rm{m}} + {\delta _{\rm{r,p}}}) + \kappa /2}}{\alpha _{\mathrm{r,p}}},
	\end{aligned}
\end{equation}
where we can assume that the coherent amplitudes are real without loss of generality.

By linearizing the nonlinear optomechanical coupling terms in the Langevin equations (Eq. (\ref{eq:lang})) around the coherent amplitudes \added{and going into the rotating frame of the mechanical oscillator ($\hat b \to \hat b e^{-i\Omega_{\rm{m}}t}$)}, we have
\begin{equation}
	\begin{aligned}
		- i{g_0}{{\hat a}^\dag }\hat a = & - i{g_0}\left\{ {\left( {\bar a_{\mathrm{r}}^*{e^{ - i(\Omega_{\rm{m}} + {\delta _{\mathrm{r}}})t}} + \bar a_{\mathrm{p}}^*{e^{ - i(\Omega_{\rm{m}} + {\delta _{\mathrm{p}}})t}}} \right)\hat a + \left( {{{\bar a}_{\mathrm{b}}}{e^{ - i(\Omega_{\rm{m}} + {\delta _{\mathrm{b}}})t}}} \right){{\hat a}^\dag }} \right\}\\
		& - i{g_0}\left\{ {\left( {{{\bar a}_{\mathrm{r}}}{e^{ + i(\Omega_{\rm{m}} + {\delta _{\mathrm{r}}})t}} + {{\bar a}_{\mathrm{p}}}{e^{ + i(\Omega_{\rm{m}} + {\delta _{\mathrm{p}}})t}}} \right){{\hat a}^\dag } + \left( {\bar a_{\mathrm{b}}^*{e^{ + i(\Omega_{\rm{m}} + {\delta _{\mathrm{b}}})t}}} \right)\hat a} \right\},\\
		- i{g_0}(\hat b + {{\hat b}^\dag })\hat a =  &- i{g_0}\left\{ {\left( {{{\bar a}_{\mathrm{r}}}{e^{ + i{\delta _{\mathrm{r}}}t}} + {{\bar a}_{\mathrm{p}}}{e^{ + i{\delta _{\mathrm{p}}}t}}} \right)\hat b + \left( {{{\bar a}_{\mathrm{b}}}{e^{ - i{\delta _{\mathrm{b}}}t}}} \right){{\hat b}^\dag }} \right\}\\
		&- i{g_0}\left\{ {\left( {{{\bar a}_{\mathrm{b}}}{e^{ - i(2\Omega_{\rm{m}}  + {\delta _{\mathrm{b}}})t}}} \right)\hat b + \left( {{{\bar a}_{\mathrm{r}}}{e^{ + i(2\Omega_{\rm{m}}  + {\delta _{\mathrm{r}}})t}} + {{\bar a}_{\mathrm{p}}}{e^{ + i(2\Omega_{\rm{m}}  + {\delta _{\mathrm{p}}})t}}} \right){{\hat b}^\dag }} \right\}.
	\end{aligned}
	\label{eq:nonlin}
\end{equation}
By neglecting the fast oscillating terms based on rotating wave approximation, Eq. (\ref{eq:nonlin}) can be simplified as
\begin{equation}
	\begin{aligned}
		- i{g_0}{{\hat a}^\dag }\hat a  & \simeq  - i{g_0}\left\{ {\left( {\bar a_{\mathrm{r}}^*{e^{ - i(\Omega_{\rm{m}} + {\delta _{\mathrm{r}}})t}} + \bar a_{\mathrm{p}}^*{e^{ - i(\Omega_{\rm{m}} + {\delta _{\mathrm{p}}})t}}} \right)\hat a + \left( {{{\bar a}_{\mathrm{b}}}{e^{ - i(\Omega_{\rm{m}} + {\delta _{\mathrm{b}}})t}}} \right){{\hat a}^\dag }} \right\},\\
		- i{g_0}(\hat b + {{\hat b}^\dag })\hat a &\simeq  - i{g_0}\left\{ {\left( {{{\bar a}_{\mathrm{r}}}{e^{ + i{\delta _{\mathrm{r}}}t}} + {{\bar a}_{\mathrm{p}}}{e^{ + i{\delta _{\mathrm{p}}}t}}} \right)\hat b + \left( {{{\bar a}_{\mathrm{b}}}{e^{ - i{\delta _{\mathrm{b}}}t}}} \right){{\hat b}^\dag }} \right\}.
	\end{aligned}
	\label{eq:lin}
\end{equation}
Thus, we have the linearized Langevin equations:
\begin{equation}
	\begin{aligned}
		\frac{{d\hat a}}{{dt}} &=  - \frac{\kappa }{2}\hat a - i\left\{ {\left( {{g_{\mathrm{r}}}{e^{ + i{\delta _{\mathrm{r}}}t}} + {g_{\mathrm{p}}}{e^{ + i{\delta _{\mathrm{p}}}t}}} \right)\hat b + \left( {{g_{\mathrm{b}}}{e^{ - i{\delta _{\mathrm{b}}}t}}} \right){{\hat b}^\dag }} \right\} + \sqrt {{\kappa _{{\rm{ex}}}}} {{\hat a}_{\rm{ex}}^{\rm{in}}}(t) + \sqrt {{\kappa _{\rm{0}}}} {{\hat a}_0^{\rm{in}}}(t),\\
		\frac{{d\hat b}}{{dt}} &=  - \frac{{{\Gamma _{\rm{m}}}}}{2}\hat b - i\left\{ {\left( {{g_{\mathrm{r}}}{e^{ - i{\delta _{\mathrm{r}}}t}} + {g_{\mathrm{p}}}{e^{ - i{\delta _{\mathrm{p}}}t}}} \right)\hat a + \left( {{g_{\mathrm{b}}}{e^{ - i{\delta _{\mathrm{b}}}t}}} \right){{\hat a}^\dag }} \right\} + \sqrt {{\Gamma _{\rm{m}}}} {{\hat b}_{\rm{0}}^{\rm{in}}}(t),
	\end{aligned}
	\label{eq:lang_lin}
\end{equation}
where $g_{\mathrm{p,r,b}}=g_0\bar a_{\mathrm{p,r,b}}$ is the linearized optomechanical coupling rate. 
Then, we take the Fourier transform of the time derivative equations, leading to
\begin{equation}
	\begin{aligned}
		- i\omega \hat a(\omega ) &=  - \frac{\kappa }{2}\hat a(\omega ) - i\left\{ {\left( {{g_{\mathrm{r}}}\hat b(\omega  + {\delta _{\mathrm{r}}}) + {g_{\mathrm{p}}}\hat b(\omega  + {\delta _{\mathrm{p}}})} \right) + \left( {{g_{\mathrm{b}}}{{\hat b}^\dag }( - (\omega  - {\delta _{\mathrm{b}}}))} \right)} \right\} + \sqrt {{\kappa _{{\rm{ex}}}}} {{\hat a}_{\rm{ex}}^{\rm{in}}}(\omega ) + \sqrt {{\kappa _{\rm{0}}}} {{\hat a}_0^{\rm{in}}}(\omega ),\\
		- i\omega \hat b(\omega ) &=  - \frac{{{\Gamma _{\rm{m}}}}}{2}\hat b(\omega ) - i\left\{ {\left( {{g_{\mathrm{r}}}\hat a(\omega  - {\delta _{\mathrm{r}}}) + {g_{\mathrm{p}}}\hat a(\omega  - {\delta _{\mathrm{p}}})} \right) + \left( {{g_{\mathrm{b}}}{{\hat a}^\dag }( - (\omega  - {\delta _{\mathrm{b}}}))} \right)} \right\} + \sqrt {{\Gamma _{\rm{m}}}} {{\hat b}_{\rm{0}}^{\rm{in}}}(\omega ),
	\end{aligned}
	\label{eq:fourier_main}
\end{equation}
where
\begin{equation}
	\begin{aligned}
		\hat b(\omega ) \equiv &\mathcal{F}\{ \hat b(t)\}  = \frac{1}{{\sqrt {2\pi } }}\int_{ - \infty }^{ + \infty } {\hat b(t){e^{i\omega t}}dt}, \\
		\hat b(t) = &{\mathcal{F}^{ - 1}}\{ \hat b(\omega )\}  = \frac{1}{{\sqrt {2\pi } }}\int_{ - \infty }^{ + \infty } {\hat b(\omega ){e^{ - i\omega t}}d\omega } ,\\
		{{\hat b}^\dag }(\omega ) = &{\left( {\mathcal{F}\{ \hat b(t)\} } \right)^\dag }.
	\end{aligned}
	\label{eq:lang_out}
\end{equation}

							By substituting $\hat a(\omega)$ into $\hat b(\omega)$ in Eq. (\ref{eq:fourier_main}), we obtain
							\begin{equation}
								\begin{aligned}
									\frac{{\hat b(\omega )}}{{{\chi _{\mathrm{m}}}}} =  &- \left( {{\chi _{\mathrm{r}}}g_{\mathrm{r}}^2 + {\chi _{\mathrm{p}}}g_{\mathrm{p}}^2 - {\chi _{\mathrm{b}}}g_{\mathrm{b}}^2} \right)\hat b(\omega ) + \\
									&- i{g_{\mathrm{r}}}\sqrt {{\kappa _{{\rm{ex}}}}} {\chi _{\mathrm{r}}}{{\hat a}_{\rm{ex}}^{\rm{in}}}(\omega  - {\delta _{\mathrm{r}}}) - i{g_{\mathrm{r}}}\sqrt {{\kappa _{\rm{0}}}} {\chi _{\mathrm{r}}}{{\hat a}_0^{\rm{in}}}(\omega  - {\delta _{\mathrm{r}}})\\
									&- i{g_{\mathrm{p}}}\sqrt {{\kappa _{{\rm{ex}}}}} {\chi _{\mathrm{p}}}{{\hat a}_{\rm{ex}}^{\rm{in}}}(\omega  - {\delta _{\mathrm{p}}}) - i{g_{\mathrm{p}}}\sqrt {{\kappa _{\rm{0}}}} {\chi _{\mathrm{p}}}{{\hat a}_0^{\rm{in}}}(\omega  - {\delta _{\mathrm{p}}})\\
									&- i{g_{\mathrm{b}}}\sqrt {{\kappa _{{\rm{ex}}}}} {\chi _{\mathrm{b}}}\hat a_{{\rm{ex}}}^{\rm{in}\dag} ( - (\omega  - {\delta _{\mathrm{b}}})) - i{g_{\mathrm{b}}}\sqrt {{\kappa _{\rm{0}}}} {\chi _{\mathrm{b}}}\hat a_{{\rm{0}}}^{\rm{in}\dag} ( - (\omega  - {\delta _{\mathrm{b}}}))\\
									&+\sqrt{\Gamma_{\rm{m}}}\hat b_0^{\rm{in}}(\omega)\\
									&- {\chi _{\mathrm{r}}}\left\{ {{g_{\mathrm{r}}}{g_{\mathrm{p}}}\hat b(\omega  + {\delta _{\mathrm{p}}} - {\delta _{\mathrm{r}}}) + {g_{\mathrm{r}}}{g_{\mathrm{b}}}{{\hat b}^\dag }( - \omega  + {\delta _{\mathrm{b}}} + {\delta _{\mathrm{r}}})} \right\}\\
									&- {\chi _{\mathrm{p}}}\left\{ {{g_{\mathrm{p}}}{g_{\mathrm{r}}}\hat b(\omega  + {\delta _{\mathrm{r}}} - {\delta _{\mathrm{p}}}) + {g_{\mathrm{p}}}{g_{\mathrm{b}}}{{\hat b}^\dag }( - \omega  + {\delta _{\mathrm{b}}} + {\delta _{\mathrm{p}}})} \right\}\\
									&+ {\chi _{\mathrm{b}}}\left\{ {{g_{\mathrm{b}}}{g_{\mathrm{r}}}{{\hat b}^\dag }( - \omega  + {\delta _{\mathrm{b}}} + {\delta _{\mathrm{r}}}) + {g_{\mathrm{b}}}{g_{\mathrm{p}}}{{\hat b}^\dag }( - \omega  + {\delta _{\mathrm{b}}} + {\delta _{\mathrm{p}}})} \right\},
								\end{aligned}
								\label{eq:b_omega}
							\end{equation}
							where
							\begin{equation}
								\chi _{\mathrm{p}}^{ - 1}(\omega ) =  - i(\omega  - {\delta _{\mathrm{p}}}) + \kappa /2,\;\;\;\;\;\;\chi _{\mathrm{r}}^{ - 1}(\omega ) =  - i(\omega  - {\delta _{\mathrm{r}}}) + \kappa /2,\;\;\;\;\;\;\chi _{\mathrm{b}}^{ - 1}(\omega ) =  - i(\omega  - {\delta _{\mathrm{b}}}) + \kappa /2,\;\;\;\;{\rm{and}}\;\;\chi _{\mathrm{m}}^{ - 1}(\omega ) =  - i\omega  + {\Gamma _{\rm{m}}}/2.
							\end{equation}
							\added{
								Since the coupling rates are small compared to detuning, i.e., $|{\chi_i}{g_i}{g_{j\neq i}}| \simeq \sqrt {\Gamma _{{\rm{opt}}}^i\Gamma _{{\rm{opt}}}^{j\neq i}} /2 \ll |{\delta _i} \pm {\delta _{j\neq i}}|$, where $\Gamma_{\rm{opt}}^{k}=4g_{k}^2/\kappa$ is the optomechanical damping rate induced by each microwave drive and $i,j,k \in \left\{ {{\rm{r}},{\rm{b}},{\rm{p}}} \right\}$, we can safely neglect the terms for the mechanical annihilation operator with the detuning ($\delta_{\mathrm{p,r,b}}$) and find a solution with respect to $\hat{b}(\omega)$:
							}
							
							\begin{equation}
								\begin{aligned}
									\frac{{\hat b(\omega )}}{{{\chi _{{\rm{eff}}}}}} \simeq  &- i{g_{\rm{p}}}\left( {\sqrt {{\kappa _{\rm{0}}}} {\chi _{\rm{p}}}\hat a_0^{{\rm{in}}}(\omega  - {\delta _{\rm{p}}}) + \sqrt {{\kappa _{{\rm{ex}}}}} {\chi _{\rm{p}}}\hat a_{{\rm{ex}}}^{{\rm{in}}}(\omega  - {\delta _{\rm{p}}})} \right)\\
									&- i{g_{\rm{r}}}\left( {\sqrt {{\kappa _{\rm{0}}}} {\chi _{\rm{r}}}\hat a_0^{{\rm{in}}}(\omega  - {\delta _{\rm{r}}}) + \sqrt {{\kappa _{{\rm{ex}}}}} {\chi _{\rm{r}}}\hat a_{{\rm{ex}}}^{{\rm{in}}}(\omega  - {\delta _{\rm{r}}})} \right)\\
									&- i{g_{\rm{b}}}\left( {\sqrt {{\kappa _{\rm{0}}}} {\chi _{\rm{b}}}\hat a_{\rm{0}}^{{\rm{in}}\dag }( - (\omega  - {\delta _{\rm{b}}})) + \sqrt {{\kappa _{{\rm{ex}}}}} {\chi _{\rm{b}}}\hat a_{{\rm{ex}}}^{{\rm{in}}\dag }( - (\omega  - {\delta _{\rm{b}}}))} \right)\\
									&+ \sqrt {{\Gamma _{\rm{m}}}} \hat b_{\rm{0}}^{{\rm{in}}}(\omega )
								\end{aligned}
								\label{eq:lang_mech}
							\end{equation}
							where $\chi _{{\rm{eff}}}^{ - 1} = \chi _{\rm{m}}^{ - 1} + {\chi _{\rm{p}}}g_{\rm{p}}^2 + {\chi _{\rm{r}}}g_{\rm{r}}^2 - {\chi _{\rm{b}}}g_{\rm{b}}^2$ is the effective mechanical susceptibility.
							Using Wiener–Khinchin theorem and correlation relations in Eq.~(\ref{eq:noise}), the power spectral density of the mechanics is calculated as
							\begin{equation}
								\begin{aligned}
									{S_{\hat b\hat b}}(\omega ) =& {\left| {{\chi _{{\rm{eff}}}}} \right|^2}\left( {\left( {g_{\rm{p}}^2|{\chi _{\rm{p}}}{|^2}({\kappa _0}n_{\rm{c}}^{{\rm{th}}}) + {\Gamma _{\rm{m}}}n_{\rm{m}}^{{\rm{th}}}} \right) + \left( {g_{\rm{r}}^2|{\chi _{\rm{r}}}{|^2}({\kappa _0}n_{\rm{c}}^{{\rm{th}}}) + g_{\rm{b}}^2|{\chi _{\rm{b}}}{|^2}({\kappa _0}n_{\rm{c}}^{{\rm{th}}} + \kappa )} \right)} \right)\\
									\simeq& \frac{{{\Gamma _{{\rm{tot}}}}}}{{|(\Gamma _{{\rm{opt}}}^{\mathrm{p}} + \Gamma _{{\rm{opt}}}^{\mathrm{r}} - \Gamma _{{\rm{opt}}}^{\mathrm{b}})/2 + (1 - i2\omega /\kappa )(\Gamma_{\rm{m}} /2 - i\omega ){|^2}}} \times \left( {\frac{{\Gamma _{{\rm{opt}}}^{\mathrm{p}}{{n}_{\mathrm{c}}} + {\Gamma _{\rm{m}}}(1 + \frac{{4{\omega ^2}}}{{{\kappa ^2}}})n_{\rm{m}}^{{\rm{th}}}}}{{{\Gamma _{{\rm{tot}}}}}} + \frac{{\Gamma _{{\rm{opt}}}^{\mathrm{r}}{{n}_{\mathrm{c}}} + \Gamma _{{\rm{opt}}}^{\mathrm{b}}({{n}_{\mathrm{c}}} + 1)}}{{{\Gamma _{{\rm{tot}}}}}}} \right).
								\end{aligned}
							\end{equation}
							Here, we approximate ${\chi _{\mathrm{p}}}/{\chi _0} \simeq {\chi _{\mathrm{b}}}/{\chi _0} \simeq {\chi _{\mathrm{r}}}/{\chi _0} \simeq 1$, where $\chi_0^{-1}(\omega)=-i\omega+\kappa/2$, which is valid as long as the detunings are sufficiently smaller than the cavity linewidth. We also define the total mechanical damping rate as ${\Gamma _{{\rm{tot}}}} = \Gamma _{{\rm{opt}}}^{\mathrm{p}} + \Gamma _{{\rm{opt}}}^{\mathrm{r}} - \Gamma _{{\rm{opt}}}^{\mathrm{b}} + {\Gamma _{\rm{m}}}$, where $\Gamma_{\rm{opt}}^{\rm{p,r,b}}=4g_{\rm{p,r,b}}^2/\kappa$ is the optomechanical damping rate induced by each microwave drive. Furthermore, we define the cavity thermal occupation as $n_{\rm{c}} = \kappa_0n_{\rm{c}}^{\rm{th}}/\kappa$, which is derived later in Eq.~(\ref{eq:cav_th}). 
							
							If we assume a flat cavity response around the sidebands, which is valid in the weak coupling regime ${{\Gamma _{{\rm{tot}}}}} \ll \kappa$, we obtain a Lorentzian function around $\omega = 0$ with linewidth $\Gamma _{{\rm{tot}}}$, which is expressed by
							\begin{equation}
								{S_{\hat b\hat b}}(\omega ) \simeq \frac{{{\Gamma _{{\rm{tot}}}}}}{{{\omega ^2} + (\Gamma _{{\rm{tot}}}/2)^2}}\left( {\frac{{\Gamma _{{\rm{opt}}}^{\mathrm{p}}{{n}_{\mathrm{c}}} + {\Gamma _{\rm{m}}}n_{\rm{m}}^{{\rm{th}}}}}{{{\Gamma _{{\rm{tot}}}}}} + \frac{{\Gamma _{{\rm{opt}}}^{\mathrm{r}}{{n}_{\mathrm{c}}} + \Gamma _{{\rm{opt}}}^{\mathrm{b}}({{n}_{\mathrm{c}}} + 1)}}{{{\Gamma _{{\rm{tot}}}}}}} \right).
								\label{eq:mech_lor}
							\end{equation}
							By integrating Eq. (\ref{eq:mech_lor}), we can calculate the steady-state phonon occupation:
							\begin{equation}
								{{n}_{\rm{m}}} \equiv \left\langle {{{\hat b}^\dag }\hat b} \right\rangle = \frac{{\Gamma _{{\rm{opt}}}^{\mathrm{p}}{{n}_{\mathrm{c}}} + {\Gamma _{\rm{m}}}n_{\rm{m}}^{{\rm{th}}}}}{{{\Gamma _{{\rm{tot}}}}}} + \frac{{\Gamma _{{\rm{opt}}}^{\mathrm{r}}{{n}_{\mathrm{c}}} + \Gamma _{{\rm{opt}}}^{\mathrm{b}}({{n}_{\mathrm{c}}} + 1)}}{{{\Gamma _{{\rm{tot}}}}}}.
								\label{eq:mech_phn}
							\end{equation}
							By increasing the cooling power, i.e. increasing $\Gamma_{\rm{opt}}^{\rm{p}}$ , the mechanical occupation will decrease as long as the system is in the weak coupling regime, and it finally converges to $n_{\mathrm{c}}$. We also see spurious effects induced by the blue and red probes, which are negligible as long as ${\Gamma _{{\rm{opt}}}^{\mathrm{b}}} \ll \Gamma _{{\rm{opt}}}^{\mathrm{p}}$ and ${\Gamma _{{\rm{opt}}}^{\mathrm{b,r}}}n_{\mathrm{c}} \ll \Gamma _{{\rm{opt}}}^{\mathrm{p}}$. 
							Note that even if the cavity thermal occupation is zero, the blue probe adds an additional phonon occupation, called by a quantum back-action, which is given by ${\Gamma _{{\rm{opt}}}^{\mathrm{b}}}/{{\Gamma _{{\rm{tot}}}}}$. 
							
							By substituting Eq.~(\ref{eq:lang_mech}) into the first equation of Eq.~(\ref{eq:fourier_main}), we find a solution for $\hat a(\omega)$:
							\begin{equation}
								\begin{aligned}
									\frac{{\hat a(\omega )}}{{{\chi _0}}} =& \sqrt {{\kappa _{\rm{0}}}} {{\hat a}_0^{\rm{in}}}(\omega ) + \sqrt {{\kappa _{{\rm{ex}}}}} {{\hat a}_{\rm{ex}}^{\rm{in}}}(\omega )\\		
									&  - {g_{\mathrm{p}}^2}{{{{\chi _{{\rm{eff}}}}}(\omega  + {\delta _{\mathrm{p}}})}}{\chi _{\mathrm{p}}}(\omega  + {\delta _{\mathrm{p}}})\left( {\sqrt {{\kappa _{\rm{0}}}} {{\hat a}_0^{\rm{in}}}(\omega  ) + \sqrt {{\kappa _{{\rm{ex}}}}} {{\hat a}_{\rm{ex}}^{\rm{in}}}(\omega )} \right)\\
									&- {g_{\mathrm{r}}^2}{{\chi _{{\rm{eff}}}(\omega  + {\delta _{\mathrm{r}}})}}{\chi _{\mathrm{r}}}(\omega  + {\delta _{\mathrm{r}}})\left( {\sqrt {{\kappa _{\rm{0}}}} {{\hat a}_0^{\rm{in}}}(\omega ) + \sqrt {{\kappa _{{\rm{ex}}}}} {{\hat a}_{\rm{ex}}^{\rm{in}}}(\omega )} \right)\\
									&+ {{g_{\mathrm{b}}^2}}{{\chi _{{\rm{eff}}}(\omega  - {\delta _{\mathrm{b}}})}}{\chi _{\mathrm{b}}}(\omega  - {\delta _{\mathrm{b}}})\left( {\sqrt {{\kappa _{\rm{0}}}} {{\hat a}_0^{\rm{in}}}(\omega ) + \sqrt {{\kappa _{{\rm{ex}}}}} {{\hat a}_{\rm{ex}}^{\rm{in}}}(\omega )} \right)\\
									&- i\sqrt {{\Gamma _{\rm{m}}}} \left( {{{g_{\mathrm{p}}}}}{{{\chi _{{\rm{eff}}}}(\omega  + {\delta _{\mathrm{p}}})}}{{\hat b}_{\rm{0}}^{\rm{in}}}(\omega  + {\delta _{\mathrm{p}}}) \right. \\
									& \ \ \ \ \ \ \ \ \ \ \ \ + {{{g_{\mathrm{r}}}}}{{{\chi _{{\rm{eff}}}}(\omega  + {\delta _{\mathrm{r}}})}}{{\hat b}_{\rm{0}}^{\rm{in}}}(\omega  + {\delta _{\mathrm{r}}}) \\
									& \ \ \ \ \ \ \ \ \ \ \ \ +\left. {{{g_{\mathrm{b}}}}}{{{\chi _{{\rm{eff}}}}(\omega  - {\delta _{\mathrm{b}}})}}\hat b_0^{{\rm{in}\dag}} ( - (\omega  - {\delta _{\mathrm{b}}})) \right).
									\label{eq:aomega}
								\end{aligned}
							\end{equation}
							Therefore, we can also obtain the microwave cavity thermal occupation using Eq.~(\ref{eq:aomega}), where the first two terms are multiplied by only the cavity susceptibility, which has the cavity linewidth $\kappa$, while the other terms are multiplied by both the cavity and the mechanical susceptibilities, where the latter has a very narrow linewidth ($\Gamma_\mathrm{tot}$) compared to $\kappa$. Since the terms with the mechanical susceptibility play a minor role in the cavity thermal photon number, the power spectral density of the cavity can be simply calculated as
							\begin{equation}
								\begin{aligned}
									{S_{\hat a\hat a}}(\omega )& \simeq {\left| {{\chi _{\rm{0}}}} \right|^2}{\kappa _0}n_{\rm{c}}^{{\rm{th}}}\\
									&= \frac{\kappa }{{{\omega ^2} + {\kappa/2}^2}}\left( {\frac{{{\kappa _0}n_{\rm{c}}^{{\rm{th}}}}}{\kappa }} \right).
								\end{aligned}
							\end{equation}
							By taking the integral of the above equation, we obtain the steady-state cavity thermal occupation:
							\begin{equation}
								{n_{\rm{c}}} \equiv \left\langle {{{\hat a}^\dag }\hat a} \right\rangle  = \frac{{{\kappa _0}n_{\rm{c}}^{{\rm{th}}}}}{\kappa }.
								\label{eq:cav_th}
							\end{equation}
							The cavity thermal occupation can be interpreted as the averaged photon occupation of the intrinsic and external baths, weighted with the respective rates, where the temperature of the external bath is assumed to be zero.
							
							Using the input-output relation for the external bath: ${{\hat a}_{{\rm{ex}}}^{\rm{out}}(\omega)} = {{\hat a}_{\rm{ex}}^{\rm{in}}(\omega)} - \sqrt {{\kappa _{{\rm{ex}}}}} \hat a(\omega)$, we can calculate the symmetrized noise power spectral density of the output microwave field as \[{{\bar S}}(\omega ) = \frac{1}{2}\int_{ - \infty }^{ + \infty } {\left\langle {{{\hat a}_{{\rm{ex}}}^{\rm{out}\dagger} }(\omega '){\hat a}_{{\rm{ex}}}^{\rm{out}}(\omega ) + {\hat a}_{{\rm{ex}}}^{\rm{out}}(\omega '){{\hat a}_{{\rm{ex}}}^{\rm{out}\dagger} }(\omega )} \right\rangle d\omega '}.\]
							For simplicity, we assume that the sideband signals are well separated in the frequency space and that the linewidths of the sidebands are much smaller than the cavity linewidth, which is the case for our experiment.
							In this case, we can neglect the cross terms between the different sideband signals, and describe the spectrum as $\bar S(\omega ) \simeq \frac{1}{2} + {{\bar S}_{{\rm{c}}}}(\omega ) + {{\bar S}_{\mathrm{p}}}(\omega ) + {{\bar S}_{\mathrm{r}}}(\omega ) + {{\bar S}_{\mathrm{b}}}(\omega )$, where
							\begin{equation}
								\begin{aligned}
									{{\bar S}_{{\rm{c}}}}(\omega ) = & 4\frac{{{\kappa _{{\rm{ex}}}}}}{\kappa }\frac{{{{n}_{\mathrm{c}}}}}{{1 + \frac{{4{\omega ^2}}}{{{\kappa ^2}}}}},
								\end{aligned}
								\label{eq:sym_spec_c}
							\end{equation}
							\begin{equation}
								\begin{aligned}
									{{\bar S}_{\mathrm{p}}}(\omega ) = &\frac{{{\kappa _{{\rm{ex}}}}}}{\kappa }\frac{{\Gamma _{{\rm{opt}}}^{\mathrm{p}}{\Gamma _{{\rm{tot}}}}}}{{|(\Gamma _{{\rm{opt}}}^{\mathrm{p}} + \Gamma _{{\rm{opt}}}^{\mathrm{r}} - \Gamma _{{\rm{opt}}}^{\mathrm{b}})/2 + (1 - i2\omega /\kappa )(\Gamma_{\rm{m}} /2 - i(\omega  + {\delta _{\mathrm{p}}})){|^2}}}\left( {\frac{1}{{1 + \frac{{4{\omega ^2}}}{{{\kappa ^2}}}}}} \right)\Biggl\{ \\
									&\frac{{\Gamma _{{\rm{opt}}}^{\mathrm{p}}{{n}_{\mathrm{c}}} + \Gamma _{{\rm{opt}}}^{\mathrm{r}}{{n}_{\mathrm{c}}} + \Gamma _{{\rm{opt}}}^{\mathrm{b}}\left( {{{n}_{\mathrm{c}}} + 1} \right) + {\Gamma _{\rm{m}}}\left( {1 + \frac{{4{\omega ^2}}}{{{\kappa ^2}}}} \right)n_{\rm{m}}^{{\rm{th}}}}}{{{\Gamma _{{\rm{tot}}}}}}\\
									&- \left( {\left( {1 - \frac{{4\omega (\omega  + {\delta _{\mathrm{p}}})}}{{\kappa {\Gamma _{{\rm{tot}}}}}}} \right)\left( {2{{n}_{\mathrm{c}}} + 1} \right) - \left( {1/2 - \frac{{4\omega (\omega  + {\delta _{\mathrm{p}}})}}{{\kappa {\Gamma _{{\rm{tot}}}}}}} \right)} \right)\\
									&+ \frac{{\Gamma _{{\rm{opt}}}^{\mathrm{p}} + \Gamma _{{\rm{opt}}}^{\mathrm{r}} - \Gamma _{{\rm{opt}}}^{\mathrm{b}} + {\Gamma _{\rm{m}}}\left( {1 + \frac{{4{\omega ^2}}}{{{\kappa ^2}}}} \right)}}{{2{\Gamma _{{\rm{tot}}}}}}\Biggr\},
								\end{aligned}
								\label{eq:sym_spec_p}
							\end{equation}
							\begin{equation}
								\begin{aligned}
									{{\bar S}_{\mathrm{r}}}(\omega ) = &\frac{{{\kappa _{{\rm{ex}}}}}}{\kappa }\frac{{\Gamma _{{\rm{opt}}}^{\mathrm{r}}{\Gamma _{{\rm{tot}}}}}}{{|(\Gamma _{{\rm{opt}}}^{\mathrm{p}} + \Gamma _{{\rm{opt}}}^{\mathrm{r}} - \Gamma _{{\rm{opt}}}^{\mathrm{b}})/2 + (1 - i2\omega /\kappa )(\Gamma_{\rm{m}} /2 - i(\omega  + {\delta _{\mathrm{r}}})){|^2}}}\left( {\frac{1}{{1 + \frac{{4{\omega ^2}}}{{{\kappa ^2}}}}}} \right)\Biggl\{ \\
									&\frac{{\Gamma _{{\rm{opt}}}^{\mathrm{p}}{{n}_{\mathrm{c}}} + \Gamma _{{\rm{opt}}}^{\mathrm{r}}{{n}_{\mathrm{c}}} + \Gamma _{{\rm{opt}}}^{\mathrm{b}}\left( {{{n}_{\mathrm{c}}} + 1} \right) + {\Gamma _{\rm{m}}}\left( {1 + \frac{{4{\omega ^2}}}{{{\kappa ^2}}}} \right)n_{\rm{m}}^{{\rm{th}}}}}{{{\Gamma _{{\rm{tot}}}}}}\\
									&- \left( {\left( {1 - \frac{{4\omega (\omega  + {\delta _{\mathrm{r}}})}}{{\kappa {\Gamma _{{\rm{tot}}}}}}} \right)\left( {2{{n}_{\mathrm{c}}} + 1} \right) - \left( {1/2 - \frac{{4\omega (\omega  + {\delta _{\mathrm{r}}})}}{{\kappa {\Gamma _{{\rm{tot}}}}}}} \right)} \right)\\
									&+ \frac{{\Gamma _{{\rm{opt}}}^{\mathrm{p}} + \Gamma _{{\rm{opt}}}^{\mathrm{r}} - \Gamma _{{\rm{opt}}}^{\mathrm{b}} + {\Gamma _{\rm{m}}}\left( {1 + \frac{{4{\omega ^2}}}{{{\kappa ^2}}}} \right)}}{{2{\Gamma _{{\rm{tot}}}}}}\Biggr\}, 
								\end{aligned}
								\label{eq:sym_spec_r}
							\end{equation}
							\begin{equation}
								\begin{aligned}
									{{\bar S}_{\mathrm{b}}}(\omega ) = & \frac{{{\kappa _{{\rm{ex}}}}}}{\kappa }\frac{{\Gamma _{{\rm{opt}}}^{\mathrm{b}}{\Gamma _{{\rm{tot}}}}}}{{|(\Gamma _{{\rm{opt}}}^{\mathrm{p}} + \Gamma _{{\rm{opt}}}^{\mathrm{r}} - \Gamma _{{\rm{opt}}}^{\mathrm{b}})/2 + (1 - i2\omega /\kappa )(\Gamma_{\rm{m}} /2 - i(\omega  - {\delta _{\mathrm{b}}})){|^2}}}\left( {\frac{1}{{1 + \frac{{4{\omega ^2}}}{{{\kappa ^2}}}}}} \right)\Biggl\{ \\
									&\frac{{\Gamma _{{\rm{opt}}}^{\mathrm{p}}{{n}_{\mathrm{c}}} + \Gamma _{{\rm{opt}}}^{\mathrm{r}}{{n}_{\mathrm{c}}} + \Gamma _{{\rm{opt}}}^{\mathrm{b}}\left( {{{n}_{\mathrm{c}}} + 1} \right) + {\Gamma _{\rm{m}}}\left( {1 + \frac{{4{\omega ^2}}}{{{\kappa ^2}}}} \right)n_{\rm{m}}^{{\rm{th}}}}}{{{\Gamma _{{\rm{tot}}}}}}\\
									&+ \left( {\left( {1 - \frac{{4\omega (\omega  - {\delta _{\mathrm{b}}})}}{{\kappa {\Gamma _{{\rm{tot}}}}}}} \right)\left( {2{{n}_{\mathrm{c}}} + 1} \right) - \left( {1/2 - \frac{{4\omega (\omega  - {\delta _{\mathrm{b}}})}}{{\kappa {\Gamma _{{\rm{tot}}}}}}} \right)} \right)\\
									&+ \frac{{\Gamma _{{\rm{opt}}}^{\mathrm{p}} + \Gamma _{{\rm{opt}}}^{\mathrm{r}} - \Gamma _{{\rm{opt}}}^{\mathrm{b}} + {\Gamma _{\rm{m}}}\left( {1 + \frac{{4{\omega ^2}}}{{{\kappa ^2}}}} \right)}}{{2{\Gamma _{{\rm{tot}}}}}}\Biggr\},
								\end{aligned}
								\label{eq:sym_spec_b}
							\end{equation}
							where ${{\bar S}_{{\rm{c}}}}$ is the noise power spectral density of the cavity thermal emission and $\bar S_{\mathrm{p,b,r}}$ is the noise power spectral density of the sideband generated by the cooling pump, red probe, and blue probe, respectively. 
							
							
							Here, we can further simplify the above equation by assuming a flat cavity response around the sidebands, i.e. $\omega/\kappa \simeq 0$, i.e.,
							\begin{equation}
								\begin{aligned}
									\bar S_{\mathrm{p}}(\omega) \simeq &\frac{{{\kappa _{{\rm{ex}}}}}}{\kappa }\frac{{\Gamma _{{\rm{opt}}}^{\mathrm{p}}{\Gamma _{{\rm{tot}}}}}}{{\Gamma _{{\rm{tot}}}^2/4 + {(\omega+\delta_{\mathrm{p}}) ^2}}}\left( {{{n}_{\rm{m}}} - 2{{n}_{\mathrm{c}}}} \right)
								\end{aligned}
								\label{eq:sym_spec_p_simp}
							\end{equation}
							\begin{equation}
								\begin{aligned}
									\bar S_{\mathrm{r}}(\omega) \simeq &\frac{{{\kappa _{{\rm{ex}}}}}}{\kappa }\frac{{\Gamma _{{\rm{opt}}}^{\mathrm{r}}{\Gamma _{{\rm{tot}}}}}}{{\Gamma _{{\rm{tot}}}^2/4 + {(\omega+\delta_{\mathrm{r}}) ^2}}}\left( {{{n}_{\rm{m}}} - 2{{n}_{\mathrm{c}}}} \right)
								\end{aligned}
								\label{eq:sym_spec_r_simp}
							\end{equation}
							\begin{equation}
								\begin{aligned}
									\bar S_{\mathrm{b}}(\omega) \simeq &\frac{{{\kappa _{{\rm{ex}}}}}}{\kappa }\frac{{\Gamma _{{\rm{opt}}}^{\mathrm{b}}{\Gamma _{{\rm{tot}}}}}}{{\Gamma _{{\rm{tot}}}^2/4 + {(\omega-\delta_{\mathrm{b}}) ^2}}}\left( {{{n}_{\rm{m}}} + 2{{n}_{\mathrm{c}}} + 1} \right).
								\end{aligned}
								\label{eq:sym_spec_b_simp}
							\end{equation}
							The full noise power spectral density contains the three Lorentzian peak of the sidebands with a linewidth of $\Gamma_{\rm{tot}}$ with a slight frequency spacing on top of the Lorentzian peak of the cavity thermal emission with a linewidth of $\kappa$, which are individually accessible due to the frequency spacing among the sidebands, as well as the large linewidth difference between the sidebands and the cavity emission. 

							\subsection{Optomechanical amplification}{\label{sec:amplificattion}}
							To characterize the thermal decoherence of our mechanical oscillator, we use a time-domain protocol where the mechanical oscillator is first prepared in either a vacuum or squeezed state and then measured after a certain free-evolution time. When we measure the mechanical oscillator, we apply a microwave pump blue-detuned by the mechanical frequency to induce a two-mode squeezing process between the mechanical oscillator and the microwave cavity, corresponding to a phase-insensitive amplification of the mechanical quadratures. By measuring the optomechanical sideband signals induced by the pump field, we can obtain both the mechanical quadratures amplified in a nearly quantum-limited manner. This intrinsic optomechanical amplification technique was introduced by Reed~\textit{et.al.}~\cite{SI_reed2017faithful,SI_delaney2019measurement}. Here, we theoretically describe the quantum measurement of the mechanical quadratures based on optomechanical amplification.
							
							The Langevin equations of the mechanical oscillator and the microwave cavity with the blue-detuned pump are given by
							\begin{eqnarray}
								\frac{d\hat a}{dt} &=& + i\Omega_\mathrm{m} \hat{a} - \frac{\kappa}{2}\hat{a} -ig_\mathrm{b}(\hat{b}+\hat{b}^\dag)+ \sqrt{\kappa _\mathrm{ex}} \hat{a}_\mathrm{ex}^\mathrm{\:in}(t) + \sqrt{\kappa _\mathrm{0}} \hat{a}_\mathrm{0}^\mathrm{\:in}(t)\\
								\frac{d\hat b}{dt} &=& - i\Omega_\mathrm{m} \hat{b} - \frac{\Gamma_\mathrm{m}}{2}\hat{b} -ig_\mathrm{b}(\hat{a}+\hat{a}^\dag)+ \sqrt{\Gamma_\mathrm{m}} \hat{b}_0^\mathrm{\:in}(t),
							\end{eqnarray}
							where $g_\mathrm{b}$ is the linearized optomechanical coupling rate.
							Note that the equation for the microwave cavity is described in the rotating frame of the pump frequency.
							By going to the rotating frames with the respective frequencies, we have
							\begin{eqnarray}
								\frac{d\hat a}{dt} &=& - \frac{\kappa}{2}\hat{a} -ig_\mathrm{b}\hat{b}^\dag+ \sqrt{\kappa _\mathrm{ex}} \hat{a}_\mathrm{ex}^\mathrm{\:in}(t) + \sqrt{\kappa _\mathrm{0}} \hat{a}_\mathrm{0}^\mathrm{\:in} \label{eq:dadt}(t)\\
								\frac{d\hat b}{dt} &=& - \frac{\Gamma_\mathrm{m}}{2}\hat{b} -ig_\mathrm{b}\hat{a}^\dag+ \sqrt{\Gamma_\mathrm{m}} \hat{b}_0^\mathrm{\:in}(t), \label{eq:dbdt}
							\end{eqnarray}
							where the rotating wave approximations are applied for neglecting the fast-oscillating terms. In the parameter regime of our experiment, it can be assumed that the cavity dynamics is much faster than that of the mechanical oscillator. Using Eq.~(\ref{eq:dadt}) under $\frac{d\hat a}{dt}=0$, we can therefore obtain a quasi steady state of the microwave cavity as
							\begin{equation}
								\label{eq:qssa}
								\hat{a} = -i \frac{2g_\mathrm{b}}{\kappa} \hat{b}^\dag  + \frac{2\sqrt{\kappa _\mathrm{ex}}}{\kappa} \hat{a}_\mathrm{ex}^\mathrm{\:in}(t) + \frac{2\sqrt{\kappa _\mathrm{0}}}{\kappa}\hat{a}_\mathrm{0}^\mathrm{\:in}(t).
							\end{equation}
							By substituting Eq.~(\ref{eq:qssa}) into Eq.~(\ref{eq:dbdt}), we have the Langevin equation of the mechanical oscillator in the optomechanical amplification process, i.e.,
							\begin{equation}
								\label{eq:dbdt_amp}
								\frac{d\hat b}{dt} = \frac{\Gamma_\mathrm{amp}}{2}\hat{b} + \sqrt{\Gamma_\mathrm{m}} \hat{b}_0^\mathrm{\:in}(t) -i \sqrt{\Gamma_\mathrm{opt}^\mathrm{b}\eta_\kappa} \hat{a}_\mathrm{ex}^\mathrm{\:in\:\dag}(t) - i \sqrt{\Gamma_\mathrm{opt}^\mathrm{b}(1-\eta_\kappa)} \hat{a}_\mathrm{0}^\mathrm{\:in\:\dag}(t),
							\end{equation}
							where $\Gamma_\mathrm{amp} = \Gamma_\mathrm{opt}^\mathrm{b} - \Gamma_\mathrm{m}$ is the effective amplification rate, $\Gamma_\mathrm{opt}^\mathrm{b} = 4g_\mathrm{b}^2/\kappa$ is the optomechanical anti-damping rate, and $\eta_\kappa=\kappa_\mathrm{ex}/\kappa$ is the collection efficiency of the cavity. 
							We can solve the derivative equation from the initial time $t=0$ when the blue-detuned pump field is applied, i.e.,
							\begin{equation}
								\hat{b}(t) =  e^{\frac{\Gamma_\mathrm{amp}t}{2}} \left[\hat{b}(0) + \int_0^tdt' e^{-\frac{\Gamma_\mathrm{amp}t'}{2}}\left( \sqrt{\Gamma_\mathrm{m}} \hat{b}_0^\mathrm{\:in}(t') -i \sqrt{\Gamma_\mathrm{opt}^\mathrm{b}\eta_\kappa} \hat{a}_\mathrm{ex}^\mathrm{\:in\:\dag} (t') - i \sqrt{\Gamma_\mathrm{opt}^\mathrm{b}(1-\eta_\kappa)} \hat{a}_\mathrm{0}^\mathrm{\:in\:\dag}(t')\right)\right],
							\end{equation}
							where $\hat{b}(t)$ is the mechanical annihilation operator at time $t$.
							By combining the solution with Eq.~(\ref{eq:qssa}) and the input-output relation of the external microwave field: $\hat{a}_\mathrm{ex}^\mathrm{out}(t) = \hat{a}_\mathrm{ex}^\mathrm{\:in}(t) - \sqrt{\kappa_\mathrm{ex}}\hat{a}$, we obtain the output microwave field as
							\begin{equation}
								\begin{aligned}
									\hat{a}_\mathrm{ex}^\mathrm{out}(t)  =&(1-2\eta_\kappa)\hat{a}_\mathrm{ex}^\mathrm{\:in}(t)  - 2\sqrt{\eta_\kappa(1-\eta_\kappa)}\hat{a}_0^\mathrm{\:in}(t) \\
									&+ i\sqrt{\eta_\kappa\Gamma_\mathrm{opt}^\mathrm{b}}\:e^{\frac{\Gamma_\mathrm{amp}t}{2}}\left[\hat{b}^\dag(0) + \int_0^tdt' e^{-\frac{\Gamma_\mathrm{amp}t'}{2}}\left( \sqrt{\Gamma_\mathrm{m}} \hat{b}_0^\mathrm{\:in\:\dag}(t') +i \sqrt{\Gamma_\mathrm{opt}^\mathrm{b}\eta_\kappa} \hat{a}_\mathrm{ex}^\mathrm{\:in}(t')+ i \sqrt{\Gamma_\mathrm{opt}^\mathrm{b}(1-\eta_\kappa)}\hat{a}_\mathrm{0}^\mathrm{\:in}(t')\right)\right].
								\end{aligned}
							\end{equation}
							
							In our experiment, we measure the output microwave field after the phase-insensitive amplifications. The measured microwave field can be effectively described as
							\begin{equation}
								\hat{a}_\mathrm{ex}^\mathrm{out \:\prime}(t) = \sqrt{G}\left[\hat{a}_\mathrm{ex}^\mathrm{out}(t) + \hat{c}^\mathrm{in\:\dag}(t)\right],
							\end{equation}
							where $G$ is the total microwave gain and $\hat{c}^\mathrm{in}(t)$ is the annihilation operator of an ancillary mode describing the effective added noise of the microwave measurement chain, which is normally dominated by the HEMT amplifier noise. 
							
							In order to maximize the signal-to-noise ratio in the measurement of the mechanical quadratures, we integrate the measured output microwave signal over a normalized matched filter function defined as $m(t) = \sqrt{\frac{\Gamma_\mathrm{amp}}{ e^{\Gamma_\mathrm{amp}\tau}-1}} e^{\frac{\Gamma_\mathrm{amp}t}{2}}$ ($0\leq t \leq\tau$), where $\tau$ is the final time of the integral.
							\added{Note that the filter function can be complex-valued for a more general case.}
							Namely, the time-independent complex amplitude of the output microwave field is given by
							\begin{equation}
								\label{eq:d}
								\hat{A} = \int_0^\tau dt\: m^*(t)\hat{a}_\mathrm{ex}^\mathrm{out \:\prime} (t).
							\end{equation}
							In the large gain limit ($e^{\Gamma_\mathrm{amp}\tau} \gg 1$), where $\Gamma_\mathrm{opt}^\mathrm{b} \gg \Gamma_\mathrm{m}$ can be assumed, the microwave complex amplitude is described as
							\begin{equation}
								\begin{aligned}
									\hat{A} \:\: =  &\sqrt{G^\mathrm{opt}}\left\{i\hat{b}^\dag(0) -  \sqrt{\eta_\kappa} \hat{a}'_\mathrm{ex}  - \sqrt{(1-\eta_\kappa)}\:\hat{a}'_0 \right.\\
									&\left.+i \sqrt{\frac{\Gamma_\mathrm{m}}{\Gamma_\mathrm{amp}}}\:\hat{b}_0^{\prime \:\dag}  + \frac{1}{\sqrt{\eta_\kappa e^{\Gamma_\mathrm{amp}\tau}}}\:\hat{c}^\dag  + \frac{1}{\sqrt{\eta_\kappa e^{\Gamma_\mathrm{amp}\tau}}}\left[(1-2\eta_\kappa)\hat{a}_\mathrm{ex} - 2\sqrt{\eta_\kappa(1-\eta_\kappa)}\hat{a}_\mathrm{0}\right]\right\},
								\end{aligned}
							\end{equation}
							where $G^\mathrm{opt} = G\eta_\kappa e^{\Gamma_\mathrm{amp}\tau}$ is the total scaling factor in the optomechanical amplification process. Here, we define time-independent convoluted annihilation operators for $\hat{a}_\mathrm{ex}^\mathrm{\; in}$, $\hat{a}_0^\mathrm{\; in}$, $\hat{b}_0^\mathrm{\; in}$, and $\hat{c}^\mathrm{in}$ respectively, to satisfy the bosonic commutation relations in the large gain limit, i.e., $\hat{O} = \int_0^\tau dt\: m^*(t)\hat{O}^\mathrm{in}(t)$  and $\hat{O}'= \frac{\Gamma_\mathrm{m}^{\frac{3}{2}}}{e^{\Gamma_\mathrm{amp}\tau}} \int_0^\tau dt \int_0^t ds\: e^{\Gamma_\mathrm{amp}(t-\frac{s}{2})}\hat{O}^\mathrm{in}(s)$. 
							Moreover, in order to straightforwardly convert the microwave complex amplitude to that of the mechanical oscillator, we redefine the microwave complex amplitude as $i\hat{A}^{\dag}\rightarrow\hat{A}$, i.e.,
							\begin{equation}
								\label{eq:alpha}
								\hat{A} = \sqrt{G^\mathrm{opt}}\left(\hat{b}(0) + \hat{c}_\mathrm{opt}^\dag \right),
							\end{equation}
							where an ancillary mode effectively describing all the contributions of the added noises is defined as
							\begin{equation}
								\hat{c}_\mathrm{opt} = i\sqrt{\eta_\kappa} \hat{a}'_\mathrm{ex}  +i \sqrt{(1-\eta_\kappa)}\:\hat{a}'_0 + \sqrt{\frac{\Gamma_\mathrm{m}}{\Gamma_\mathrm{amp}}}\:\hat{b}'_\mathrm{m}  -\frac{i}{\sqrt{\eta_\kappa e^{\Gamma_\mathrm{amp}\tau}}}\:\hat{c}^\dag  - \frac{i}{\sqrt{\eta_\kappa e^{\Gamma_\mathrm{amp}\tau}}}\left[(1-2\eta_\kappa)\hat{a}_\mathrm{ex} - 2\sqrt{\eta_\kappa(1-\eta_\kappa)}\hat{a}_\mathrm{0}\right].
							\end{equation}
							Importantly note that the first and second terms dominate the total added noise.
							The third term is the added noise due to the thermal decoherence of the mechanical oscillator during the amplification process, which is negligible in the large amplification rate limit, i.e., $\Gamma_\mathrm{amp} \gg \Gamma_\mathrm{th} = \Gamma_\mathrm{m}n_\mathrm{m}^\mathrm{th}$.
							The fourth term is the added noise from the microwave measurement noise, dominated by the HEMT amplifier noise for our experiment, which is also negligible when the optomechanical amplification gain is sufficiently large, i.e.\ $e^{\Gamma_\mathrm{amp}\tau} \gg n_\mathrm{add}^\mathrm{H}$, where $n_\mathrm{add}^\mathrm{H}$ is the added noise of the HEMT amplifier.
							The rest terms are the quantum noises for the microwave cavity, which are suppressed by the optomechanical amplification gain, and can be safely neglected.
							Therefore, the annihilation operator of the ancilla mode is approximated by 
							\begin{equation}
								\hat{c}_\mathrm{opt} \approx i\sqrt{\eta_\kappa} \: \hat{a}'_\mathrm{ex} +i \sqrt{1-\eta_\kappa} \: \hat{a}'_0,
							\end{equation}
							where the ancilla mode can be understood as the hybridized mode of the convoluted microwave external and internal noise operators. 
							
							To characterize the thermalization of mechanical vacuum or squeezed states, we need  to measure the variances of the mechanical quadratures at $t=0$, when the blue-detuned pump is turned on.
							As defined in Eq.~(\ref{eq:alpha}), we are able to directly measure both the quadratures of the output microwave field. Here, the measured microwave quadratures are defined as 
							\begin{equation}
								\begin{aligned}
									\hat{I} &=& \frac{\hat{A}+\hat{A}^\dag}{\sqrt{2}},\\
									\hat{Q}&=& \frac{\hat{A}-\hat{A}^\dag}{\sqrt{2}i},
								\end{aligned}
							\end{equation}
							while the mechanical quadratures to be measured are defined as
							\begin{equation}
								\begin{aligned}
									\hat{X}_1 &=& \frac{\hat{b}(0)+\hat{b}^\dag(0)}{\sqrt{2}}\\
									\hat{X}_2 &=& \frac{\hat{b}(0)-\hat{b}^\dag(0)}{\sqrt{2}i}.
								\end{aligned}
							\end{equation}
							
							Using Eq.~(\ref{eq:alpha}), the expectation value of the variance of the microwave quadratures are described as
							\begin{equation}
								\begin{aligned}
									\label{eq:IQX1X2}
									\langle\hat{I}^2\rangle &= G^\mathrm{opt} \left(\langle\hat{X}_1^2\rangle + n_\mathrm{add}^\mathrm{opt} +\frac{1}{2}\right),\\
									\langle\hat{Q}^2\rangle &= G^\mathrm{opt} \left(\langle\hat{X}_2^2\rangle + n_\mathrm{add}^\mathrm{opt} +\frac{1}{2}\right),
								\end{aligned}
							\end{equation} 
							where it is assumed that the ancilla mode is not initially correlated with the mechanical oscillator, while it is in the thermal state with the effective added noise, given by $\langle \hat{c}_\mathrm{opt}^\dag \hat{c}_\mathrm{opt}\rangle = n_\mathrm{add}^\mathrm{opt}$.
							Note that when the ancilla mode is in the vacuum states, i.e., $n_\mathrm{add}^\mathrm{opt}=0$, the added noise becomes $1/2$, which corresponds to the ideal simultaneous measurements of both the mechanical quadratures in the quantum limit. Using Eqs.~(\ref{eq:IQX1X2}), we can obtain the variance of the mechanical quadratures as
							\begin{equation}
								\begin{aligned}
									\label{eq:X1mX2m}
									\langle\hat{X}_1^2\rangle &= \frac{\langle\hat{I}^2\rangle}{G^\mathrm{opt}}- n_\mathrm{add}^\mathrm{opt} -\frac{1}{2},\\
									\langle\hat{X}_2^2\rangle &= \frac{\langle\hat{Q}^2\rangle}{G^\mathrm{opt}}- n_\mathrm{add}^\mathrm{opt} -\frac{1}{2}.
								\end{aligned}
							\end{equation} 
							Furthermore, using Eq.~(\ref{eq:X1mX2m}), we can obtain the phonon occupation as the average of the variances of both the mechanical quadratures, subtracted by the half quanta, i.e.,
							\begin{equation}
								\begin{aligned}
									n_\mathrm{m} =\langle \hat{b}^\dag(0)\hat{b}(0)\rangle &= \frac{\langle\hat{X}_1^2\rangle+\langle\hat{X}_2^2\rangle}{2} - \frac{1}{2}\\
									&= \frac{1}{G^\mathrm{opt}}\frac{\langle\hat{I}^2\rangle + \langle\hat{Q}^2\rangle}{2}- n_\mathrm{add}^\mathrm{opt} -1.
								\end{aligned}
							\end{equation}

							\section{Design and simulation}
							Here we discuss the mechanical properties of the drumhead resonator used in this work and calculate the optomechanical coupling rate based on geometrical and material parameters of the system.
							\subsection{Mechanical properties of the drumhead resonator}
							\label{sec:mechanics}
							For sufficiently small deformation of the drumhead, we can approximate the displacement of each element in time as a harmonic oscillation along the vertical axis. The oscillation amplitude of the drum at $(r,\phi)$ in polar coordinate is described by $z(t)u(r,\varphi)$, where $z(t)$ with the dimension of length obeys the harmonic oscillation equation: $\ddot z =  - \Omega _{\mathrm{m}}^2z$, while $u(r,\varphi)$ is the unit-less mode shape of the drum, normalized such that the amplitude at the origin is 1. For a circular drum with radius of $R$, we have
							\begin{equation}
								\begin{aligned}
									u(r,\varphi)=&J_n(\frac{\alpha_{n,m}}{R}r)\cos(n\varphi), \\
									\Omega_{\mathrm{m}}^{(n,m)} =& \frac{\alpha_{n,m}}{R} \sqrt{\frac{\sigma_\mathrm{m}}{\rho}},
								\end{aligned}
								\label{eq:omega_m}
							\end{equation}
							where $\alpha_{n,m}$ is the $m^\mathrm{th}$ root of the $n^\mathrm{th}$ order Bessel function of the first kind, $J_n$, and  $\sigma_\mathrm{m}$ and $\rho$ are the mechanical stress and density of the material, respectively, in our case aluminum with $\rho_\mathrm{Al} = 2700 \ \mathrm{ kg}/\mathrm{m}^3$. For the fundamental mode ($n=0$, $m=1$), which we use in this work, $\alpha_{0,1} \simeq 2.4$. The small holes in the actual device slightly deviate the mode shape from that of a uniform drum. The radius of the drumhead used in the main work is $R=75\mu$m. We extracted $\sigma_\mathrm{Al}\simeq 350$~MPa tensile stress at cryogenic temperatures by measuring mechanical frequency of several drums with different radii.
							\begin{figure*}[!]
								\includegraphics[scale=1]{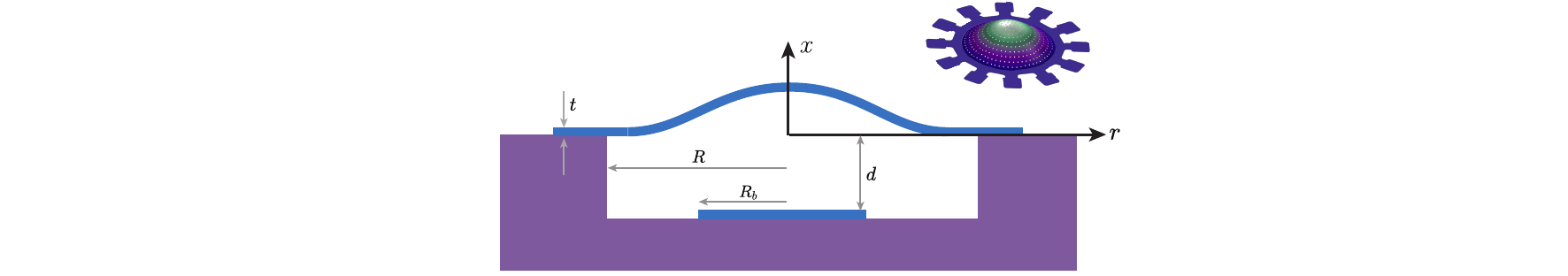}
								\caption{\textbf{Physical parameters of the vacuum-gap capacitor.} The radius of the bottom layer is $R_b$, the radius of the drumhead is $R$, the thickness of the drumhead is $t$, and the distance between the top and bottom layer is $d$. The FEM simulation of the drumhead is shown on the top right.}
								\label{fig:SI_VG}
							\end{figure*}
							
							The effective mass of such a mechanical oscillator can be defined as the total kinetic energy over the velocity squared:
							\begin{equation}
								\begin{aligned}
									{U_{{\rm{kinetic}}}} = &\int {\frac{1}{2}\rho |\dot z(t)u(r,\varphi){|^2} trdrd\varphi} \\
									\equiv &\frac{1}{2}{m_{{\rm{eff}}}}\dot z{(t)^2},
								\end{aligned}
							\end{equation}	
							where $t$ is the thickness of the drumhead resonator. For the fundamental mode of a circular drum this reduces to
							\begin{equation}
								\begin{aligned}
									{m_{{\rm{eff}}}} = &2\pi \rho t\int_0^R {r|u(r){|^2}dr} \\
									= &{m_{{\rm{phys}}}}{\xi_{\rm{mass}}},
								\end{aligned}
							\end{equation}
							where ${m_{{\rm{phys}}}}$ is the physical mass of the drum and ${\xi_{\rm{mass}}}$ is the ratio between the effective and physical mass:
							\begin{equation}
								\begin{aligned}
									{m_{{\rm{phys}}}} =& \rho  \times \pi {R^2}t,\\
									{\xi_{\rm{mass}}} = &\frac{2}{{{R^2}}}\int_0^R {r|u(r){|^2}dr}.
								\end{aligned}
							\end{equation}
							For the fundamental mode, we have $\xi_{\rm{mass}} \simeq 0.27$. Having the effective mass and the frequency of the mechanical oscillator, we can find its zero-point fluctuation:
							\begin{equation}
								x_{\rm{ZPF}} = \sqrt{\frac{\hbar}{2m_{\rm{eff}}\Omega_{\mathrm{m}}}}.
							\end{equation}
							
							For the drumhead used in this work, the effective mass of the fundamental mode can be calculated as $m_\mathrm{eff} = 2.3$~ng and the zero-point-fluctuation of motion as $x_\mathrm{ZPF} = 1.4$~fm.
							
							The loss mechanisms in a macroscopic mechanical oscillator can be separated as intrinsic (bending loss, surface loss, etc.) and extrinsic (gas damping, acoustic radiation, clamp loss, etc.)~\cite{SI_fedorov2020mechanical}. The traditional design of drumhead capacitors was resulting in a non-flat geometry of the drumhead, specifically sharp edges at the clamping point. Such sharp structure induces the bounding mechanical dissipation at the clamping points, i.e. phonon tunneling to the substrate~\cite{SI_cole2011phonon,SI_wilson2011high}. 
							The conventional circuit optomechanical devices mainly suffer from such radiative loss, which can be confirmed by the fact that there is a saturation in the temperature dependence of the mechanical damping rate since the clamp loss normally does not have temperature dependence, while the intrinsic mechanical dissipation is expected to decrease for a lower temperature~\cite{SI_cattiaux2021macroscopic}. 
							The flat geometry in our design may result in smaller clamp losses, as observed in the temperature sweep experiment (see Fig.~\ref{fig:SI_T_sweep}) where mechanical damping rate shows strong dependency to temperature (while mechanical frequency barely shift exhibiting constant stress and Young's modulus in that temperature range).
							
\begin{figure*}[!]
								\includegraphics[scale=1]{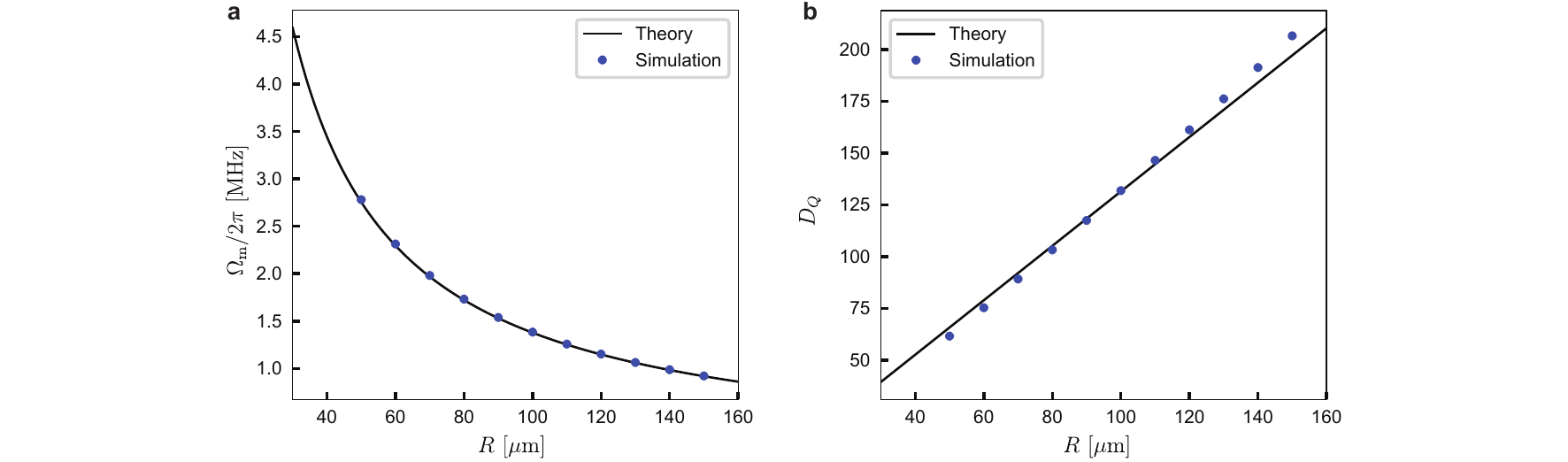}
								\caption{\textbf{FEM simulation of the drumhead resonator.} \textbf{a, b,} Simulated and theoretically calculated mechanical frequency and loss dilution factor versus drum's radius considering $t=180$~nm and $\sigma_\mathrm{Al}\simeq 350$~MPa.}
								\label{fig:SI_mechanics}
							\end{figure*}

							The intrinsic mechanical quality factor in a nanomechanical string or membrane can be described as
							\begin{equation}
								Q_\mathrm{m} = Q_0 \times D_Q,
								\label{eq:DQ}
							\end{equation}
							where $Q_0$ is the material's bulk quality factor and $D_Q$ is the dissipation dilution factor~\cite{SI_fedorov2019generalized,SI_fedorov2020mechanical}. $D_Q = \frac{\langle W_\mathrm{total}\rangle}{\langle W_\mathrm{lossy}\rangle}$ is defined as the ratio of the dynamic elastic energy averaged over
							the vibrational period ($\langle W_\mathrm{total}\rangle$) over its lossy part ($\langle W_\mathrm{lossy}\rangle$). It can be theoretically shown~\cite{SI_fedorov2019generalized,SI_fedorov2020mechanical} that $D_Q = (A \lambda + B \lambda^2)^{-1}$, where $A$ and $B$ are unit-less geometrical factors and $\lambda = \frac{t}{2R}\sqrt{\frac{Y}{12\sigma_\mathrm{m}}}$ with $Y$ presenting Young's modulus (for aluminum at low temperatures $Y \simeq 75$~GPa~\cite{SI_ekin2006experimental}). Considering system parameters the range of $\lambda$ parameter for our drumhead resonators is $\mathcal{O}(10^{-2})$ which validates the approximation of $D_Q\propto \lambda^{-1}$. We simulated the drumhead resonator using $\text{COMSOL}^\text{®}$ with finite element method (FEM) and numerically calculate $D_Q$ values. The result shows a good agreement with the theory of loss dilution as shown in Fig.~\ref{fig:SI_mechanics}. The loss dilution factor for the device discussed in the main text is estimated as $D_Q\simeq100$. Considering measured quality factor of $Q_\mathrm{m} = 40\times10^6$, we extract aluminum's bulk quality factor at 10~mk as $Q_0\simeq4\times10^5$ for the device we studied in the main text. It is worth noting that we observed the mechanical quality factor of our drumhead resonators are very sensitive to temperature shocks and thermal cycles, i.e., a fast increase of the fridge's temperature even in a few kelvin range may degrade the mechanical quality factor, even though their mechanical frequencies are not affected. This may be due to the microscopic crack formation on the clamping points in an abrupt stress shock, as a possible explanation. Because of this issue, we omit the pulse-precooling step in the cooldown process of the dry dilution fridge, which may help maintaining high quality factors for mechanical devices under test. We note that a controlled slow temperature sweep, similar to what is shown in Fig.~\ref{fig:SI_T_sweep}, are reproducible and do not reduce the mechanical quality factors.
							
							\subsection{Single photon optomechanical coupling rate ($g_0$)}{\label{sec:g0}}
							In this section, we calculate the theoretical value for $g_0$ based on the physical parameters of the vacuum-gap capacitor (Fig.~\ref{fig:SI_VG}). Knowing the displacement function of the top plate (Sec.~\ref{sec:mechanics}), we can find the total capacitance as
							\begin{equation}
								\begin{aligned}
									{C_{{\rm{tot}}}} &= \int {\frac{{{\varepsilon _0}rdrd\varphi}}{{d + z\cdot u(r,\varphi)}}}+C_{\rm{par}} \\
									&= \int_0^{{R_\mathrm{b}}} {\frac{{{\varepsilon _0}2\pi rdr}}{{d + z\cdot u(r)}}}+C_{\rm{par}} \\
									&\simeq \int_0^{{R_\mathrm{b}}} {\frac{{{\varepsilon _0}2\pi rdr}}{d}} \left( {1 - \frac{z}{d}u(r)} \right)+C_{\rm{par}}\\
									&= {C_0} - \frac{{2\pi {\varepsilon _0}}}{{{d^2}}}z\int_0^{{R_\mathrm{b}}} {ru(r)dr} + C_{\rm{par}},
								\end{aligned}
							\end{equation}
							where $R_\mathrm{b}$ is the radius of the bottom plate, $C_0 = \varepsilon_0\pi R_\mathrm{b}^2/d$ is the unmodulated capacitance of the vacuum-gap, and $C_{\rm{par}}$ is the total parasitic capacitance of the other elements of circuit (wires and spiral inductor). Here, we have considered the mode without angular dependency and assumed $z\ll d$. The frequency of the microwave cavity is then given by
							\begin{equation}
								\begin{aligned}
									{\omega _{\mathrm{c}}}(z)& = \frac{1}{{\sqrt {L{C_{{\rm{tot}}}}} }}\\
									&= \frac{1}{{\sqrt {L\left( {{C_0} + {C_{{\rm{par}}}}} \right)\left( {1 - \frac{{2\pi {\varepsilon _0}}}{{{d^2}\left( {{C_0} + {C_{{\rm{par}}}}} \right)}}z\int_0^{{R_\mathrm{b}}} {ru(r)dr} } \right)} }}\\
									&\simeq \frac{1}{{\sqrt {L{\left( {{C_0} + {C_{{\rm{par}}}}} \right)}} }}\left( {1 + \frac{1}{2}\frac{{2\pi {\varepsilon _0}/{d^2}}}{{{{{C_0} + {C_{{\rm{par}}}}}}}}z\int_0^{{R_\mathrm{b}}} {ru(r)dr} } \right)\\
									&= {\omega _{\mathrm{c}}} + \frac{{{\omega _{\mathrm{c}}}}}{{2d}}{\xi _{\rm{cap}}}{\xi _{\rm{par}}}z,
								\end{aligned}
							\end{equation}
							where
							\begin{equation}
								\begin{aligned}
									{\xi  _{\rm{cap}}} &= \frac{2}{{R_\mathrm{b}^2}}\int_0^{{R_\mathrm{b}}} {ru(r)dr}, \\
									{\xi  _{\rm{par}}} &= \frac{C_0}{C_0+C_{\rm{par}}}, \\
									{\omega _{\mathrm{c}}}& = \frac{1}{{\sqrt {L\left( {{C_0} + {C_{{\rm{par}}}}} \right)} }}.
								\end{aligned}
							\end{equation}
							We can interpret $\xi_{\rm{cap}}$ as the geometrical mode shape contribution factor and $\xi_{\rm{par}}$ as the participation ratio of the modulated capacitor to the total capacitance of the microwave cavity. For the fundamental mode $u(r) = J_0(\frac{\alpha_{0,1}}{R}r)$, $\xi_{\rm{cap}}$ is dependent to the $R_\mathrm{b}/R$. In our design, $R_\mathrm{b}=23\;\mu \rm{m}$ and $R=75\;\mu \rm{m}$, which results in $\xi_{\rm{cap}}\simeq 0.93$. 
							The participation ration of the vacuum-gap capacitance to the total capacitance can be extracted in FEM simulations. Simulating our circuit design in $\text{SONNET}^\text{®}$ results in $\xi_{\rm{par}} \simeq 0.8 $. 
							
							The single photon optomechanical coupling rate is then given by
							\begin{equation}
								g_0 = \frac{{{\omega _{\mathrm{c}}}}}{{2d}}{\xi _{\rm{cap}}}\xi_{\rm{par}}{x_{{\rm{ZPF}}}},
							\end{equation}
							where $x_{\rm{ZPF}} = \sqrt{\frac{\hbar}{2m_{\rm{eff}}\Omega_{\mathrm{m}}}}$ is the zero point fluctuation, and $m_{\rm{eff}}$ and $\Omega_{\rm{m}}$ were already derived in Sec.~\ref{sec:mechanics}. 
							
							The final expression for $g_0$ based on the system parameters for the fundamental mode then is given by
							\begin{equation}
								{g_0} = 0.37\sqrt \hbar  \frac{{{\omega _{\rm{c}}}}}{{2d}}{\left( R^2t^2\rho \sigma_\mathrm{m}  \right)^{ - 1/4}}.
							\end{equation}
							As it can be seen from the above formula, for a constant microwave cavity frequency $g_0$ scales with $d^{-1}$, $R^{-0.5}$, $t^{-0.5}$, $\rho^{-0.25}$, and $\sigma_\mathrm{m}^{-0.25}$. The theoretically expected $g_0$ for the device discussed in the main test is calculated $g_0/2\pi \simeq 14$~Hz, which is in a good agreement with the experimentally measured value (Sec.~\ref{sec:g0_meas}).
							
							To sum up, we provide a scaling rules table (table \ref{table:scaling}) showing relations between optomechanical properties and physical system parameters.
							\begin{table}[h!]
								
								\centering
								\caption{Scaling rules in drum-head capacitor based circuit optomechanics.}
									\begin{tabularx}{0.65\textwidth} { 
											| >{\centering\arraybackslash}X
											| >{\centering\arraybackslash}X
											| >{\centering\arraybackslash}X
											| >{\centering\arraybackslash}X
											| >{\centering\arraybackslash}X| }
										\hline
										& $R$ & $\sigma_\mathrm{m}$ & $t$ & $d$  \\ [0.8ex] \hline\hline 
										
										$\Omega_\mathrm{m}$ & $\frac{1}{R}$ & $\sqrt{\sigma_\mathrm{m}}$ & - & -  \\ [0.8ex] \hline
										$\Gamma_\mathrm{m}$ & $\frac{1}{R^2}$ & - & $t$  & -  \\ \hline
										$Q_\mathrm{m}, 1/\Gamma_\mathrm{th}$ & $R$ & $\sqrt{\sigma_\mathrm{m}}$ & $t^{-1}$ & -  \\ \hline
										$g_0$ & $\frac{1}{\sqrt{R}}$ & $\sigma_\mathrm{m}^{-\frac{1}{4}}$ & $\frac{1}{\sqrt{t}}$  & $\frac{1}{d}$  \\ \hline
										$\mathcal{C}_0 \equiv 4g_0^2/\kappa\Gamma_\mathrm{m} $ & $R$ & $\sigma_\mathrm{m}^{-\frac{1}{2}}$ & $\frac{1}{t^2}$  & $\frac{1}{d^2}$  \\ \hline
									\end{tabularx}
									\label{table:scaling}
								\end{table}
								
								\section{Fabrication}{\label{sec:fab}}
								In this section, we briefly present the challenge in the conventional fabrication technique used in circuit optomechanics which was limiting the coherence of drumhead mechanical resonators, and then provide the detailed fabrication process we introduced to overcome such challenges.
								\subsection{Challenges and limitations of the conventional nanofabrication process}
\begin{figure}[h]
									\centering
									\includegraphics[width=\textwidth]{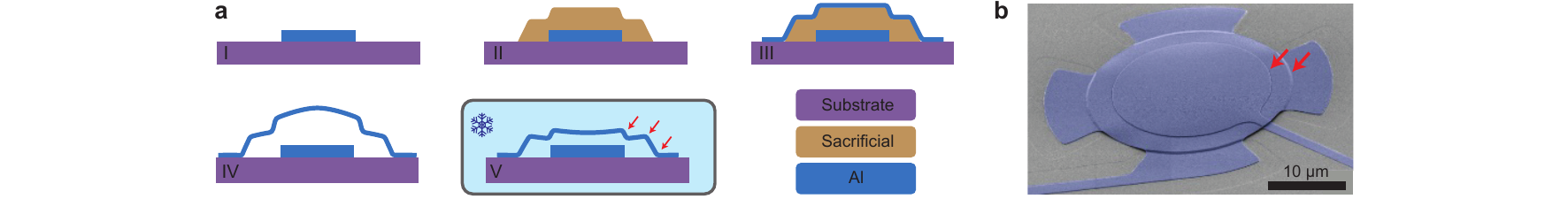}
									\caption{\textbf{Non-flat geometry of the mechanical resonator in the conventional fabrication process.} \textbf{a}, The conventional nanofabrication process used to make mechanically compliant vacuum-gap capacitors: a sacrificial layer used to support the top layer. Due to the non-flat topography of the sacrificial layer, the suspended plate has several edges close to the clamping point shown by red arrows. \textbf{b}, SEM image of a drumhead fabricated with the conventional process at EPFL~\cite{SI_toth2017dissipative} showing the non-flat geometery of the mechanically compliant plate.}
									\label{fig:old_PF}
								\end{figure}

								Since 2010, when the conventional nanofabrication process of making superconducting vacuum-gap capacitors was introduced by Cicak, \textit{et al.}~\cite{SI_cicak2010low}, the design and process did not have substantial change, while it was used to implement outstanding quantum experiments in optomechanics. The main steps of the conventional process (Fig.~\ref{fig:old_PF}a) consist of deposition and definition the bottom plate of the capacitor, deposition of a sacrificial layer covering the bottom layer, deposition and definition of the top capacitor plate, and finally releasing the device by removing (isotropic etching) the sacrificial layer. Following the same principle, several research groups realized circuit optomechanical systems using various set of sacrificial materials on different substrates such as Si$_3$N$_4$ on sapphire at NIST~\cite{SI_teufel2011circuit}, polymer on Si at Caltech~\cite{SI_suh2014mechanically}, SiO$_2$ on quartz at Aalto~\cite{SI_pirkkalainen2015squeezing}, and aSi on sapphire at EPFL~\cite{SI_toth2017dissipative}. Due to the non-flat geometry of the suspended plate, induced by the topography of the bottom plate, the final gap size of the capacitor is not well-controllable at cryogenic temperatures, and more importantly, the mechanical dissipation in such resonators is limited by the phonon tunneling loss through the substrate~\cite{SI_cattiaux2021macroscopic}.
								
								\subsection{The improved nanofabrication process for ultra-coherent and low-heating circuit optomechanics}
								\begin{figure*}[!]
									\includegraphics[scale=1]{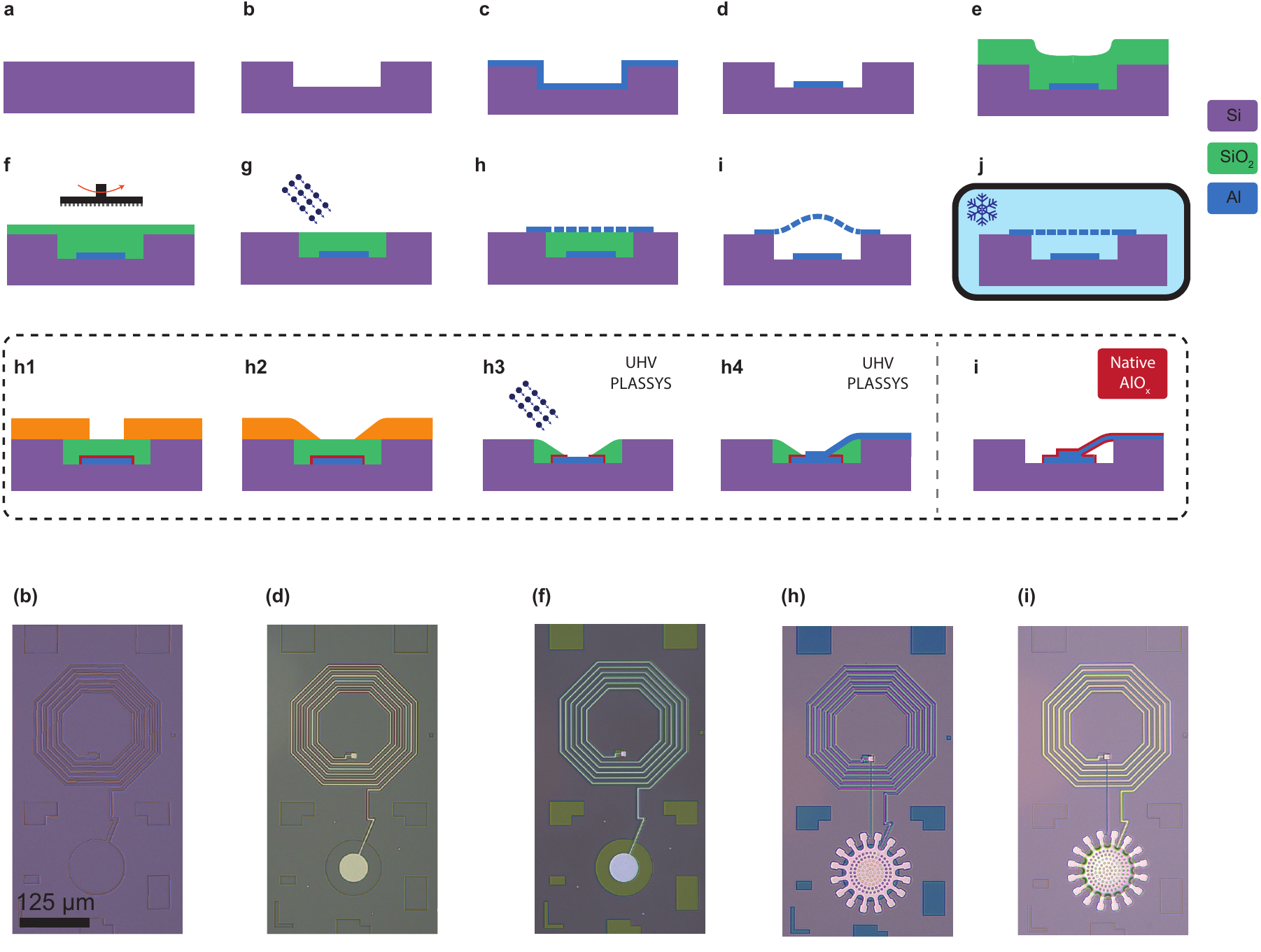}
									\caption{\textbf{Nanofabrication process for ultra-coherent circuit optomechanics.} \textbf{a}, \textbf{b}, Etching a trench in a silicon wafer (300 nm). \textbf{c}, Aluminum deposition of the bottom plate (100 nm). \textbf{d}, Patterning of Al. \textbf{e}, SiO$_2$ sacrificial layer deposition (2.5 $\mu$m). \textbf{f}, CMP planarization. \textbf{g}, Landing on the substrate using IBE etching. \textbf{h}, Opening the oxide, removing native AlOx, top Al layer deposition and patterning (180 nm). \textbf{i}, Releasing the structure using HF vapor. Due to compressive stresses, the top plate will buckle up. \textbf{j}, At cryogenic temperatures, the drumhead shrinks and flattens. The optical micrographs shows examples of selected steps of the process flow for a slightly different design as shown in the main text.}
									\label{fig:SI_fab}
								\end{figure*}
								Here we present a novel nanofabrication process (Fig.~\ref{fig:SI_fab}) to overcome the above-mentioned challenges, realizing a significant improvement in mechanical quality factors, as well as the gap control and the yield in the release process. We define a trench in a silicon substrate, followed by deposition and patterning of the bottom plate of the capacitor. The trench is then covered by a thick SiO$_2$ sacrificial layer, which inherits the same topography of the layer underneath. To remove this topography and obtain a flat surface, we use chemical mechanical polishing (CMP) to planarize the SiO$_2$ surface. We then etch back the sacrificial layer down to the substrate layer and deposit the top Al plate of the capacitor. Although after the release of the structure the drumhead will buckle up due to the compressive stress, at cryogenic temperature the high tensile stress ensures the flatness of the top plate. This will guarantee the gap size to be precisely defined by the depth of the trench and the thickness of the bottom plate. Furthermore, the flat geometry of the top plate significantly reduces the mechanical dissipation of the drumhead resonator. 
								
								\begin{figure}[h]
									\centering
									\includegraphics[width=\textwidth]{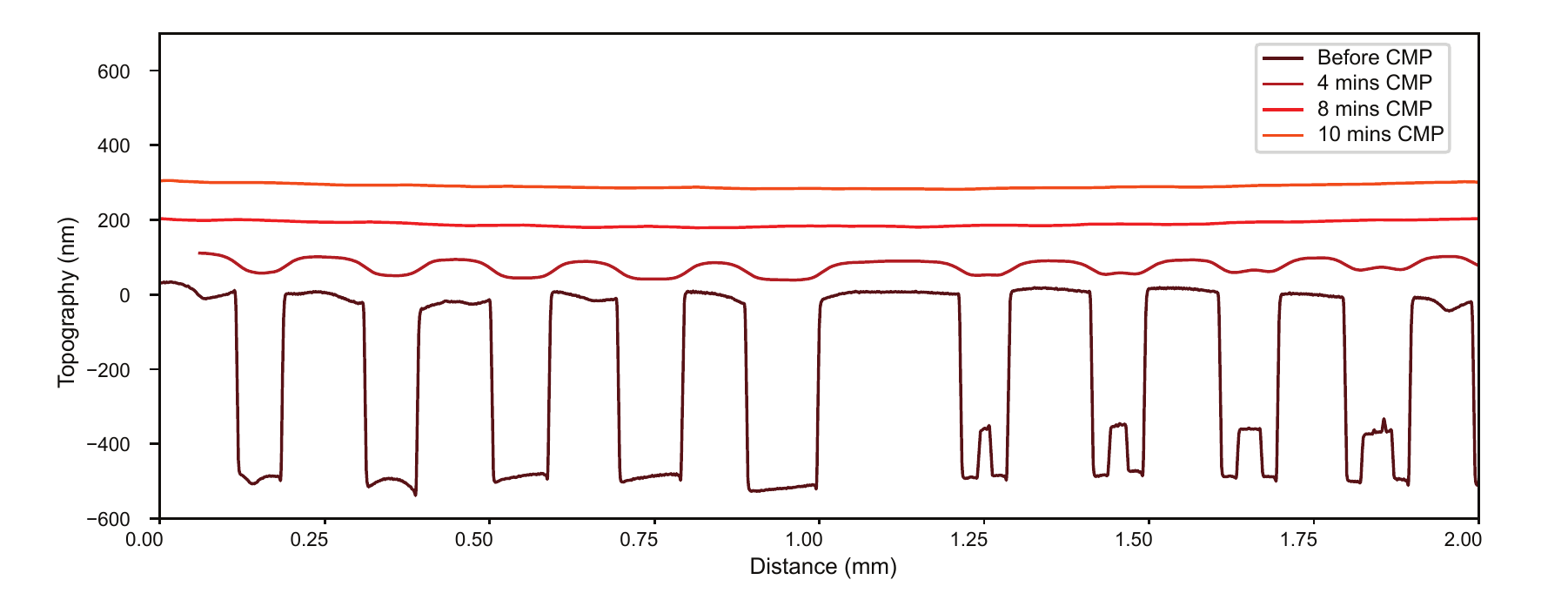}
									\caption{\textbf{Chemical mechanical polishing (CMP) for removing the surface topography}. CMP enables us to reduce the surface topography from $\sim500$ nm to below 10 nm. The figure shows the effect of sequential CMP steps on the topography measured by mechanical profilometry. The final global curve is the wafer bow.}
									\label{fig:SI_CMP}
								\end{figure}
								
								Here, we describe the fabrication process in more detail. We use a high-resistive and low-bow silicon wafer (supplied from Topsil, Fig.~\ref{fig:SI_fab}a) and define a 300 nm trench using deep reactive ion etching with $\rm{C_4F_8}$ gas (Alcatel AMS200, Fig.~\ref{fig:SI_fab}b). We then deposit 100 nm aluminum using electron beam evaporation (Alliance-Concept EVA 760, Fig.~\ref{fig:SI_fab}c) and pattern it to define the bottom part of the circuit using wet etching ($\rm{H_3PO_4}\; 85\% + \rm{CH_3COOH }\; 100\% + \rm{HNO_3 }\; 70\%$ 83:5.5:5.5, Fig.~\ref{fig:SI_fab}d). The next step is the deposition of the sacrificial layer, where we use 2.5~$\mu$m silicon oxide using low thermal oxide deposition (LTO, Fig.~\ref{fig:SI_fab}e). The surface topography, due to the presence of trench and the bottom plate, is planarized using chemical mechanical polishing (CMP)(ALPSITEC MECAPOL E 460, Fig.~\ref{fig:SI_fab}f). This reduces the surface topography to less than 10 nm (an example of the polishing process is shown in Fig.~\ref{fig:SI_CMP}). We made dummy trenches on all the empty space of the wafer to increase the uniformity in the CMP process. To remove the remaining sacrificial layer and land on silicon surface, we etch back the oxide using ion beam etching (IBE,Veeco Nexus IBE350, Fig.~\ref{fig:SI_fab}g). Prior to the deposition of top layer, we have to make an opening in the oxide to make a galvanic connection between top and bottom layers of the circuit in the spiral inductor and capacitor. To this end, the coated photoresist (Fig.~\ref{fig:SI_fab}h1) is reflowed (30 sec at 200 degree Celsius) to make slanted sidewalls to smoothly connect two layers (Fig.~\ref{fig:SI_fab}h2). Afterwards, we use deep reactive ion etching (DRIE) with $\rm{CHF_3}$ as etchant at equivalent rates for both the photoresist and $\rm{SiO_2}$ to transfer the photoresist pattern to the oxide (SPTS APS). One important step is to remove the native aluminum oxide prior to making the galvanic connection, where we observed such a thin resistive layer can significantly increase the cavity heating (see Sec.~\ref{sec:heating_treatment} for more details). To this end, we use an ultra-high vacuum electron beam evaporator (PLASSYS), which enables us to do argon milling (Fig.~\ref{fig:SI_fab}h3) to etch $\sim6$~nm aluminum oxide and then deposit 180 nm aluminum without breaking the vacuum (Fig.~\ref{fig:SI_fab}h4). The next step is patterning top aluminum layer using the aforementioned wet etching. 
								After dicing the wafer into chips, we finally release the structure using Hydrofluoric (HF) acid vapor (SPTS uEtch, Fig.~\ref{fig:SI_fab}i), which is an isotropic etch process dedicated for MEMS structuring and does not attack Al. The holes on the drumhead are designed to facilitate the release process. \added{After the release step, the chip will be glued to the copper sample box using silver paste and wire bonded to PCB ports of the sample box.} All patterning steps are performed by direct mask-less optical lithography (Heidelberg MLA 150) using 1$\mu$m photo resist (AZ ECI 3007). All coatings and developing of photoresist are processed using automatic coater/developer Süss ACS200 GEN3.
								
								\added{We note that we did not systematically investigate the minimum gap size that can be achieved using this process, but observed a lower successful release rate (in the HF vapor etching step) for gap sizes below 100~nm most probably due to the van~der~Waals force between two plates or water formation during the release step. 	Increasing the thickness of the top layer, reducing the radius, and decreasing the HF release etch rate (lowering the pressure and increasing etching time) may help to increase the release success rate. In addition, the compressive room temperature stress of the top layer of the aluminum thin film helps the buckling of the drumhead and facilitates the release. This thermally induced deposition stress can be controlled by evaporation rate and temperature.	Optimizing the release process using the methods mentioned above combined with enhancing the CMP planarization may allow us to achieve much lower effective gap sizes down to the ultimate limit of the roughness of the top and bottom aluminum films (in the order of a few nano-meters).
									It worth to mention that the fabrication induced disorder in identically designed LC circuits is measured below 1\% for mechanical frequencies and 0.5\% for microwave frequencies~\cite{SI_youssefi2021superconducting}.
								}

								\section{Experimental setup}
								Here we describe the experimental setup including room temperature and cryogenic wiring, as well as hardwares dedicated to circuit optomechanical experiments, such as microwave filter cavities and tone cancellation setup.
								
								\subsection{Wiring}{\label{sec:exp_setup}}
								\begin{figure*}[!]
									\includegraphics[scale=1]{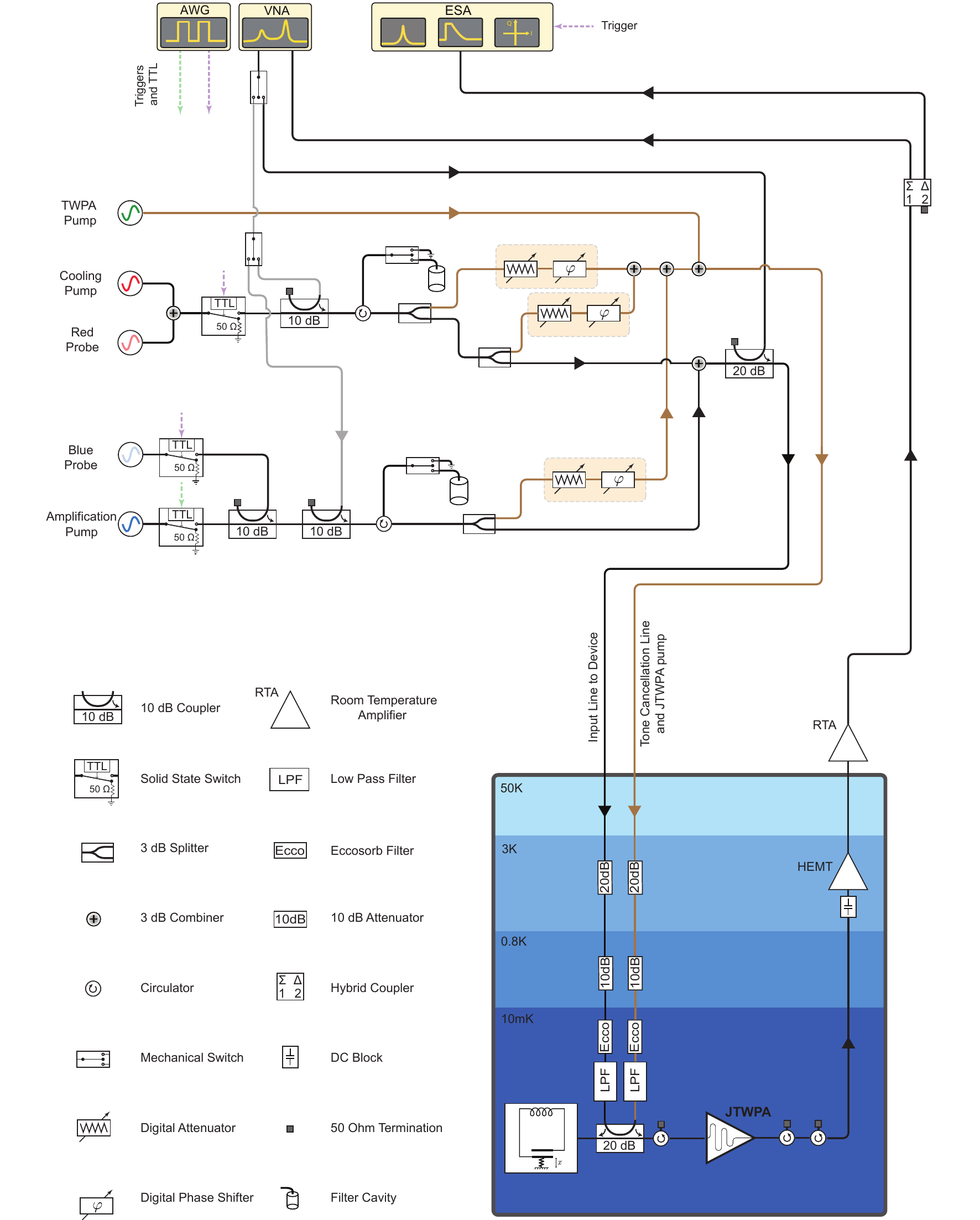}
									\caption{\textbf{Detailed experimental setup.} The black lines shows the main signal path of the control and readout. The brown lines show the tone cancellation and JTWPA pump path. The gray lines show auxiliary paths used for filter cavity tuning. The solid state switches are only used in the time domain experiments.}
									\label{fig:SI_exp}
								\end{figure*}
								The experimental setup, including room temperature (RT) and cryogenic, can be seen in Fig.~\ref{fig:SI_exp}. For the sideband cooling experiment, four sources are required: one for pumping JTWPA near its stop band (Rohde \& Schwartz, SGS100A) to activate a 4-wave mixing amplification process, one strong pump for optomechanical sideband cooling (Rohde \& Schwartz, SMA100B), which is red detuned from the cavity, and two weak probes (Keysight, N518313) to generate Stokes and anti-Stokes sidebands for sideband asymmetry calibration. All signals from sources first go through a microewave filter cavity to remove the phase noise around the cavity frequency (see Sec.~\ref{sec:filter_cavity}). The reflected signal from the filter cavity will be divided in two paths: one goes to the device in the fridge and the other one goes to a tone cancellation line, which consists of a variable digital phase shifter (Vaunix, LPS-802) and attenuator (Vaunix, LDA-602EH). 
								The signal from the tone cancellation line will be combined with the TWPA input line to realize a destructive interference and suppress strong optomechanical drives that saturate the JTWPA (see Sec.~\ref{sec:tone_cancellation}).
								A coherent signal from the Vector Network Analyzer (VNA: Rohde \& Schwartz, ZND) is also combined with the input line to measure the microwave frequency response of the device. Using mechanical microwave switches (Mini-Circuits, MSP2T-18XL+), we can redirect the main signal path to directly observe the frequency response of the filter cavities and accurately tune them. 
								
								In time domain experiments, where we observe the thermalization of the vacuum and squeezed states, we use an additional microwave source, as a blue-detuned pump for optomechanical amplification (Rohde \& Schwartz, SMA100B). The cooling pump and blue probe are also used for generating thermal states or squeezed states. All microwave pulses are generated using ultra-fast solid state switches (Planar Monolithics Industries, P1T-4G8G-75-R-SFF) with $\sim 50$~ns rising/falling time, and are controlled with an arbitrary wave generator (AWG: Tektronix, AFG325230). The output signal from the fridge divides in two parts using a hybrid coupler: one part goes to the second port of the VNA to measure scattering parameters, and the other part goes to the Electronic Spectrum Analyzer (ESA: Rohde \& Schwartz, FSW). Depending on purposes, we use different modes of ESA: for measuring PSD, e.g. in the sideband asymmetry experiment, we use frequency domain mode, for measuring signals in the time domain e.g. optomechanical amplification experiment, we use I-Q analyzer mode, and for measuring the ringdown of the mechanical oscillator, we use the zero span PSD measurement mode. In the time domain experiments, the ESA is triggered by the AWG. 
								
								The device is mounted in the mixing chamber of the dilution refrigerator (BLUEFORS, LD250). In the refrigerator, for both the input line to the device and the tone cancellation line, we use cryogenic attenuators, Eccosorb filters, and 18-GHz low pass filters. The reflected signal from the device which is combined with the tone cancellation signal is amplified using a JTWPA (provided by MIT Lincoln Lab), which is a nearly quantum limited amplifier \cite{SI_macklin2015near}. Before the JTWPA, we use a cryogenic circulator (Low Noise Factory, LNF-CIC4-12A) to remove reflections from JTWPA back to the device. We also use circulators after the JTWPA to remove hot noise penetrating down from the higher temperature stages to the JTWPA. The signal is then amplified using a High-Electron-Mobility Transistor (HEMT: Low Noise Factory, LNF-LNC4-8C) at 4 Kelvin, followed by a Room Temperature Amplifier (RTA: Mini-Circuits, ZVA-183-S+).

								\subsection{Tone cancellation with digital phases shifter/attenuator}
								\label{sec:tone_cancellation}
								\begin{figure*}[!]
									\centering
									\includegraphics[width=\textwidth]{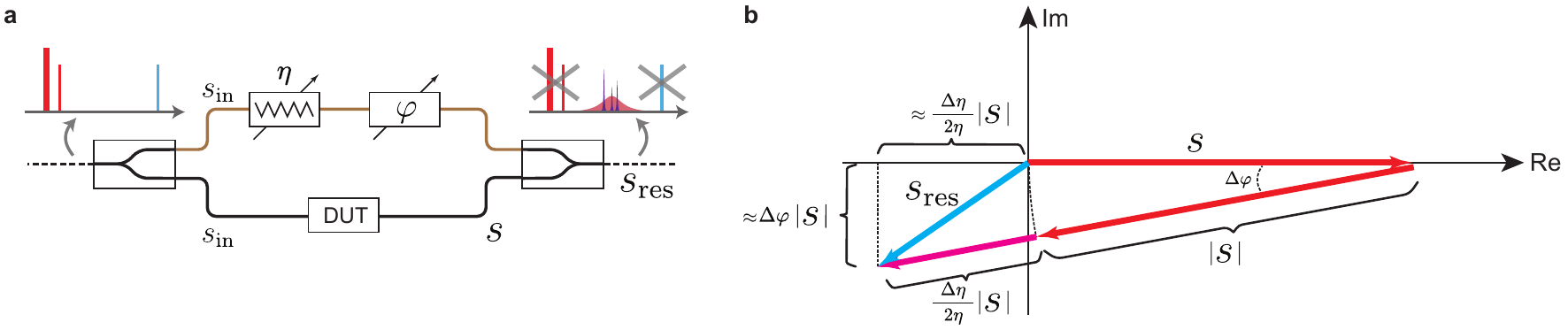}
									\caption{\textbf{Tone cancellation in the complex plane.} \textbf{a}, Splitting a signal into two paths and adding an attenuation and inverse phase using the digital attenuator and phase shifter, results in a destructive interference upon recombining them. This is used to cancel out strong optomechanical pumps to avoid saturation of the JTWPA. \textbf{b}, Due to the finite step size of the phase shifter ($\Delta \varphi$) and attenuator ($\Delta \eta$), the combined signal (red vectors plus pink vector) cannot completely vanish. The maximum possible power of the remaining signal (blue vector) determines the guaranteed limit of the tone cancellation.}
									\label{fig:SI_TC}
								\end{figure*}
								To avoid the saturation of the JTWPA by strong optomechanical pumps, we have to cancel the signal before the JTWPA by destructively interfering it with a tone cancellation signal. This is realized by splitting the signal at room temperature into two paths and adding a phase and attenuation to the signal in the tone cancellation path, and combining it with the signal right before the JTWPA. We use digital attenuators and phase-shifters for this purpose, as shown in Fig.~\ref{fig:SI_exp}.
								In this section, we calculate the guaranteed level of the tone cancellation which can be achieved using the digital phase shifters and digital attenuators.
								
								For simplicity, we define the signal after the splitter as $s_\mathrm{in}$ and the signal from the device under test as $s$ (as shown in Fig.~\ref{fig:SI_TC}a). 
								Note that we here consider the case with a balanced splitter, which can be straightforwardly extended to be general.
								In the ideal case, the attenuator in the cancellation path should add the same total attenuation as the signal in the main path is experiencing, while the phase shifter is adding a $\pi$ phase shift for the destructive interference. However, our digital attenuator and phase-shifter can only change the attenuation and phase by finite steps of 0.25~dB and $\frac{\pi}{180}$, respectively. This limits the maximum guaranteed cancellation that can be achieved in the system.
								In this case, the residual combined signal can be written as
								\begin{equation}
									s_\mathrm{res} = s+\sqrt{\eta + \Delta\eta} e^{i(\varphi+\Delta\varphi)}s_\mathrm{in},
								\end{equation}
								where $\eta$ and $\varphi$ are the optimal attenuation and phase shift to destructively interfere with the signal, resulting in $\sqrt{\eta}e^{i\varphi} s_\mathrm{in}= -s$, while $\Delta\eta$ and $\Delta\varphi$ are unwanted errors in the tone cancellation elements. An illustration in complex plane is provided in Fig.~\ref{fig:SI_TC}b.
								Using Taylor expansion, we then approximate the combined signal as
								\begin{equation}
									s_\mathrm{res} \simeq -\left( i \Delta\varphi + \frac{\Delta\eta}{2\eta}\right) s.
								\end{equation}
								Using the relative error in the attenuation, the error in dB is described as
								\begin{equation}
									\begin{aligned}
										\Delta {({\rm{Att.}})_{\rm{dB}}} &= \Delta ( - 10{\log _{10}}\eta )\\
										&=  - 10\Delta ({\log _{10}}\eta )\\
										&= \frac{{ - 10}}{{\ln 10}}\frac{{\Delta \eta }}{\eta }.
									\end{aligned}
								\end{equation}
								Hence, the guaranteed tone cancellation in dB can be calculated as 
								\begin{equation}
									10{\log _{10}}(s_\mathrm{res}/s) = 10{\log _{10}}\left( {\Delta {\varphi ^2} + {{\left( {\frac{{\ln 10}}{{20}}\Delta {{({\rm{Att.}})}_{\rm{dB}}}} \right)}^2}} \right).
								\end{equation}
								This guaranteed cancellation is independent of absolute values of $\eta$ and $\varphi$. Therefore even in the case of unbalanced signal splitter, different paths, or unbalanced coupler, where different values of attenuation or phase shift is needed to compensate the effective path different between two signals, the residual power will not be changed and only is dependent to the errors of phase shifter and attenuator.
								Considering the reported values for the Vaunix phase shifter $\Delta\varphi = \frac{\pi}{2\times180}$ and attenuator ${\Delta {{({\rm{Att.}})}_{\rm{dB}}}} = 0.25/2$ dB (division by 2 in both cases originates from their digital nature), we can achieve at least -35 dB cancellation for every tone cancellation branch. Using several tone cancellation branches multiplies the cancellation level. We use two branches for the optomechanical cooling pump cancellation corresponding to at least -70~dB cancellation. It is worth to mention that the effective bandwidth of the tone cancellation is inversely proportional to the effective microwave path length ($L_\mathrm{path}$) : $\mathrm{BW}_\mathrm{TC} \propto \frac{\pi v}{L_\mathrm{path}}$, where $v$ is the phase velocity of the signal in the microwave cables. Considering a few meters length of wiring in our setup, this bandwidth is $\mathrm{BW}_\mathrm{TC} = \mathcal{O} (100~\mathrm{kHz})$ in our setup. For this reason we use another tone cancellation branch for the blue detuned pumps which are detuned by $2\Omega_\mathrm{m} /2\pi= 3.6$~MHz from the red tone.

								\subsection{Microwave filter cavities}
								\label{sec:filter_cavity} 
\begin{figure*}[!]
									\includegraphics[scale=1]{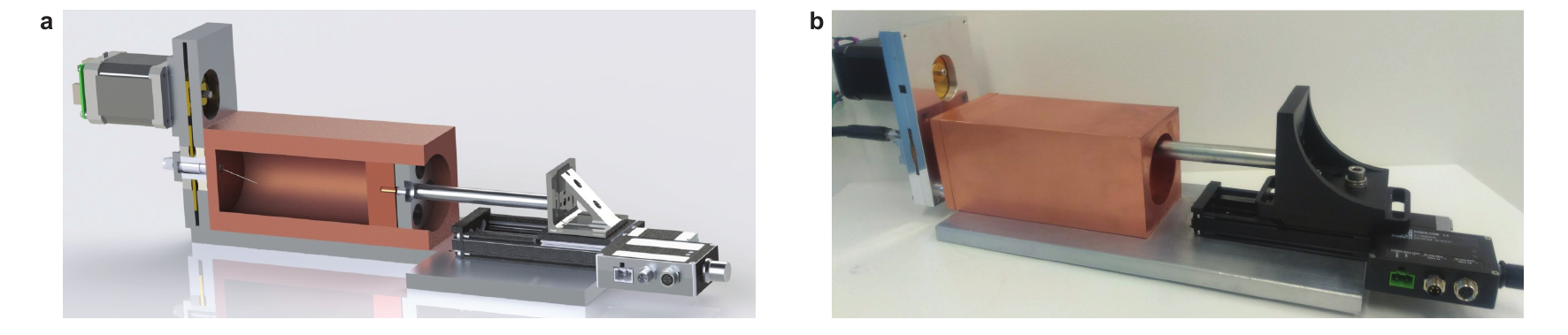}
									\caption{\textbf{Motorized filter cavity.} \textbf{a} The schematic design (cross section). \textbf{b}, The actual device. The frequency of the cavity can be tuned by changing its length using a linear micro-positioner which can be controlled remotely. (Image courtesy: Joshi,~\textit{et.al.}~2021~\cite{SI_joshi2021automated})}
									\label{fig:SI_filter_cavity}
								\end{figure*}

								The classical phase noise in a microwave pump for sideband cooling can in principle drive the mechanical resonator and impose a limitation for cooling. The lowest attainable mechanical occupation in the presence of the phase noise~\cite{SI_aspelmeyer2014cavity,SI_rabl2009phase} is given by
								\begin{equation}
									n_\mathrm{m}^\mathrm{(min)} = \sqrt {\frac{{\Omega_{\rm{m}}^2{{ n}_{{\rm{m}}}^{\rm{th}}}{\Gamma _{\rm{m}}}}}{{g_0^2}}{{\bar S}_{\varphi \varphi }}({\Omega _{\rm{m}}})},
								\end{equation}
								where ${\bar S}_{\varphi \varphi}({\Omega _{\rm{m}}})$ is the phase noise spectral density at the mechanical frequency in the rotating frame of the pump frequency. To achieve $n_\mathrm{m}^\mathrm{(min)}<1$, the theoretical limitation requires
								\begin{equation}
									{{\bar S}_{\varphi \varphi }}({\Omega _{\rm{m}}}) < \frac{{g_0^2}{n_\mathrm{m}^\mathrm{(min)}}^2}{\Omega_{\rm{m}}^2{{{ n}_{{\rm{m}}}^{\rm{th}}}{\Gamma _{\rm{m}}}}}.
								\end{equation}
								Considering our system parameters, the theoretical limit to realize $n_\mathrm{m}^\mathrm{(min)}<0.1$~quanta is -137 dBc/Hz at 1.8~MHz detuning from the microwave pump. The measured phase noise for our microwave sources (Rohde \& Schwartz, SMA100B) is below -140 dBc/Hz, so it almost fulfills this requirement. However, higher-fidelity ground-state cooling requires a further lower phase noise. This is why we use a microwave filter cavity \cite{SI_joshi2021automated} which is automated, tunable, and narrow-band (Fig.~\ref{fig:SI_filter_cavity}). It reduces the phase noise within $\sim 50$ kHz linewidth below -155~dBc/Hz on the microwave cavity frequency, which makes sure that classical phase noise cannot drive mechanical system.
								
\begin{figure*}[!]
									\includegraphics[scale=1]{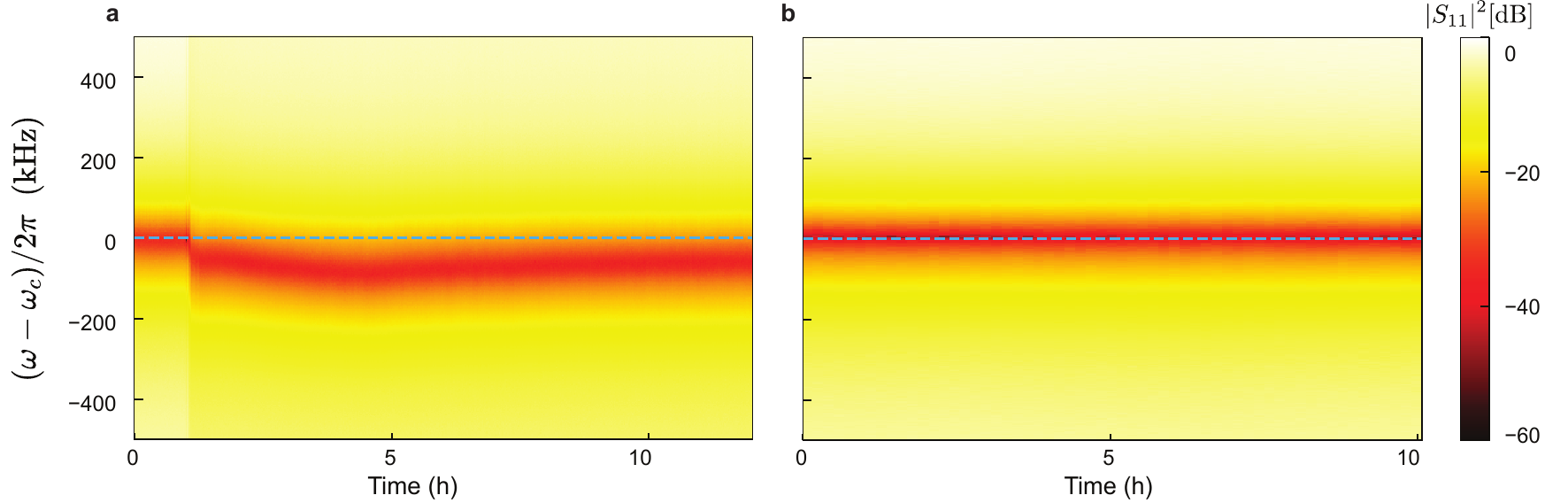}
									\caption{\textbf{Frequency stability of the motorized filter cavity.} Filter cavity response \textbf{a}, without and \textbf{b}, with adaptive frequency stabilization. The dashed blue line indicates the target filter frequency. Due to slight environmental temperature change, the cavity without stabilization can be completely detuned from the target frequency. The issue is solved using an adaptive stabilization technique.}
									\label{fig:SI_phase_noise}
								\end{figure*}

								The frequency stability of the filter cavity is essential, especially for the thermal decoherence measurements, where the experiment and calibration process continues for hours. Due to the large size ($\sim 15$ cm) and high thermal expansion rate of the copper of the filter cavity, we observe a slow frequency drift of the filter cavity frequency. In fact, we estimate 0.5 centigrade temperature difference can shift the resonance frequency by more than 50 kHz (as shown in Fig.~\ref{fig:SI_phase_noise}a). Therefore, we monitor the resonance frequencies of the filter cavities during the experiment by re-routing the VNA signal through them using mechanical switches, and adaptively tune their frequency using a sub-micron accuracy linear positioner. The frequency stability of the cavities with and without the adaptive tuning is shown in Fig.~\ref{fig:SI_phase_noise}, where the stabilization process is clearly keeping the cavity resonance on the target frequency. The design and control codes for such cavities are available online and can be found in the supplementary materials of Joshi,~\textit{et.al.}~(2021)~\cite{SI_joshi2021automated}.

								\section{Characterization}
								In this section we discuss characterization techniques used to measure and extract parameters of our circuit optomechanical system.
								
								\subsection{Measurement of single photon optomechanical coupling rate}{\label{sec:g0_meas}}
								The single photon optomechanical coupling rate, $g_0$, is an important figure of merit for an optomechanical systems, which describes the cavity resonance shift induced by the zero-point fluctuation of motion. The method we used here for measuring $g_0$ is based on the PSD measurement of the motional sidebands when pumping the system on-resonance ($\omega_{\mathrm{c}}$) with a relatively weak pump \cite{SI_bernier2019multimode}. In this case, two sidebands will appear ($\omega=\omega_{\mathrm{c}} \pm \Omega_{\rm{m}}$, Fig.~\ref{fig:SI_g0}a), where the power in the upper motional sideband is given by
								\begin{equation}
									{P_{{\rm{SB}}}} = 4g_0^2{{n}_{\rm{m}}^{\rm{th}}}\frac{{{{({\kappa _{{\rm{ex}}}}/\kappa )}^2}}}{{{\Omega _{\rm{m}}}^2 + {{(\kappa /2)}^2}}}\frac{{{\omega _{\mathrm{c}}}}}{{{\omega _{\mathrm{c}}} + {\Omega _{\rm{m}}}}}{P_{\rm{MW}}}.
								\end{equation}
								Here, ${P_{{\rm{SB}}}}$ is the scattered sideband power emitted from the device at $\omega=\omega_{\mathrm{c}}+\Omega_{\rm{m}}$, $P_{\rm{MW}}$ is the microwave input pump power to the device at $\omega=\omega_{\mathrm{c}}$, and $n_{\rm{m}}^{\rm{th}}$ is the mechanical thermal bath occupation at temperature $T$. On resonance pumping does not induce dynamical backaction (i.e. damping or anti-damping) on the mechanics; however, the back action noise can still heat up the mechanical oscillator~\cite{SI_bowen2015quantum}. Here we use microwave powers that results in negligible back-action noise of the microwave pump on the mechanical oscillator, i.e. ${{n}_{{\rm{ba}}}} \equiv \frac{{4g_0^2}}{{\kappa {\Gamma _{\rm{m}}}}}\frac{{(4{\kappa _{{\rm{ex}}}}/{\kappa ^2}){P_{\rm{MW}}}}}{{1 + {{(2{\Omega _{\rm{m}}}/\kappa )}^2}}} \ll {{n}_\mathrm{m}^\mathrm{th}}$, where ${{n}_{{\rm{ba}}}}$ is the equivalent back-action noise in units of quanta~\cite{SI_bowen2015quantum}. 
								
								While it is challenging to directly measure $P_{\rm{MW}}$ and ${P_{{\rm{SB}}}}$, we can measure the sideband at the detector (spectrum analyzer) and the pump power at the microwave source. The pump signal is attenuated with an unknown factor $\eta_{\rm{att}}$ from the source to the device, and the measured sideband is amplified with an unknown factor $G$ from the device to the detector. Since we need to use the tone cancellation of the pump to avoid the saturation of the JTWPA and also the dynamic range of the ESA is limited, we send an additional weak calibration tone placed at the upper motional sideband frequency with a small detuning ($\omega = \omega_{\mathrm{c}}+\Omega_{\rm{m}}+\delta$) that passes through the same input/output lines as the sideband signal, so we can accurately obtain the relative power of the calibration tone to the on-resonance pump. In this case we have
								\begin{equation}
									\begin{aligned}
										P_{{\rm{SB}}}^{{\rm{meas}}} =& G{P_{{\rm{SB}}}} = G\frac{{{P_{{\rm{SB}}}}}}{{{P_{\rm{MW}}}}}\eta_{\rm{att}} P_{\rm{MW}}^{{\rm{src}}}\\
										P_{{\rm{cal}}}^{{\rm{meas}}} = &G\eta_{\rm{att}} \frac{{{\Omega _{\rm{m}}}^2 + {{(({\kappa _{{\rm{ex}}}} - {\kappa _0})/2)}^2}}}{{{\Omega _{\rm{m}}}^2 + {{(({\kappa _{{\rm{ex}}}} + {\kappa _0})/2)}^2}}}P_{{\rm{cal}}}^{{\rm{src}}},
									\end{aligned}
								\end{equation}
								where $P_{{\rm{SB}}}^{{\rm{meas}}}$ is the measured sideband at the detector, $P_{\rm{MW}}^{{\rm{src}}}$ is the pump power at the microwave source, $P_{{\rm{cal}}}^{{\rm{meas}}}$ is the measure calibration pump power at the detector, and $P_{{\rm{cal}}}^{{\rm{src}}}$ is the calibration tone power at the microwave source. In the above expression, the fraction in the calibration tone comes from its interaction with the microwave cavity. 
								
								We can now eliminate the unknown parameter $G\eta_{\rm{att}}$ and arrive at: 
								\begin{equation}
									\frac{{P_{{\rm{SB}}}^{{\rm{meas}}}}}{{P_{\rm{MW}}^{{\rm{src}}}}}\frac{{P_{{\rm{cal}}}^{{\rm{src}}}}}{{P_{{\rm{cal}}}^{{\rm{meas}}}}} = 4g_0^2{{n}_{\rm{m}}}\frac{{{{({\kappa _{{\rm{ex}}}}/\kappa )}^2}}}{{{\Omega _{\rm{m}}}^2 + {{(({\kappa _{{\rm{ex}}}} - {\kappa _0})/2)}^2}}}\frac{{{\omega _{\mathrm{c}}}}}{{{\omega _{\mathrm{c}}} + {\Omega _{\rm{m}}}}}.
									\label{eq:g0_cal}
								\end{equation}
								At high temperatures, the mechanical occupation is approximately given by $n_\mathrm{m} \approx n_\mathrm{m}^\mathrm{th} = k_{\rm{B}}T/\hbar\Omega_{\rm{m}}$, where $k_{\rm{B}}$ is the Boltzmann constant, so the measured sideband power is linearly proportional to the temperature. 
								\begin{figure*}[!]
									\includegraphics[scale=1]{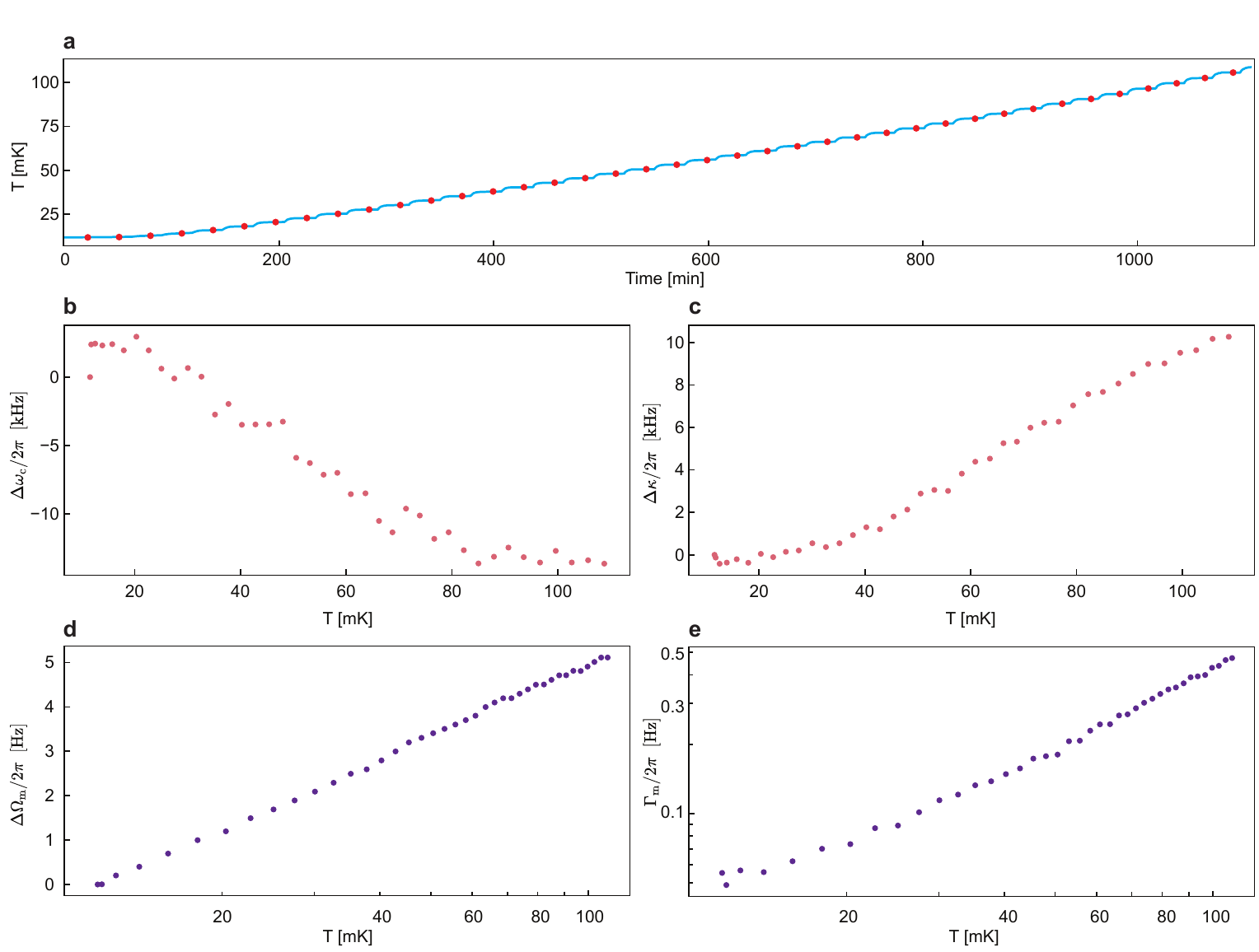}
									\caption{\textbf{Temperature dependence of system parameters. a,} Temperature of the mixing chamber versus time. Measurements are performed when the temperature is stable (red data points). \textbf{b}, Shift of the microwave cavity frequency and \textbf{c}, linewidth versus temperature. \textbf{d}, Shift of the mechanical frequency \textbf{e}, linewidth versus temperature.}
									\label{fig:SI_T_sweep}
								\end{figure*}
								
								To characterize $g_0$, we first sweep the mixing chamber temperature. After the stabilization of the fridge's temperature (Fig.~\ref{fig:SI_T_sweep}a), the mechanical and microwave parameters, such as $\kappa$ and $\omega_{\mathrm{c}}$ are measured  by taking a VNA trace (Fig.~\ref{fig:SI_T_sweep}b and c), while $\Gamma_{\rm{m}}$ is measured by a ring down experiment and $\Omega_{\rm{m}}$ is determined as the frequency difference between the optomechanical sideband and the applied microwave pump (Fig.~\ref{fig:SI_T_sweep}d and e). Both the mechanical and microwave frequencies slightly vary with temperature (a few parts per million compared to the respective frequencies). Nevertheless, we observe a strong temperature dependence in $\Gamma_\mathrm{m}$, which indicates that the mechanical dissipation in our drumhead resonator is not dominated by the clamping loss, i.e. phonon tunneling through substrate (see more discussion in Sec.~\ref{sec:mechanics}).
								
								By measuring the sideband power induced by the resonant cavity pump as a function of temperature and fitting a linear function to the results as shown in Fig.~\ref{fig:SI_g0}b, the single photon optomechanical coupling rate is found to be $g_0/2\pi = 13.4 \pm 0.5\;\rm{Hz}$. The experimentally measured value is in good agreement with the theoretically expected one (see Sec.~\ref{sec:g0}). Note that the JTWPA performance is monitored at each temperature to verify its stable gain during the temperature sweep.
								
								\added{It is worth mentioning that in principle higher $g_0$ values can be achieved by lowering the gap size without sacrificing other optomechanical parameters. We did not systematically investigate the fabrication of samples with lower gaps in this work. More information about the limitations and feasibility of reducing the gap size is provided in the fabrication section (Sec.~\ref{sec:fab}). }
								
								\begin{figure*}[!]
									\includegraphics[scale=1]{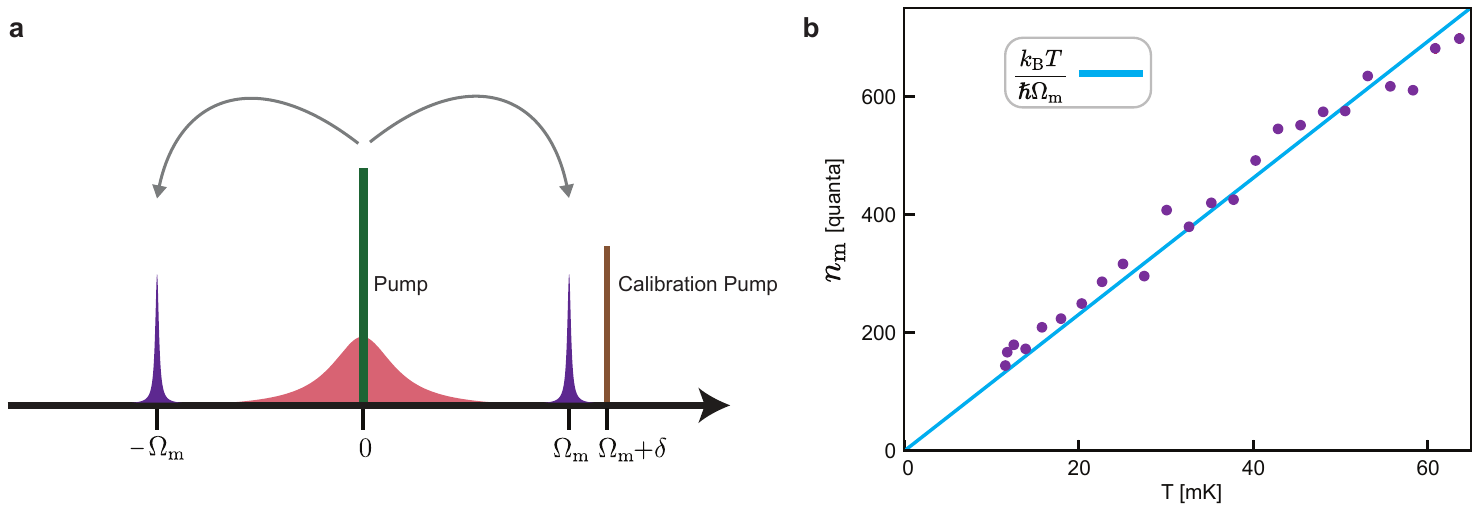}
									\caption{\textbf{$g_0$ measurement. a,} Frequency scheme for $g_0$ measurement experiment. An on-resonance pump (green line) is used to generate two motional sidebands (in purple). A calibration tone (brown) is placed close to the motional sideband to rule out the unknown parameters for the input loss/output gain. \textbf{b}, Mechanical occupations (calculated after extracting $g_0$ from the measured PSD of the motional sidebands and fitting) versus nominal temperature of the fridge.}
									\label{fig:SI_g0}
								\end{figure*}

								\subsection{Ringdown measurement}
\begin{figure*}[h!]
									\includegraphics[scale=1]{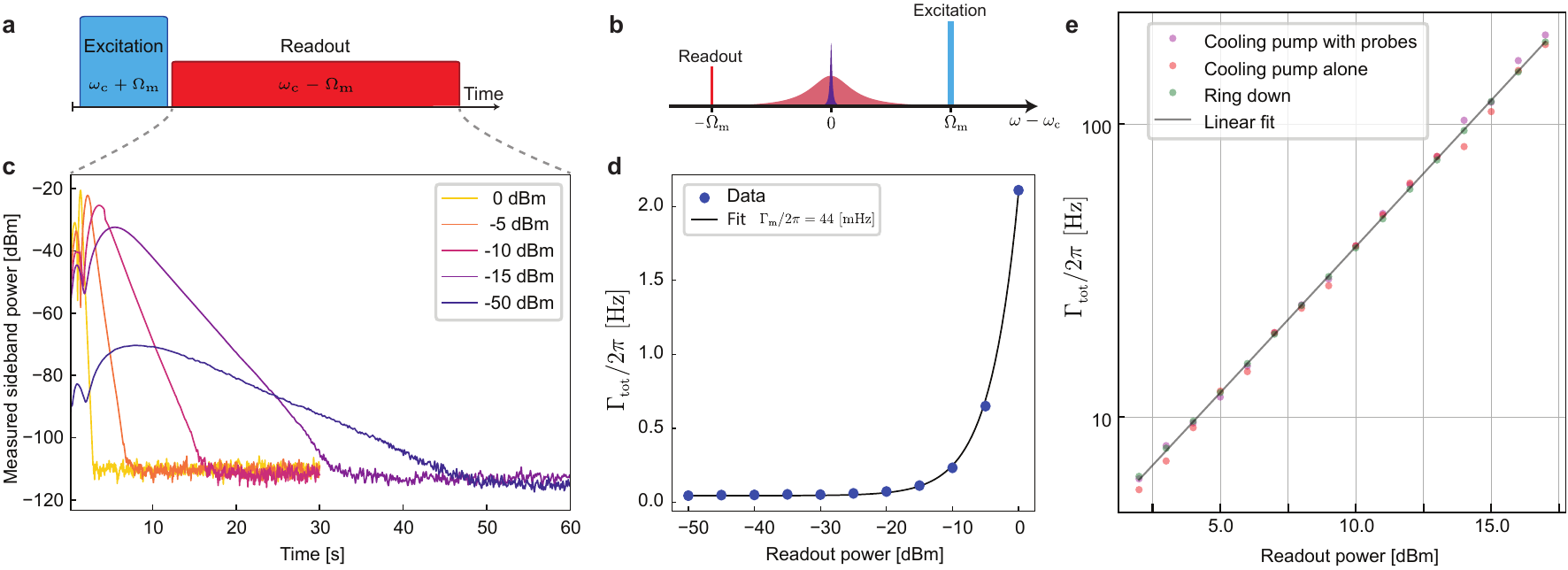}
									\caption{\textbf{Ringdown measurements.} \textbf{a, b} Pulse and frequency scheme of the ring down measurement. An strong blue detuned pulse is exciting the mechanical oscillator through optomechanical parametric instability. A red detuned readout pulse generates optomechanical sideband on resonance. \textbf{c}, Example of ringdown traces measured for different readout powers. The initial nonlinear behavior in the ringdown trace may be due to the energy exchange between different mechanical modes of the drumhead at high amplitude vibrations. For the exponential fitting we only use the low power linear part of the ring down. \textbf{d}, Effective mechanical damping rates extracted from the linear part of the ringdowns versus the readout nominal power at the source. The fit shows bare mechanical damping rate of $\Gamma_\mathrm{m}/2\pi = 44$~mHz. \textbf{e}, Effective mechanical damping rates versus nominal source powers in the optomechanical cooling experiment (measured in different room temperature setup). Mechanical linewidth is measured directly by ringdown (green) and by fitting to the measured PSD f the cooling pump's sideband when sideband asymmetry probes are on (magenta) and off (red). The fact that extracted linewidths are same in three experiments verifies that the effective dynamical back action of balanced probes is negligible compared to the cooling pump.}
									\label{fig:SI_ringdown}
								\end{figure*}

								In order to measure the bare mechanical damping rate, $\Gamma_\mathrm{m}$, we use a time-domain experiment where we first excite the mechanical oscillator by applying a strong blue pump, having the system in an optomechanical parametric instability, and then observe the energy decay by measuring the optomechanical sideband scattered from a weak red-detuned probe in time, as shown in Figs. \ref{fig:SI_ringdown}a and b. The effective mechanical damping rate in the presence of a red-detuned pump with cooperativity of $\mathcal{C}$ is $\Gamma_\mathrm{tot} = \Gamma_\mathrm{m} (1+\mathcal{C})$. Sweeping the power of the red probe enables us to directly measure $\Gamma_\mathrm{m}$ (Figs.~\ref{fig:SI_ringdown}c and d). In addition, we can accurately calibrate the cooperativity for every pump power, by measuring the effective damping rate in other experiments. The optomechanical amplification rates are also measured in a similar way, by recording the energy increase rate of the sideband generated by a blue pump.
								
								Furthermore, we use ringdown in the optomechanical cooling experiment to verify that the dynamical back-actions of the balanced probes are negligible compare to the cooling pump. This is confirmed by comparing the linewidth of the measured PSD of optomechanical sidebands when probes are on and off with the values from the ringdown experiment. As shown in Fig.~\ref{fig:SI_ringdown}e, the agreement between these three values of the total damping rate ensures that two probes are well balanced. Moreover, we can also confirm that the optomechanical coupling still behaves linearly for higher cooling cooperativities. 
								
								\subsection{Microwave cavity heating}
								\label{sec:heating_treatment}
								
								Increasing the intracavity photon number $n_\mathrm{p}$ induced by a pump field enhances the optomechanical coupling rate between a mechanical oscillator and an optical/microwave cavity, as expressed by $g = \sqrt{n_\mathrm{p}} g_0$. To realize mechanical ground-state cooling in contemporary optomechanical platforms, a large $n_\mathrm{p}$ is required -- in the order of $10^4$ to $10^7$ photons, depending on the system parameters. Such a large cavity photon number can lead to thermally heating the optical/microwave cavity, resulting in an increase in the effective temperature of the optical/microwave intrinsic bath.  In the context of sideband cooling of a mechanical oscillator, the cavity heating effect limits the lowest phonon occupation for most cases~\cite{SI_teufel2011sideband}. In this section, we address how to significantly reduce the cavity heating effect in circuit optomechanical platform.
								
								
								A microwave cavity can be coupled to several thermal bathes through different loss mechanisms such as radiation losses, dielectric losses, substrate losses, galvanic connection, etc. In a phenomenological description, we define a loss rate of $\kappa_0^{i}$ for intrinsic bath $i$ with a thermal bath occupation of $n_\mathrm{c}^{\mathrm{th}, i}$. This results in the effective intrinsic loss rate of $\kappa_0  = \sum \kappa_0^i$ and the effective thermal bath occupation of $n_\mathrm{c}^\mathrm{th} = (\sum \kappa_0^i n_\mathrm{c}^{\mathrm{th}, i}) / \kappa_0$. In principle, the thermal bath occupation can be a function of the intracavity photon number, i.e., $n_\mathrm{c}^{\mathrm{th}, i} (n_\mathrm{p})$, since the absorbed energy from the cavity can be inelastically scattered to the bath and increase its effective temperature. However, the dependency of the bath occupation on $n_\mathrm{p}$ can be different due to the different heat capacity and microscopic loss mechanism of each intrinsic bath. For example, the radiative loss thermal bath is expected to be significantly less sensitive to $n_\mathrm{p}$ compared to the dielectric and substrate losses. 
								
								In our platform, the main suspect for the cavity heating was the galvanic connection between the top and bottom aluminum plates of a LC circuit. The native aluminum oxide, which grows on the bottom layer by a few nanometers, remains when the top layer is deposited without any treatment, leading to a thin resistive layer for the LC circuit. The resistive layer dissipates the intracavity energy and heats up the intrinsic bath, inducing a finite cavity thermal photon. To address the heating effect, we perform Argon milling to remove the aluminum oxide layer, followed by the deposition of the top aluminum layer. Importantly, note that these two processes are performed successively under the ultra-high vacuum (see more detail in Sec.~\ref{sec:fab}).
								
								
								To characterize the cavity heating effect, i.e., the cavity thermal photon number induced by one or more microwave drive fields, we measure the cavity thermal emission by using the JTWPA as a nearly quantum limited amplifier.
								As shown in Eq.~(\ref{eq:sym_spec_c}), the noise power spectral density of the cavity emission is given by $\bar{S}_\mathrm{c}(\omega)$. 
								Note that the expression is general regardless of the number of the microwave drive fields.
								By integrating the power spectral density and normalizing it by the external coupling rate with $2\pi$, we obtain the cavity thermal photons as 
								\begin{equation}
									n_\mathrm{c} = \frac{\int d\omega \: \bar{S}_\mathrm{c}(\omega)}{2\pi\kappa_\mathrm{ex}}.
								\end{equation}
								
\begin{figure*}[!]
									\includegraphics[scale=1]{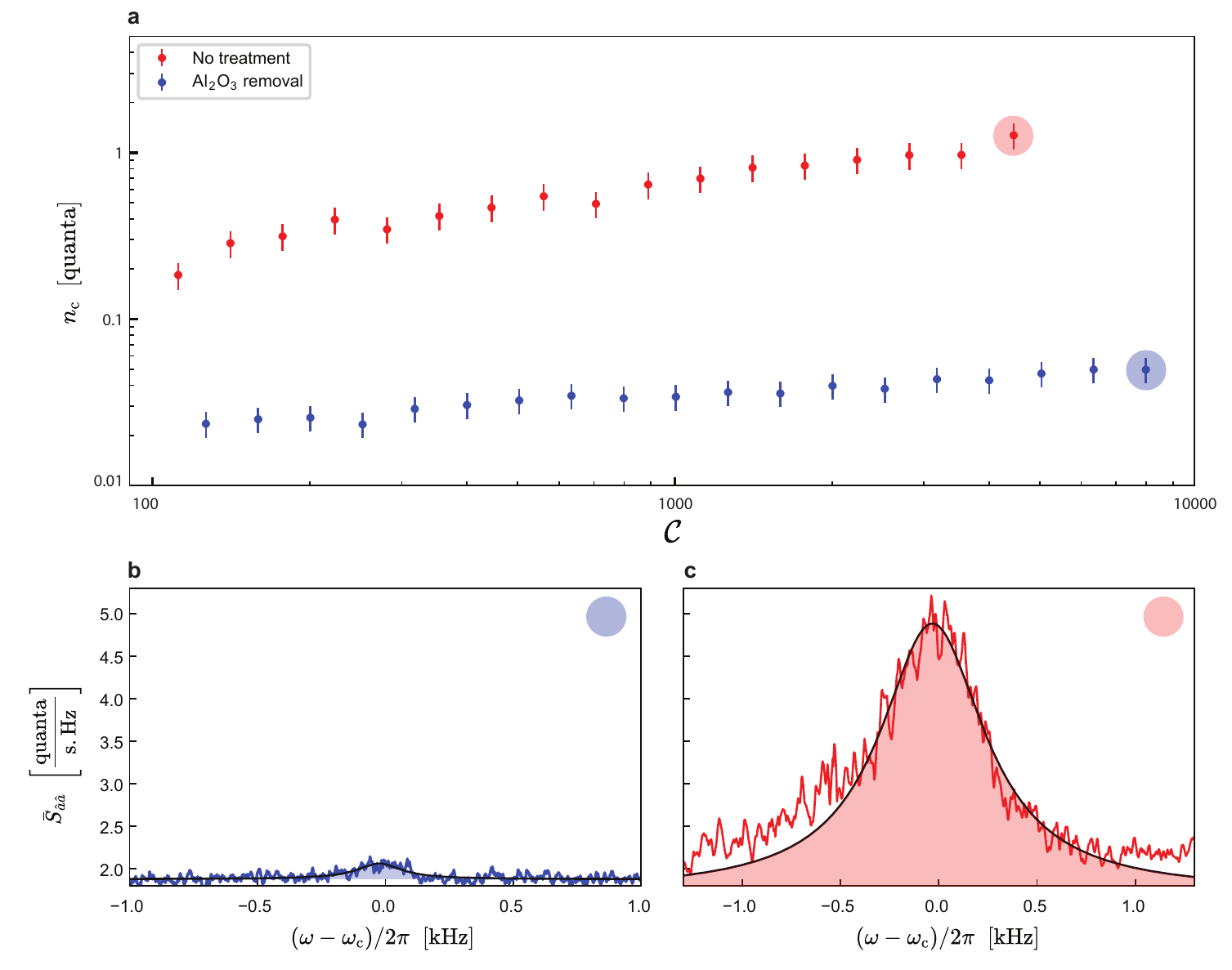}
									\caption{\textbf{Cavity heating effect. a,} Cavity thermal photon number as a function of cooling pump cooperativity for two samples with~(blue) and without~(red) the Argon miling treatment to remove the aluminum oxide layer. The shaded circles correspond to the noise power spectra respectively, shown in b and c. \added{Error bars are corresponding to standard deviations.} \textbf{b, c,}~Noise power spectrum of the cavity thermal emission with a red pump applied to the samples with~b and without~c the Argon milling treatment, respectively. The resonators of both the samples are over-coupled, resulting in the fact that the peaks correspond to $4n_{\mathrm{c}}$.
										.}
									\label{fig:SI_heating}
								\end{figure*}

								In Fig.~\ref{fig:SI_heating}, the cavity themal photon number induced by a strong pump is extracted for two samples with and without the native oxide resistive layer, respectively. By employing the Ar milling to remove the oxide layer, we could reduce the cavity heating effect by a factor of $\sim$30, resulting in a vast improvement in the ground-state cooling of the mechanical oscillator, since the final phonon occupation is usually limited by the cavity heating. 
								
								\subsection{Full data of chip characterization}
								\added{The circuit optomechanical device studied in this work is one of the 16 separate electromechanical LC circuits fabricated on a 9.5~mm$\times$6.5~mm chip (Fig.\ref{fig:SI_chip}). Those 16 LC resonators follow the same design principle as shown in the main text, but frequencies were multiplexed in a chip in the range of 5-7 GHz and ~1.5-2.5 MHz for both microwave and mechanical frequencies respectively. This was done by changing the trench radius for mechanical frequency tuning, and the capacitor bottom plate radius for microwave frequency tuning. All 16 LC circuits were magnetically coupled to a micro-strip waveguide.}
								
								In Table \ref{table:char_table}, we provide the system parameters for all \added{14} independent electromechanical LC resonators in a chip\added{ - we did not observe two LC resonators, most probably due to overlapping their frequencies with the JTWPA stop-band}. More than 50\% of the resonators exhibit more than $20\times10^6$ mechanical quality factor, which demonstrates a high yield in our new fabrication process. 
								
								The total linewidth, intrinsic loss, and external coupling rates of each microwave cavity are obtained by taking the VNA trace and applying a circle fit in the complex plane. For a few microwave cavities, we could not reliably obtain the internal and external coupling rates due to the Fano effect, which may be originated from the impedance mismatch. The microwave cavity used in the main work (No. 2) is strongly over-coupled, while it suffers from the Fano effect. Nevertheless, the total linewidth can be reliably determined by the circle fit even in the presence of the Fano effect.
								Importantly note that all the calibration methods used in the main work (e.g. sideband asymmetry experiment, optomechanical amplification and squeezing) are independent of the absolute values of $\kappa_0$ and $\kappa_{\rm{ex}}$, since our calibration method is based on the total coupling rate ($\kappa$).
								As all the microwave cavities in the same chip are fabricated together, we do not expect to observe a significant statistical deviation of the intrinsic loss rates among them. Therefore, we can assume that the internal coupling rate of our device is $\kappa_0/2\pi=47\pm17\;\rm{kHz}$, which is the average and the standard deviation of all the extracted internal coupling rates. 
								
								\begin{table}[h!]
									\begin{threeparttable}
										\centering
										\caption{Full characterization data of different resonances in a chip.}
										\begin{tabular}{|c| c| c |c |c |c |c ||c|} 
											\hline
											& $\omega_{\mathrm{c}}/2\pi$ (GHz) & $\kappa/2\pi$ (kHz) & $\kappa_{\rm{ex}}/2\pi$ (kHz) & $\kappa_{\rm{0}}/2\pi$ (kHz) & $\Omega_{\mathrm{m}}/2\pi$ (MHz) &$\Gamma_{\mathrm{m}}/2\pi$ (Hz)& $\mathbf{Q_{\mathrm{m}}}$ \textbf{(M)} \\ [0.5ex] 
											\hline\hline
											1 & 5.30 & 610 & - & - &1.48&0.056&\textbf{26.4}\\
											$2^*$ & 5.55 & 250 & - & - &1.80&0.045&\textbf{40.0}\\
											3 & 5.70 & 76 & 46 & 30 &2.10&6.6&\textbf{0.3}\\
											4 & 5.74 & 81 & 36 &45 &1.82&0.076&\textbf{23.9}\\
											5 & 5.80 & 264 & 192 &72&1.56&0.079&\textbf{19.7}\\  
											6 & 5.81 & 179 & 157 &22&1.92&0.058&\textbf{33.1}\\
											7 & 5.86 & 342 & - &-&1.63&0.040&\textbf{40.8}\\  
											8 & 5.92 & 125 & 91 &34&1.96&0.085&\textbf{23.1}\\  				
											9 & 6.09 & 288 & 232 &56&1.70&5.88&\textbf{0.3}\\  	
											10 & 6.12 & 528 & 456 &72&1.74&0.07&\textbf{24.9}\\  		
											11 & 6.20 & 221 & 162 &59&1.77&0.05&\textbf{35.4}\\  		
											12 & 6.25 & 137 & 103 &34&2.24&0.30&\textbf{7.5}\\  				
											13 & 6.36 & 360 & - &-&1.76&13.2&\textbf{0.1}\\  									
											14 & 6.43 & 256 & - &-&1.83&31.0&\textbf{0.1}\\  									
											
											\hline
											
										\end{tabular}
										\begin{tablenotes}
											\small
											\item * This resonator is used for all measurements in the main work. 
										\end{tablenotes}
										
										\label{table:char_table}
									\end{threeparttable}
								\end{table}
								\newpage
								
								\section{Measurement and calibration techniques}
								In the main text, we demonstrate the deep ground-state cooling of our mechanical oscillator using a continuous-wave protocol. This requires a calibration method for characterizing the phonon occupancy from the thermomechanical sideband signals. Furthermore, we show the low thermal decoherence rate of vacuum and squeezed states of our mechanical oscillator using a time-domain protocol. This requires a calibration method of the optomechanical amplification process for measuring the quadratures of the mechanical motion. Here we explain the calibration methods in more detail.

								\subsection{Sideband asymmetry measurement and calibration}
\begin{figure*}[h!]
									\includegraphics[scale=1]{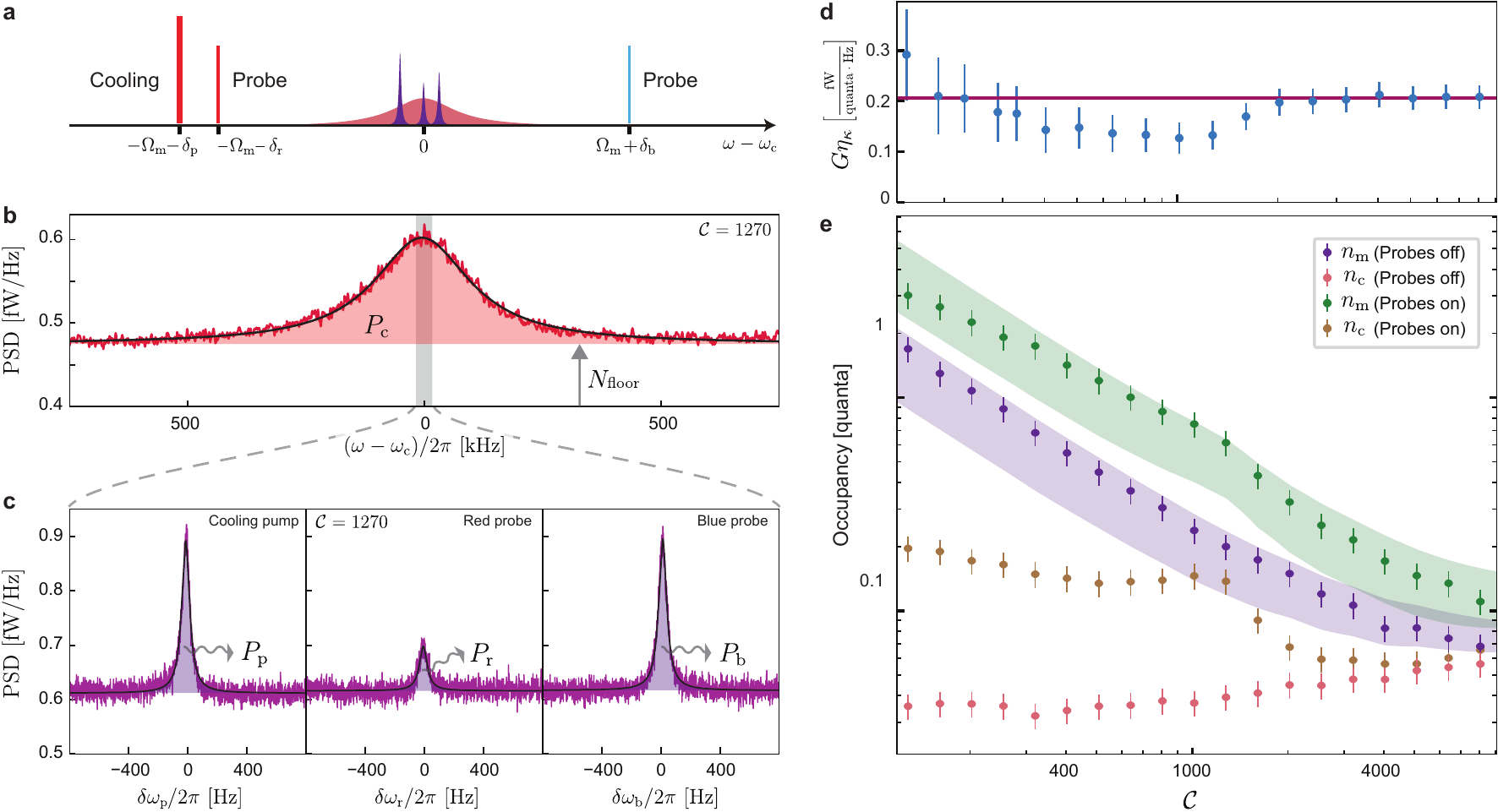}
									\caption{\textbf{Sideband asymmetry experiment} \textbf{a}, Frequency scheme of the sideband asymmetry experiment. \textbf{b, c,} An example of cavity emission and optomechanical sidebands measured for $\mathcal{C} = 1270$. The power of each signal is extracted by Voigt function fitting as shown by black lines. \textbf{d}, Calculated values of $G\eta_\kappa$ for different cooling cooperativities. The solid line shows the average of the flat part of the plot which used as the global calibration factor. This value is in a good agreement with the calibration method based on SNR improvement of JTWPA. \added{Error bars are corresponding to standard deviations.} \textbf{e}, Calibrated photon and phonon occupations of the cavity and mechanics respectively for both cases of probes on and off. \added{Error bars are corresponding to standard deviations.} The shaded area shows theoretically expected occupations in the range of system parameters errors.}
									\label{fig:SI_SB_asym}
								\end{figure*}

{\label{sec:calibration}}
								To characterize the phonon occupation cooled by a red-detuned pump in a continuous-wave measurement, we use the sideband asymmetry measurement~\cite{SI_weinstein2014observation} as an out-of-loop calibration, where two balanced probes red- and blue-detuned by the mechanical frequency are applied, and the thermomechanical sideband signals are measured (Fig.~\ref{fig:SI_SB_asym}a). This scheme requires quantum measurement of the noise power spectrum density of the output signals from the cavity. Pre-amplifying the weak signals in a quantum-limited manner at the cryogenic temperature allows us to realize such a quantum measurement.
								Therefore, we amplify the output signals using a JTWPA operated in a nearly quantum-limited phase-insensitive manner and measure them classically at room temperature (see Sec.~\ref{sec:exp_setup} for the detailed experimental setup).
								The total microwave measurement chain, including propagation losses, the gain and added noise for each amplifier, and the detection noise, can be characterized by only two parameters, an effective added noise ($n_\mathrm{add}$) and a scaling factor ($G$).
								Using these two parameters, the measured noise power spectrum density is described as
								\begin{equation}
									\bar S'(\omega) = G\left[\bar S(\omega)+\frac{1}{2} + n_\mathrm{add}\right],
								\end{equation}
								where $\bar S(\omega)$ is the noise power spectrum density from the device, which is described in Eqs.~(\ref{eq:sym_spec_c})--(\ref{eq:sym_spec_b_simp}).
								Note that the $1/2$ quanta added noise is inevitable even with an ideal phase-insensitive amplification process, and is distinguished from $n_\mathrm{add}$ that may be induced by any kinds of imperfections, such as propagation losses and amplification noises. Namely, the ideal measurement of the noise power spectral density is realized by $n_\mathrm{add}=0$ in our definition.
								In the sideband asymmetry measurement with a cooling pump, there are three thermomechanical sideband peaks and one thermal cavity emission peak in the full noise power spectrum density.
								By integrating each Lorentzian noise peak after subtraction of the corresponding noise floor, we can obtain the photon flux for the respective peak, which is analytically given by
								\begin{eqnarray}
									P_\mathrm{p} &=& G\eta_\kappa \left[2\pi \Gamma_\mathrm{opt}^\mathrm{p} (n_\mathrm{m}-2n_\mathrm{c})\right], \\
									P_\mathrm{r} &=& G\eta_\kappa \left[2\pi \Gamma_\mathrm{opt}^\mathrm{r} (n_\mathrm{m}-2n_\mathrm{c})\right], \\
									P_\mathrm{b} &=& G\eta_\kappa \left[2\pi \Gamma_\mathrm{opt}^\mathrm{b} (n_\mathrm{m}+1+2n_\mathrm{c})\right],\\
									P_\mathrm{c} &=& G\eta_\kappa \left[2\pi \kappa \: n_\mathrm{c}\right],
								\end{eqnarray}
								where $P_\mathrm{p}$, $P_\mathrm{r}$, and $P_\mathrm{b}$ are the noise photon fluxes of the sidebands induced by the cooling pump and the red- and blue-detuned probes respectively, and $P_\mathrm{c}$ is the noise photon flux from the cavity. Here, the collection efficiency of the cavity is defined as $\eta_\kappa = \kappa_\mathrm{ex}/\kappa$. 
								For convenience, we normalize the photon fluxes by the corresponding rates and define the scaled phonon and photon numbers as
								\begin{eqnarray}
									N_\mathrm{p} &=&P_\mathrm{p}/(2\pi \Gamma_\mathrm{opt}^\mathrm{p}) = G\eta_\kappa \left(n_\mathrm{m}-2n_\mathrm{c}\right), \label{eq:nc} \\
									N_\mathrm{r} &=& P_\mathrm{r}/(2\pi \Gamma_\mathrm{opt}^\mathrm{r}) = G\eta_\kappa \left(n_\mathrm{m}-2n_\mathrm{c}\right), \label{eq:nr}\\
									N_\mathrm{b} &=& P_\mathrm{b}/(2\pi \Gamma_\mathrm{opt}^\mathrm{b}) = G\eta_\kappa \left(n_\mathrm{m}+1+2n_\mathrm{c}\right),\label{eq:nb}\\
									N_\mathrm{c} &=& P_\mathrm{c}/(2\pi \kappa)  = G\eta_\kappa \: n_\mathrm{c} \label{eq:ncav}.
								\end{eqnarray}
								Note that $\Gamma_\mathrm{opt}^\mathrm{p}$, $\Gamma_\mathrm{opt}^\mathrm{r}$, $\Gamma_\mathrm{opt}^\mathrm{b}$, and $\kappa$ can be determined from independent experiments, allowing us to experimentally obtain all the scaled occupation, $N_\mathrm{p}$, $N_\mathrm{r}$, $N_\mathrm{b}$, and $N_\mathrm{c}$.
								Furthermore, we can obtain the noise floor of the power spectrum density, which is described as
								\begin{equation}
									\label{eq:nfloor}
									N_\mathrm{floor} = G(1+n_\mathrm{add}).
								\end{equation}
								
								Here, our task is to extract the four unknown parameters, $n_\mathrm{m}$, $n_\mathrm{c}$, $G$, and $n_\mathrm{add}$ from Eqs. (\ref{eq:nc})--(\ref{eq:nfloor}) including the experimentally accessible parameters, $N_\mathrm{p}$, $N_\mathrm{r}$, $N_\mathrm{b}$, $N_\mathrm{c}$, and $N_\mathrm{floor}$.
								Given the fact that $N_\mathrm{p}$ and $N_\mathrm{r}$ are equivalent, we have four independent equations and four unknown parameters, enabling us to analytically obtain the solutions for the four unknown parameters.
								This shows that the sideband asymmetry measurement is useful even when there is a squashing (or a fake asymmetry) in the thermomechanical sideband signals induced by the cavity heating, i.e., when $n_\mathrm{c}>0$.
								
								For our experiment, there is a Fano effect in the cavity reflection spectrum, preventing us from reliably determining the collection efficiency $\eta_\kappa$.  
								Nevertheless, we find that it is still possible to extract the three unknown parameters $n_\mathrm{m}$, $n_\mathrm{c}$, and the effective scaling factor $G\eta_\kappa$ from the three equations, Eqs. (\ref{eq:nb}), (\ref{eq:ncav}) and (\ref{eq:nr}). 
								This is because the collection efficiency effectively modifies only the scaling factor in the measurement chain for all the signals emitted from the cavity~[see Eqs.~(\ref{eq:nc})--(\ref{eq:ncav})].
								By using Eqs. (\ref{eq:nr}), (\ref{eq:nb}), and (\ref{eq:ncav}), we have the analytical solutions described as
								\begin{eqnarray}
									n_\mathrm{m} &=& \frac{R_\Gamma(N_\mathrm{r}+N_\mathrm{b})+2(R_N+R_\Gamma)N_\mathrm{p}}{(R_N - R_\Gamma)(N_\mathrm{r}+N_\mathrm{b})- 4(R_N+R_\Gamma)N_\mathrm{p}} \label{eq:nm_2p}\\
									n_\mathrm{c} &=& \frac{R_Nn_\mathrm{m} - R_\Gamma (n_\mathrm{m}+1)}{2(R_N+R_\Gamma)} \label{eq:nc_2p}\\
									G\eta_\kappa &=& \frac{(N_\mathrm{r}+N_\mathrm{b})/2}{n_\mathrm{m}+1/2} \label{eq:Geta_2p},
								\end{eqnarray}
								where we define $R_N = N_\mathrm{b}/N_\mathrm{r}$ and $R_\Gamma = \Gamma_\mathrm{opt}^\mathrm{b}/\Gamma_\mathrm{opt}^\mathrm{r}$. 
								Note that we use $N_\mathrm{p}$ instead of $N_\mathrm{r}$ in our analysis since $N_\mathrm{p}$ can be obtained with a higher signal-to-noise ratio for a high cooling pump power.
								
								It is known that there is a quantum-backaction induced by the blue-detuned probe that may dominantly heat the mechanical oscillator in the sideband asymmetry measurement even when the red- and blue- detuned probes are perfectly balanced (see Eq. (\ref{eq:mech_phn})). Also, the two probes may induce an additional heating effect on the cavity, which limits the lowest phonon occupancy achievable by sideband cooling. 
								
								Here we propose to characterize the phonon occupation cooled by a cooling pump without using the two probes used for the sideband asymmetry measurement. 
								In this case, the measured noise power spectral density contains the thermomechanical sideband induced by the cooling pump and the cavity thermal emission. Thus, the photon fluxes of the sideband and the cavity emission, $P_\mathrm{p}$ and $P_\mathrm{c}$, can be obtained independently by integrating each peak. Then, we obtain the scaled phonon and photon numbers by normalizing the photon fluxes by the corresponding emission rates, which are analytically described as
								\begin{eqnarray}
									N_\mathrm{p} &=&P_\mathrm{p}/(2\pi \Gamma_\mathrm{opt}^\mathrm{p}) = G\eta_\kappa \left(n_\mathrm{m}-2n_\mathrm{c}\right) \label{eq:nc_woprobe} \\
									N_\mathrm{c} &=& P_\mathrm{c}/(2\pi \kappa)  = G\eta_\kappa \: n_\mathrm{c} \label{eq:ncav_woprobe}.
								\end{eqnarray}
								The key idea is that we use the effective scaling factor, $G\eta_\kappa$, which has already been calibrated in a reliable way based on the sideband asymmetry measurement. 
								Importantly, note that we can experimentally confirm that the effective scaling factor does not depend on how strong total power is applied to the device in our measurement.
								With $G\eta_\kappa$, we can first obtain the thermal cavity photon number as
								\begin{equation}
									\label{eq:nc_Geta}
									n_\mathrm{c} = \frac{N_\mathrm{c}}{G\eta_\kappa}.
								\end{equation} 
								Using $n_\mathrm{c}$, we then obtain the phonon occupation as
								\begin{equation}
									\label{eq:nm_Geta}
									n_\mathrm{m} = \frac{N_\mathrm{p}}{G\eta_\kappa} + 2n_\mathrm{c}.
								\end{equation} 
								
								In the experiment, we set the optomechanical damping and anti-damping rates of the balanced probes to $\Gamma_\mathrm{opt}^\mathrm{r} /2\pi = \Gamma_\mathrm{opt}^\mathrm{b} / 2\pi = 12.9$~Hz, inducing sidebands with frequency detuning of $\delta_\mathrm{r}/2\pi = 0$ and $\delta_\mathrm{b}/2\pi = 10$~kHz at the cavity frequency. The cooling pump is applied to cool down the mechanical oscillator and generate the sideband with $\delta_\mathrm{p}/2\pi = 25$~kHz, and its power is swept up to the maximum output power of the microwave source, corresponding to cooperativity of $\mathcal{C} = 8000$. 
								
								For the calibration based on the sideband asymmetry measurement, we apply the cooling pump together with the red- and blue-detuned balanced probes. 
								Figures \ref{fig:SI_SB_asym}b and c show the measured noise power spectrum density of the cavity thermal emission and the respective optomechanical sidebands. 
								The power of the cavity thermal emission is obtained as the area of a Lorentzian fitted to the measured spectrum density, while the power of each optomechanical sideband is obtained by fitting a Voigt function (a convolution of a Lorentzian with a Gaussian filter with a 1 Hz resolution bandwidth) to the spectrum density and extracting the area of the Lorentzian. 
								Note that the residual imbalance of the probe powers (less than 2\%) is directly measured and compensated in the data analysis. 
								Using the obtained powers together with Eqs.~(\ref{eq:nm_2p})--(\ref{eq:Geta_2p}), we obtain and show the scaling factor $G\eta_\kappa$ as a function of the cooling cooperativity in Fig.~\ref{fig:SI_SB_asym}d. 
								We find that there is an unknown heating effect on the cavity (see the cavity occupation with the two probes in Fig.~\ref{fig:SI_SB_asym}e) when the cooling pump is nearly balanced with the blue probe, which distorts the scaling factor $G\eta_\kappa$. Nevertheless, we reliably obtain the scaling factor to be   $G\eta_\kappa = 0.21 (\pm 0.03)$~[fW/quanta$\cdot$Hz], shown with the red line in Fig.~\ref{fig:SI_SB_asym}d, by using a cooperativity region where the heating effect disappears when the cooling pump is sufficiently large ($\mathcal{C} \gtrapprox2000$). 
								
								We experimentally confirm that the gain and SNR improvement of the JTWPA for a weak coherent tone is not varying for different cooling powers, with and without the two probes, ensuring the scaling factor $G\eta_\kappa$ is constant for the entire measurements. 
								Using the well-calibrated $G\eta_\kappa$, we can therefore directly extract the microwave and mechanical occupations from the cavity thermal emission and the optomechanical sideband induced by the cooling pump respectively by using Eqs.~(\ref{eq:nc_Geta}) and (\ref{eq:nm_Geta}).
								Figure \ref{fig:SI_SB_asym}e shows both the extracted occupations when probes are on and off. 
								Due to additional heating effects induced by the two probes, the cavity occupation with the probes is higher than that without the probes, which limits the minimum phonon occupation that can be achieved by sideband cooling. Here, we obtain the minimal phonon occupation to be $n_\mathrm{m} = 6.8 (\pm 0.9) \times 10^{-2}$~quanta when the probes are switched off, which is approaching the limitation imposed by the cavity heating effect.
								
								\added{It is worth mentioning that the power spectral density of sidebands scattered from the red probe or the cooling pump can show a dip, instead of a peak, when $n_\mathrm{m}-2n_\mathrm{c}<0$. As shown in Fig. 3d of the main text, a Lorentzian dip of the cooling pump sideband is indeed observed for $\mathcal{C}=6400$, implying that  the mechanical occupation is reaching the minimum limit imposed by the cavity heating.}
								
								Furthermore, using Eq.~(\ref{eq:nfloor}), we determine the effective added noise of the microwave measurement chain to be $n_\mathrm{add} = 0.9 (\pm 0.2)$~quanta in agreement with the independently calibrated added noise using the SNR improvement of the JTWPA discussed in section~\ref{sec:TWPA_SNRI}.
								
								\added{We note that power spectral densities are measured by an electrical spectrum analyzer (see Sec.\ref{sec:exp_setup}) in the fast Fourier transform (FFT) mode. The signal flow in spectrum analyzer in FFT mode is described by the following: The microwave signal gets down converted to intermediate frequency (IF) range and will be sampled by an analog to digital converter (with 200 MHz sampling clock in our case). The total record length of a single time trace is defined by the resolution band width (RBW), e.g. for $\mathrm{RBW}=1$~Hz the record length is $\propto1/\mathrm{RBW}\sim2$~s. Then the FFT of the digital time trace is numerically calculated, with arbitrary number of points in the frequency range. The resolution of the frequency spectrum is limited by RBW, i.e. the record length. The truncation of the digital time trace data results in the convolution of the original signal frequency spectrum by a filter (in our case Gaussian) with the linewidth of RBW. 
									A PSD measured with a given resolution bandwidth~(RBW) corresponds to the convolution of an actual PSD $S(\omega)$ with a Gaussian filter $f(\omega)= e^{-(\omega/\sqrt{2}\sigma_\mathrm{RBW})^2}$, where $\sigma_\mathrm{RBW} = \mathrm{RBW}/\sqrt{2\pi}$. 
									This is described as
									\begin{equation}
										\tilde{S}(\omega) \equiv \int_{-\infty}^{+\infty} \: S(\omega')\times f(\omega-\omega') d\omega'.
									\end{equation}
									For a Lorentzian spectrum, the convolution with a Gaussian filter results in a Voigt function: 
									\begin{equation}
										V(\omega ;\Gamma,\sigma_\mathrm{RBW}) \equiv \int_{-\infty}^{+\infty} \frac{\Gamma}{\pi (\omega'^2 + \Gamma^2)} \times  e^{-\frac{(\omega-\omega')^2}{2\sigma^2_\mathrm{RBW}}} \; d\omega' 
									\end{equation}
									The Voigt function is used to numerically fit on PSD of sidebands.
									When the linewidth of a sideband signal is smaller or comparable with the RBW, the convolution cannot be negligible for obtaining the sideband power, i.e., the integration of the measured PSD does not give the actual sideband power.
									Therefore, we fit the measured PSD to a Voigt function with a given RBW and extract the height and linewidth of the Lorentzian PSD to calculate the actual sideband power.
									The inset of Fig.1H in the main text shows the measured PSD of a mechanical sideband, compared with the ideal Gaussian core of the Voigt function ($\mathrm{RBW}=1$~Hz~$\gg \Gamma_\mathrm{m} = 45$~mHz). The linewidth of the mechanical sideband- which in this case includes any unwanted frequency fluctuation - can be then extracted by fitting a Voigt function to the measured PSD.}
								
								\subsection{Verification of quantum efficiency with JTWPA SNR improvement}{\label{sec:TWPA_SNRI}}
								\begin{figure*}[h!]
									\includegraphics[scale=1]{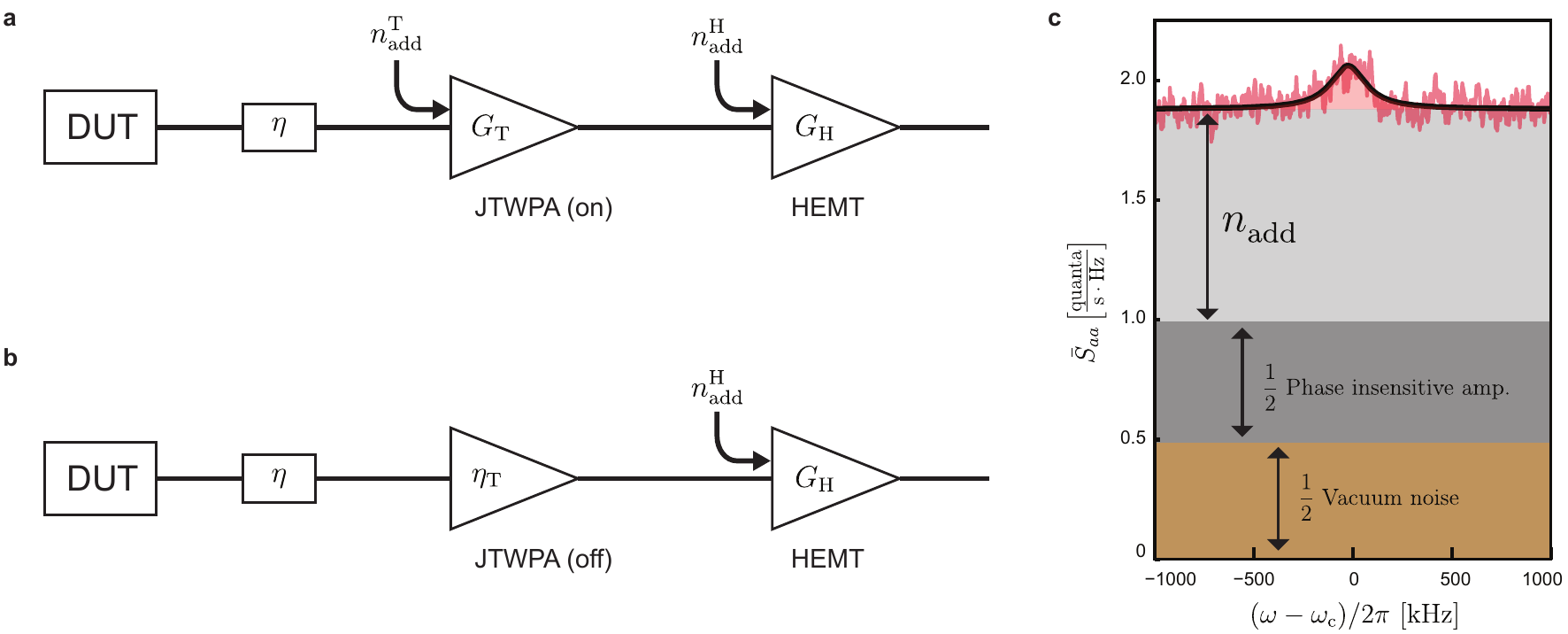}
									\caption{\textbf{Measurement chain.} \textbf{a,b,} Schematic model of the measurement chain for two cases of JTWPA's pump on (amplification) and off (attenuation). \textbf{c}, An example of calibrated PSD of the microwave cavity noise emission. In the ideal phase-insensitive amplification -\added{50}\% quantum efficiency- a $\frac{1}{2}$~quanta/s$\cdot$Hz noise will be added to the vacuum noise level. In practice, the finite quantum efficiency of the measurement chain can be measured and the actual noise background including total added noise referred to the device can be calculated as shown by $\frac{1}{2}+\frac{1}{2} + n_\mathrm{add}$.  }
									\label{fig:SI_chain_noise}
								\end{figure*}
								In the previous section, we provided an out-of-loop calibration technique to measure the quantum efficiency and conversion factor in our measurement chain. That method can be cross-checked and verified independently by measuring the signal-to-noise improvement (SNRI) of the JTWPA. Here we provide calculations connecting SNRI to the total added noise of the measurement chain. 
								We consider a simplified model for the output chain as shown in Fig.~\ref{fig:SI_chain_noise}. The emitted signal from DUT is experiencing a total attenuation of $\eta$ before reaching to JTWPA, our near-quantum-limited amplifier operated at 10 mK. The JTWPA is amplifying the signal in a four-wave mixing process in the presence of a microwave pump, resulting in amplified signal and idler, with a typical gain of $\sim25$ dB. When the pump of the JTWPA is off it acts as an attenuator, $\eta_\mathrm{T}$. When the JTWPA's pump is on, it amplifies the signal by the gain of $G_\mathrm{T}$, and adds an added noise of $\frac{1}{2} + n_\mathrm{add}^\mathrm{T}$ to the signal (referred to the input). The signal is amplified again at 3~K stage by HEMT, with an effective added noise of $\frac{1}{2} +n_\mathrm{add}^\mathrm{H}$ (including the attenuation between HEMT and JTWPA) and the gain of $G_\mathrm{H} \sim 40$~dB. Because of the high gain of HEMT, the noise figure of the measurement chain is dominated by HEMT (and JTWPA, if its pump is on), and will not be considerably affected by further attenuation or amplification in the room temperature setup. 
								
								We can consider a signal, $s$, plus the vacuum noise of $\frac{1}{2}$ emitted from the DUT. The total output of the HEMT amplifier can be calculated when JTWPA is on or off:
								
								\begin{equation}
									\text{JTWPA off: } G_\mathrm{H} \left(\frac{1}{2}+n_\mathrm{add}^\mathrm{H}+\eta_\mathrm{T}\left(\eta s+\frac{1}{2}\right) + (1-\eta_\mathrm{T})\frac{1}{2}\right) = G_\mathrm{H} \eta_\mathrm{T} \eta \left(s + \frac{1+n_\mathrm{add}^\mathrm{H}}{\eta \eta_\mathrm{T}}\right),
								\end{equation}
								
								\begin{equation}
									\text{JTWPA on: } G_\mathrm{H} \left(\frac{1}{2}+n_\mathrm{add}^\mathrm{H}+G_\mathrm{T}\left(\frac{1}{2} + n_\mathrm{add}^\mathrm{T}+\eta s+\frac{1}{2}\right)\right)  = G_\mathrm{H} G_\mathrm{T} \eta \left(s + \frac{1+n_\mathrm{add}^\mathrm{T}}{\eta} + \frac{\frac{1}{2}+n_\mathrm{add}^\mathrm{H}}{\eta G_\mathrm{T}}\right).
								\end{equation}
								
								In each case, the SNR can be simplified as:
								\begin{equation}
									\text{SNR}^\mathrm{off} = \frac{s}{\frac{1}{\eta\eta_\mathrm{T}}  (1 + n_\mathrm{add}^\mathrm{H})},
								\end{equation}
								
								\begin{equation}
									\text{SNR}^\mathrm{on} = \frac{s}{\frac{1}{\eta}  (1 + n_\mathrm{add}^\mathrm{T} + \frac{\frac{1}{2}+n_\mathrm{add}^\mathrm{H}}{G_\mathrm{T}})},
								\end{equation}
								
								and the SNRI can be written as:
								\begin{equation}
									\text{SNRI} = \frac{1+n_\mathrm{add}^\mathrm{H}}{ (1+n_\mathrm{add}^\mathrm{T}) \eta_\mathrm{T} + \frac{\eta_\mathrm{T}(\frac{1}{2}+n_\mathrm{add}^\mathrm{H})}{G_\mathrm{T}}}.
								\end{equation}
								
								The second term in the denominator can be neglected compared to the first term because of the high gain of JTWPA. Therefore, the JTWPA added noise can be extracted as:
								\begin{equation}
									n_\mathrm{add}^\mathrm{T} = \frac{1}{\eta_\mathrm{T}}\frac{1+n_\mathrm{add}^\mathrm{H}}{\text{SNRI}} - 1.
								\end{equation}
								
								With this the background noise level of the measured signal referred to the DUT, $1+n_\mathrm{add}$, can be expressed by:
								\begin{equation}
									1 + n_\mathrm{add} = \frac{1}{\eta \eta_\mathrm{T}}\frac{1+n_\mathrm{add}^\mathrm{H}}{\text{SNRI}}.
								\end{equation}
								
								In the ground state cooling experiment, we measured SNRI=11.3~dB and a loss compensated gain of $\frac{G_\mathrm{T}}{\eta_\mathrm{T}} = 25.7$~dB. The HEMT's added noise is extracted from the data sheet as $n_\mathrm{add, nominal}^\mathrm{H} \simeq 6.9$~quanta. The attenuation of the cryogenic circulators are reported 0.25~dB at low temperatures based on their data-sheet. Considering two circulators between JTWPA and HEMT, 0.5~dB cable loss (1~dB/m flexible coaxial cable loss), and lossless superconducting lines from circulators to the HEMT, the effective HEMT's added noise is assumed $n_\mathrm{add}^\mathrm{H} \simeq 8.7$~quanta. The attenuation of the JTWPA is provided by its data-sheet as $\eta_\mathrm{T}=2.5$~dB at 5.5 ~GHz. The total attenuation between DUT and the JTWPA is composed of: 0.25~dB (cryogenic circulator) + 0.3~dB (directional coupler) + 1.0~dB (1 dB/m flexible coaxial cable loss). This results in the total estimated noise background of $1+n_\mathrm{add} \simeq 1.9$~quanta/s$\cdot$Hz, and JTWPA added noise of $n_\mathrm{add}^\mathrm{T} \simeq 0.3$~quanta, in agreement with the results from sideband asymmetry calibration.
								
								\subsection{Optomechanical amplification measurement and calibration}
								To observe the thermal decoherence rate of our mechanical oscillator, we use a time-domain protocol, where we first prepare the mechanical oscillator to either a vacuum state or a squeezed state, let it freely evolve for a certain time ($\tau_\mathrm{ev}$), and then measure the mechanical quadrature by using the optomechanical amplification process induced by a blue-detuned pump~\cite{SI_reed2017faithful,SI_delaney2019measurement} (see Sec.~\ref{sec:amplificattion}). 
								We measure the optomechanical induced amplification sideband signal and obtain the phase-insensitive amplified mechanical quadratures. Repeating such a time domain protocol and collecting the measured output microwave quadrature data (a pair of $I$ and $Q$ electrical quadratures in units of $\mu$V extracted from each recorded time trace) allow us to construct the quadrature probability density function (PDF) of the measured state and calculate its standard deviations, which in case of a thermal (symmetric Gaussian) state are equal to each other $\sigma^2 = \langle I^2\rangle = \langle Q^2\rangle$. Using this intrinsic optomechanical amplifier, we achieve $G^\mathrm{opt}=\exp{(+\tau_\mathrm{amp} \Gamma_\mathrm{amp})}\simeq50$ dB gain in the microwave optomechanical sideband signal where $\tau_\mathrm{amp} = 22$~ms and $\Gamma_\mathrm{amp} = 2\pi \times85$~Hz are the blue pulse length and total optomechanical anti-damping rates, respectively. It is worth to note that in each cycle, the length of the preparation (cooling or squeezing) pulse should be long enough to damp the mechanics which was amplified in the previous amplification sequence ($\Gamma_\mathrm{prep}\tau_\mathrm{prep} \gg \Gamma_\mathrm{amp}\tau_\mathrm{amp}$) to avoid instability of the time domain protocol. We set the $\tau_\mathrm{prep}>100$~ms to ensure $\sim50$~dB higher cooling rate than the amplification rate in the sequence. After each sequence, we turn off all pulses for 10 ms to collect the traces from the measurement device, which is also sufficient to thermalize the cavity and microwave thermal bath to their initial states in the absence of pumps. The total repetition rate is around 5 sequences per second in our experiment, slightly varying based on the preparation time needed for different cooling rates.
								
								As discussed in the theory section (Sec.~\ref{sec:amplificattion}), the original mechanical state is added with an effective noise referred to input, $n_\mathrm{add}^\mathrm{opt}$, in the amplification process. The measured PDF normalized with $\sqrt{G^\mathrm{opt}}$ shows the convolution of the quadrature PDF of the original state (thermal or squeezed state) with the added noise of a Gaussian PDF. 
								By calculating the average of the standard deviations of both the mechanical quadratures, we can obtain the scaled phonon occupation including the vacuum noises and the input-referred added noise in the optomechanical amplification process.
								For a thermal state, the measured phonon occupation is described by
								\begin{equation}
									\sigma^2 = \frac{\langle I^2\rangle + \langle Q^2\rangle}{2} =  G^\mathrm{opt}\left(n_\mathrm{m} + \frac{1}{2} + n_\mathrm{add}^\mathrm{opt}+ \frac{1}{2}\right),
								\end{equation}
								where $n_\mathrm{m}\geq0$ is the thermal phonon occupation of the prepared state, $G^\mathrm{opt}$ and $n_\mathrm{add}^\mathrm{opt}$ are the conversion factor and the added noise in the optomechanical phase-insensitive amplification, respectively.
								Note that the $1$ quanta is the sum of the vacuum noise of the mechanical oscillator and the added noise of the ideal phase-insensitive amplification.
								To precisely calibrate the conversion factor and added noise of the amplification process, we prepare several different thermal states which are independently calibrated with the sideband asymmetry experiment based on continuous waves, and measure them through the optomechanical readout. Then, we can relate the measured microwave variances, $\sigma^2$, to the actual phonon occupation of the prepared state, $n_\mathrm{m}$, by fitting a linear function to the variance as a function of the phonon occupation (as shown in Fig. 3 in the main text). With this, we are able to determine the added noise to be $n^\mathrm{opt}_\mathrm{add} = 0.80 (\pm0.09)$~quanta and the conversion factor to be $G^\mathrm{opt} = 1.13 \pm (0.04) \ [\mu\mathrm{V}^2/\text{quanta}]$, respectively.
								
								\begin{figure*}[h!]
									\includegraphics[scale=1]{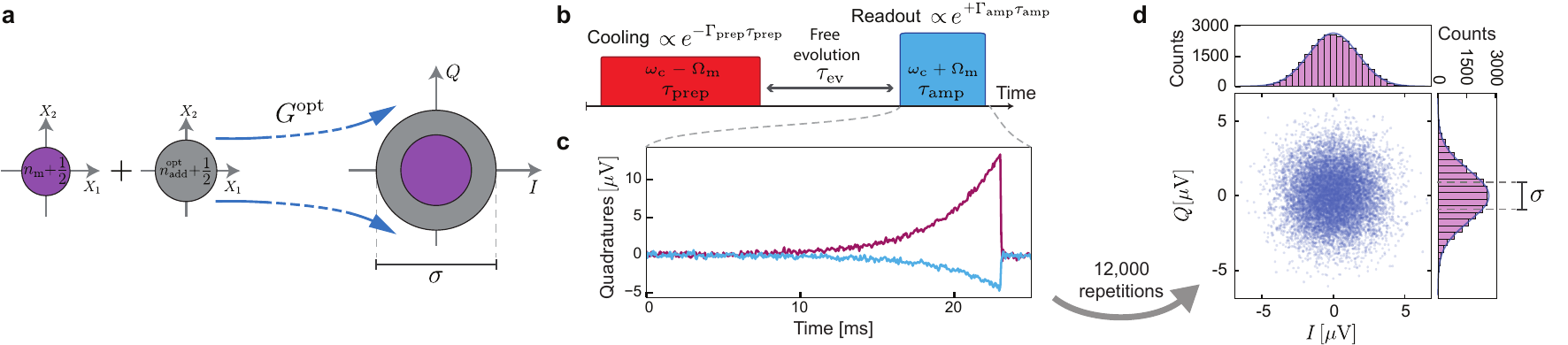}
									\caption{\textbf{Optomechanical amplification.} \textbf{a}, Schematic diagram showing optomechanical amplification process. A desired mechanical mode to be measured is amplified intrinsically in the optomechanical system by activating two-mode-squeezing interaction between mechanics and microwave. During this phase-insensitive amplification, a constant noise is added to the state shown by $n_\mathrm{add} + \frac{1}{2}$ referred to the input. \textbf{b}, The pulse sequence of optomechanical cooling and amplification. A red-detuned pulse prepares the mechanical oscillator in its ground-state. After leaving the state to freely evolve for $\tau_\mathrm{ev}$, a blue-detuned pulse amplifies the thermalized mechanical state to allow low-noise measurement of quadratures of motion. The sequence is repeated ($\sim 12,000$ times) to extract the quadrature scatter plots. \textbf{c}, an example of the time trace collected during optomechanical amplification process. Both quadratures of the microwave field corresponding to the optomechanical sidebands are exponentially growing and being simultaneously recorded. One pair of $I$ and $Q$ is extracted from each trace. \textbf{d}, An example of the scatter plot of measured quadratures. Marginal histograms verifies the Gaussian distribution.}
									\label{fig:SI_amplification}
								\end{figure*}

								\subsection{Generation and calibration of mechanical squeezed states}
\begin{figure*}[h!]
									\includegraphics[scale=1]{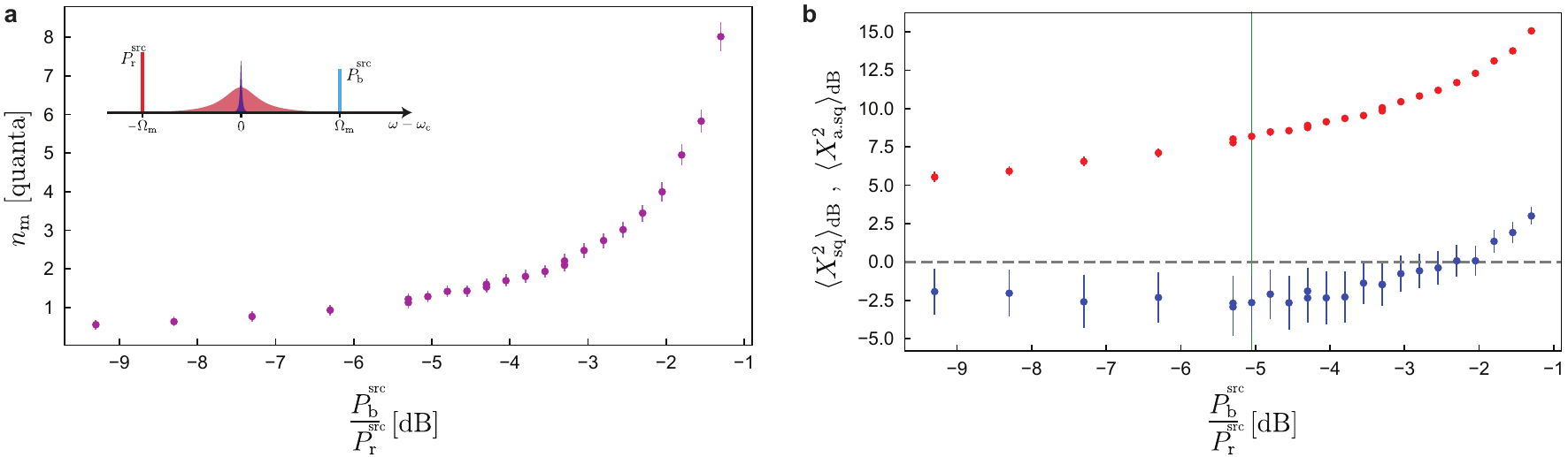}
									\caption{\textbf{Optomechanical squeezing.} \textbf{a (inset)}, Frequency landscape for optomechanical squeezing. two symmetrically detuned blue and red pumps with nominal powers of $P_\mathrm{b}^\mathrm{src}$, $P_\mathrm{r}^\mathrm{src}$ are applied to the cavity. \textbf{a}, Measured total phonon occupation versus the ratio between blue and red pump powers. While the blue pump power increases, The system will be closer to optomechanical instability threshold. \added{Error bars are corresponding to standard deviations.} \textbf{b}, Measured squeezed and anti-squeezed quadratures of motion versus the blue pump relative power. The optimally selected squeezing point for the thermalization experiment is shown by the green line. \added{Error bars are corresponding to standard deviations.}}
									
									\label{fig:SI_squeezing_opt}
								\end{figure*}

								We use optomechanical dissipative squeezing technique to squeeze one quadrature of motion below the zero-point-fluctuation~\cite{SI_kronwald2013arbitrarily,SI_wollman2015quantum,SI_pirkkalainen2015squeezing}. As shown in Fig.~\ref{fig:SI_squeezing_opt}a inset, simultaneously applying two symmetrically red- and blue-detuned pumps from the microwave frequency with optomechanical damping and anti-damping rates of $\Gamma_\mathrm{opt}^\mathrm{r}$ and $\Gamma_\mathrm{opt}^\mathrm{b}$ respectively results in a beam-splitter interaction Hamiltonian:
								\begin{equation}
									\hat{H}_\mathrm{int} = -\hbar \mathcal{G} \ \hat{a}^\dagger \ \hat{\beta} \ + \text{H.c.},
									\label{eq:hamil_squeez}
								\end{equation}
								where $\hat{\beta} = \sqrt{\frac{\Gamma_\mathrm{opt}^\mathrm{r}}{\Gamma_\mathrm{opt}^\mathrm{r}-\Gamma_\mathrm{opt}^\mathrm{b}}} \ \hat{b} + \sqrt{\frac{\Gamma_\mathrm{opt}^\mathrm{b}}{\Gamma_\mathrm{opt}^\mathrm{r}-\Gamma_\mathrm{opt}^\mathrm{b}}} \ \hat{b}^\dagger$ is the Bogoliubov's mode annihilation operator and $\mathcal{G} = \sqrt{\kappa/4}\sqrt{\Gamma_\mathrm{opt}^\mathrm{r}-\Gamma_\mathrm{opt}^\mathrm{b}}$ is the coupling rate. This beam-splitter Hamiltonian allows us to reach the ground-state of the Bogoliubov operator, which is a squeezed state of motion with 
								\begin{equation}
									\langle \hat{X}_\mathrm{sq}^2\rangle = \frac{1}{2} e^{-2r} \ , \ \langle \hat{X}_\mathrm{a.sq}^2\rangle = \frac{1}{2} e^{+2r},
									\label{eq:noise_subtraction}
								\end{equation}
								where $\tanh(r) = \sqrt{\Gamma_\mathrm{opt}^\mathrm{b}/\Gamma_\mathrm{opt}^\mathrm{r}}$. In practice, the non-zero occupation of the thermal bath prevents us from reaching to such a pure squeezed state. A more comprehensive analytical derivation on the purity of optomechanical squeezing is discussed by Kronwald~\textit{et.al.}~(2013)~\cite{SI_kronwald2013arbitrarily}.
								
\begin{figure*}[h!]
									\includegraphics[scale=1]{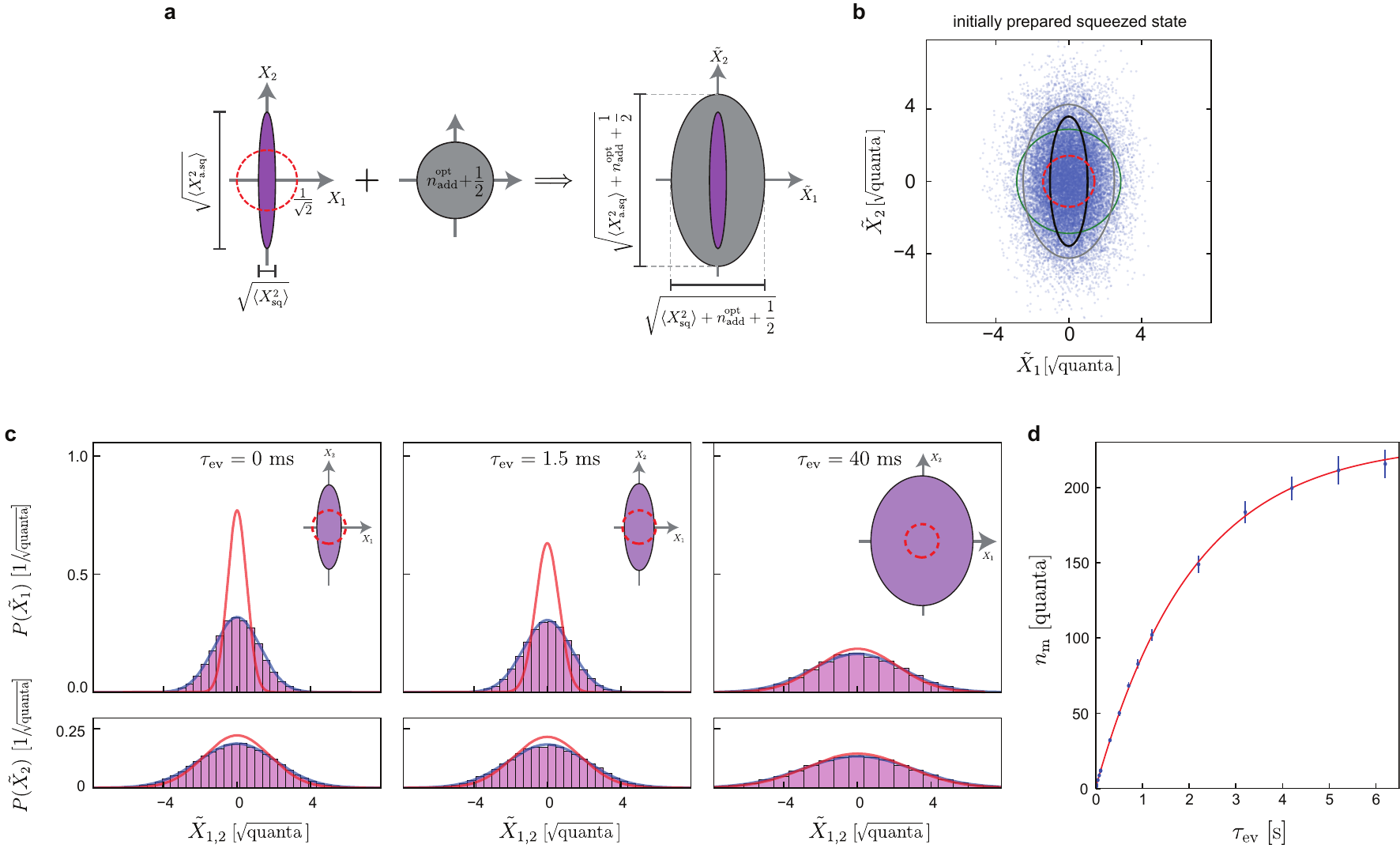}
									\caption{\textbf{Measurement of squeezing.} \textbf{a}, Schematic diagram showing the prepared squeezed state added with the input referred noise of amplification process. The measured state has a quadrature density function proportional to convolution of the initial state with the added noise. Precise calibration of the Gaussian added noise allows us to extract the initial state quadrature variances. \textbf{b}, Quadrature scatter plot for the initially prepared squeezed state. The gray ellipse represent standard deviation \added{contours scaled by factor of 2 (for better visualization)} of the Gaussian density function. The green circle shows the same for the measured ground state. The black ellipse shows the same for the noise-subtracted state demonstrating the initial squeezing. The red dashed line shows the same for the zero-point-fluctuation (ideal ground-state). \textbf{c}, Quadrature marginal histograms are shown by purple bars and calculated Gaussian marginal density functions by thick blue lines verifying measurement of Gaussian squeezed state. The red lines show the added noise subtracted density functions of the initial squeezed state. Insets depict the extracted squeezed state versus the zero-point-fluctuation. \textbf{d}, Phonon occupation of the squeezed state evolution in long time scales. The red line shows the exponential fit. \added{Error bars are corresponding to standard deviations.}}
									\label{fig:SI_squeezing}
								\end{figure*}

								We use optomechanical amplification to measure quadrature PDF of the amplified state ($\tilde{X}_1 = I/\sqrt{G^\mathrm{opt}}$, $\tilde{X}_2 = Q/\sqrt{G^\mathrm{opt}}$ scaled to $[\sqrt{\mathrm{quanta}}]$ unit) as a convolution of the initial state with the precisely calibrated Gaussian added noise during amplification as shown in Fig.~\ref{fig:SI_squeezing}a. Subtracting the added noise from the variance of the scaled quadrature PDF results in the squeezed and anti-squeezed quadrature variances of the input mechanical state for the optomechanical amplification:
								\begin{equation}
									\langle \hat{X}_\mathrm{sq, a.sq}^2\rangle = \langle \hat{X}_{1,2}^2\rangle - n_\mathrm{add}^\mathrm{opt} - \frac{1}{2} .
									\label{eq:noise_subtraction}
								\end{equation}
								Figures \ref{fig:SI_squeezing_opt}a and b show the measured phonon occupation and quadratures of the generated squeezed state prepared with different blue pump powers. We used a damping rate of $\Gamma^\mathrm{r}_\mathrm{opt} /2\pi = 75$~Hz. The optimal power of the blue pump for the thermalization of the squeezed state is found 5 dB below the red pump, where the maximum quadrature squeezing is achieved, as shown in Fig.~\ref{fig:SI_squeezing_opt}b. 
								Increasing the relative power of the blue pump results in optomechanical heating. In particular, when the power of pumps close to the balanced condition, optomechanical instabilities, such as two-tone instability, will occur in the presence of a slight frequency detuning~\cite{SI_shomroni2019two}. 
								With the optimal pump powers, we obtain $\langle X_\mathrm{sq}^2\rangle_\mathrm{dB} = 10 \log_{10}\left(\langle X_\mathrm{sq}^2\rangle/\frac{1}{2}\right) =  -2.7^{+1.4}_{-2.3}$~dB squeezing of one quadrature of motion below the vacuum fluctuation and $\langle X_\mathrm{a.sq}^2\rangle_\mathrm{dB} = 8.1^{+0.3}_{-0.3}$~dB anti-squeezing in the other quadrature, as shown in Fig.~\ref{fig:SI_squeezing}b. 
								The achieved squeezing value is close to the theoretical squeezing limit, corresponding to $2\langle \hat{X}_\mathrm{sq}^2\rangle = \sqrt{(1+2n_\mathrm{m}^\mathrm{th})/\mathcal{C}} = -3.5$~dB~\cite{SI_kronwald2013arbitrarily}.
								
								Afterward, we record the free evolution of the prepared squeezed state, as shown in the main text and Fig.~\ref{fig:SI_squeezing}c. The marginal PDF of both the measured quadratures, as well as the noise subtracted squeezed and anti-squeezed mechanical quadratures are shown for different evolution times. One quadrature of motion remains below the zero-point-fluctuation up to 2 ms. At the longer evolution times, the state is thermalized to an isotropic thermal state (Fig.~\ref{fig:SI_squeezing}d). 
								
								\subsection{Dephasing effect on thermalization of squeezed states}
								A mechanical squeezed state has a phase coherence between the Fock states, resulting in the sensitivity to the pure dephasing. By observing the free evolution of such a phase-sensitive state, we can characterize the pure dephasing rate of our mechanical oscillator. 
								When the mechanical state is initialized in an isotropic Gaussian state, such as the vacuum or thermal states, the decoherence rates for the quadrature variances in all phases are identical regardless of the amount of the pure dephasing. 
								However, when the mechanical state is initialized in a squeezed state, the decoherence rate of the quadrature variance in the squeezed axis is greater than that in the anti-squeezed axis if the pure dephasing rate is finite.
								Thus, we can use this property to extract the pure dephasing rate of the mechanical oscillator. 
								
								In the large phonon bath occupation limit ($n_\mathrm{m}^\mathrm{th}\gg1$), which is the case for our experiment, the free evolution of a mechanical oscillators with a pure dephasing is described by a master equation,
								\begin{equation}
									\label{eq:master}
									\frac{d\hat{\rho}}{dt} = \Gamma_\mathrm{th}\mathcal{D}[\hat{b}]\hat{\rho}  + \Gamma_\mathrm{th}\mathcal{D}[\hat{b}^\dag]\hat{\rho} +2\Gamma_\varphi\mathcal{D}[\hat{b}^\dag\hat{b}]\hat{\rho},
								\end{equation}
								where $\Gamma_\varphi$ is the pure dephasing rate, $\hat{\rho}$ is the density matrix, and $\mathcal{D}[\hat{O}]\rho = \hat{O} \hat{\rho} \hat{O}^\dag - (\hat{O}^\dag\hat{O}\hat{\rho} + \hat{\rho}\hat{O}^\dag\hat{O})/2$ is the Lindblad dissipator.
								By calculating the time evolution of the density matrix, we can numerically obtain the time evolution of the quadrature variances and the phonon occupation as $\langle \hat{X}_1(t)^2\rangle = \mathrm{Tr}[\rho(t) \hat{X}_1^2]$, $\langle \hat{X}_2(t)^2\rangle = \mathrm{Tr}[\rho(t) \hat{X}_2^2]$, and $n_\mathrm{m} = \langle\hat{b}^\dag\hat{b}\rangle$, respectively. 
								Note that we here consider only a mechanical state with $\mathrm{Tr}[\rho \hat{X}_1]=\mathrm{Tr}[\rho \hat{X}_2]=0$, e.g. a vacuum, thermal, and squeezed states.
								
\begin{figure*}[h!]
									\includegraphics[scale=1]{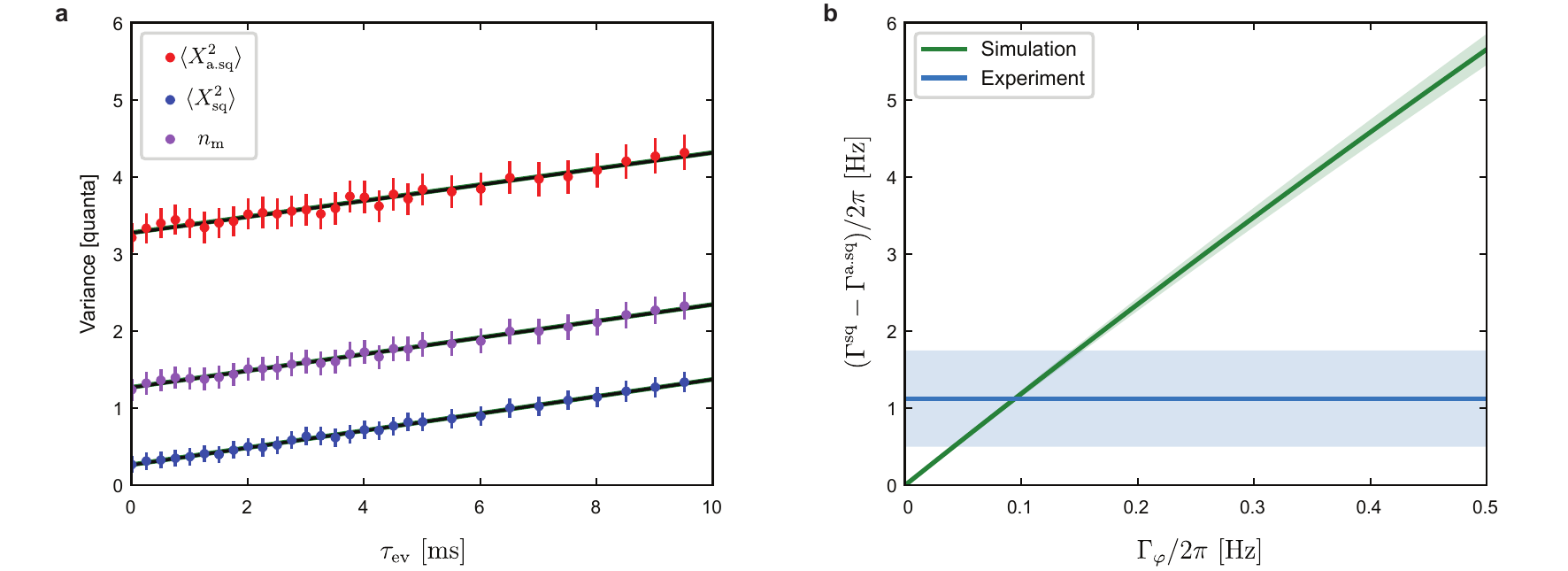}
									\caption{\textbf{Free evolution of a mechanical squeezed state.} 
										\textbf{a},~Quadrature variances and average phonon number of a squeezed state as a function of the free-evolution time. The blue, red and purple circles are the data for the quadrature variances in the squeezed and anti-squeezed axes, and the average phonon number, respectively. The black dotted lines are linear fits, while the green lines are the numerical simulation results. \added{Error bars are corresponding to standard deviations.} 
										\textbf{b},~The difference of the thermal decoherence rates for the quadrature variances in the squeezed and anti-squeezed axes as a function of the pure dephasing rate. The green line shows the numerical simulation results, while the blue line is the experimentally obtained value. The shaded regions showing the errors, respectively.
									}
									\label{fig:SI_ThrermalizedSqueezing}
								\end{figure*}

\begin{figure*}[!]
									\includegraphics[scale=1]{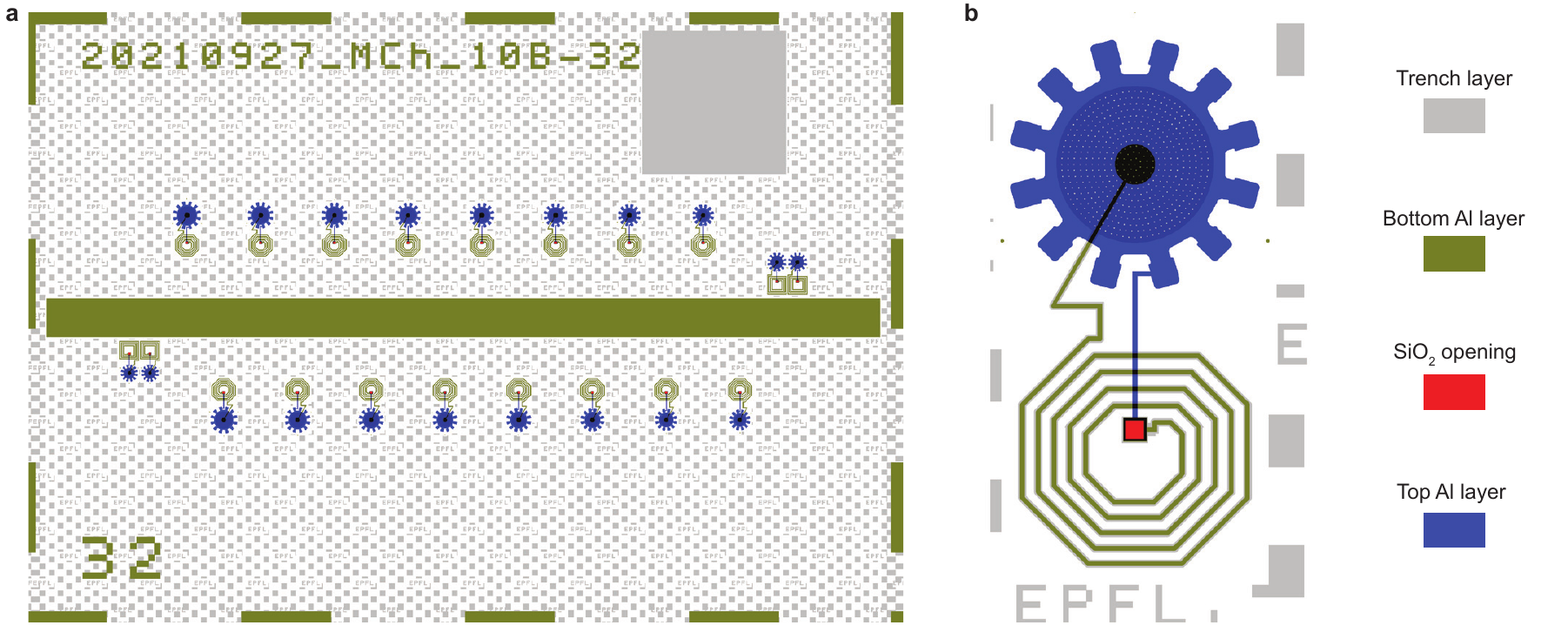}
									\caption{\textbf{Chip layout.} \textbf{a}, The designed layout of the chip used in the experiment. Each chip includes 16 separate frequency multiplexed LC resonators as well as two dimer LCs which are not studied in this work. \textbf{b}, magnification of the layout of one LC resonator. The color code explains different lithography layers. }
									\label{fig:SI_chip}
								\end{figure*}

								The blue and red circles in Fig.~\ref{fig:SI_ThrermalizedSqueezing}a show the experimental results of the quadrature variances, $\langle X_\mathrm{sq}^2\rangle$ in the squeezed axis and $\langle X_\mathrm{a.sq}^2\rangle$ in the anti-squeezed axis as a function of the free-evolution time respectively, while the purple circles show the time evolution of the phonon occupation, $n_\mathrm{m} = (\langle X_\mathrm{sq}^2\rangle +\langle X_\mathrm{a.sq}^2\rangle)/2-1/2$.
								Here, we define the decoherence rate as a slope of the time evolution, i.e., $\Gamma^\mathrm{sq} =\frac{d\langle X_\mathrm{sq}^2\rangle}{dt}$, $\Gamma^\mathrm{a.sq} =\frac{d\langle X_\mathrm{a.sq}^2\rangle}{dt}$, and $\Gamma_\mathrm{th} = \frac{d n_\mathrm{m}}{dt}$.
								By linearly fitting the results (see the black dotted lines in Fig.~\ref{fig:SI_ThrermalizedSqueezing}a), the decoherence rates are found to be $\Gamma^\mathrm{sq}/2\pi = 17.7\pm0.7$~Hz, $\Gamma^\mathrm{a.sq}/2\pi = 16.5\pm0.6$~Hz, and $\Gamma_\mathrm{th}/2\pi = 17.1\pm0.6$~Hz.
								Furthermore, the difference between the decoherence rates in the squeezed and anti-squeezed axes is obtained as $(\Gamma^\mathrm{sq}-\Gamma^\mathrm{a.sq})/2\pi = 1.1 \pm 0.6$~Hz.
								Importantly note that the thermal decoherence rate can be obtained by observing the thermalization of a squeezed state, instead of the vacuum state, i.e., $\Gamma_\mathrm{th} = (\Gamma^\mathrm{sq} + \Gamma^\mathrm{a.sq})/2$, enabling us to determine the pure dephasing rate from the difference between the decoherence rates in the squeezed and anti-squeezed axes.
								
								The sensitivity of a squeezed state to the pure dephasing depends on its squeezing level. 
								To accurately extract the pure dephasing rate, we therefore need to numerically determine the initial squeezed state using the experimentally obtained quadrature variances, i.e., $\langle X_\mathrm{sq}^2\rangle=0.27 \pm 0.1$ and $\langle X_\mathrm{a.sq}^2\rangle=3.27 \pm0.2$. We can consider the prepared squeezed state as a squeezed thermal state, which is defined as $S\rho_\mathrm{th}S^\dag$, where $S=\exp\left[\frac{r}{2}(\hat{b}^2-\hat{b}^{\dag\:2})]\right]$ is a squeezing operator with a squeezing parameter of $r$ and $\rho_\mathrm{th}$ is a thermal state with an average phonon number of $n_\mathrm{th}$. 
								We can extract the thermal phonon number and the squeezing parameter as $n_\mathrm{th} = \sqrt{\langle X_\mathrm{sq}^2\rangle\langle X_\mathrm{a.sq}^2\rangle}- 1/2=0.4\pm0.2$ and $r = -\frac{1}{4}\log\left(\langle X_\mathrm{sq}^2\rangle/\langle X_\mathrm{a.sq}^2\rangle\right) = 0.6 \pm 0.1$.

								Using the initial squeezed state, we calculate the time evolution of the quadrature variances for different pure dephasing rates and obtain the difference between the thermal decoherence rates of the squeezed and anti-squeezed variances. The green line in Fig.~\ref{fig:SI_ThrermalizedSqueezing}b shows the numerical results of the thermal decoherence rate difference. As the dephasing rate increases, the difference becomes larger.
								We compare the numerical results with the experimentally obtained value, which are shown with the blue line.
								From this, we find that the pure dephasing rate of our mechanical oscillator is found to be $\Gamma_\varphi/2\pi = 0.09\pm0.05$~Hz.
								
								Finally, we show the numerical results of the time evolution of the quadrature variances and the phonon occupation with a dephasing rate of $\Gamma_\varphi/2\pi = 0.09$~Hz in Fig.~\ref{fig:SI_ThrermalizedSqueezing}a (see the green lines). The numerical results reproduce well the experimental ones.

								\newpage
								\newpage
								\bigskip
								
								\section*{Supplementary References}
								\bigskip


\begin{thebibliography}{47}%
	\makeatletter
	\providecommand \@ifxundefined [1]{%
		\@ifx{#1\undefined}
	}%
	\providecommand \@ifnum [1]{%
		\ifnum #1\expandafter \@firstoftwo
		\else \expandafter \@secondoftwo
		\fi
	}%
	\providecommand \@ifx [1]{%
		\ifx #1\expandafter \@firstoftwo
		\else \expandafter \@secondoftwo
		\fi
	}%
	\providecommand \natexlab [1]{#1}%
	\providecommand \enquote  [1]{``#1''}%
	\providecommand \bibnamefont  [1]{#1}%
	\providecommand \bibfnamefont [1]{#1}%
	\providecommand \citenamefont [1]{#1}%
	\providecommand \href@noop [0]{\@secondoftwo}%
	\providecommand \href [0]{\begingroup \@sanitize@url \@href}%
	\providecommand \@href[1]{\@@startlink{#1}\@@href}%
	\providecommand \@@href[1]{\endgroup#1\@@endlink}%
	\providecommand \@sanitize@url [0]{\catcode `\\12\catcode `\$12\catcode
		`\&12\catcode `\#12\catcode `\^12\catcode `\_12\catcode `\%12\relax}%
	\providecommand \@@startlink[1]{}%
	\providecommand \@@endlink[0]{}%
	\providecommand \url  [0]{\begingroup\@sanitize@url \@url }%
	\providecommand \@url [1]{\endgroup\@href {#1}{\urlprefix }}%
	\providecommand \urlprefix  [0]{URL }%
	\providecommand \Eprint [0]{\href }%
	\providecommand \doibase [0]{https://doi.org/}%
	\providecommand \selectlanguage [0]{\@gobble}%
	\providecommand \bibinfo  [0]{\@secondoftwo}%
	\providecommand \bibfield  [0]{\@secondoftwo}%
	\providecommand \translation [1]{[#1]}%
	\providecommand \BibitemOpen [0]{}%
	\providecommand \bibitemStop [0]{}%
	\providecommand \bibitemNoStop [0]{.\EOS\space}%
	\providecommand \EOS [0]{\spacefactor3000\relax}%
	\providecommand \BibitemShut  [1]{\csname bibitem#1\endcsname}%
	\let\auto@bib@innerbib\@empty
	\bibitem [{\citenamefont {Aasi}\ \emph {et~al.}(2013)\citenamefont {Aasi},
		\citenamefont {Abadie}, \citenamefont {Abbott}, \citenamefont {Abbott},
		\citenamefont {Abbott}, \citenamefont {Abernathy}, \citenamefont {Adams},
		\citenamefont {Adams}, \citenamefont {Addesso}, \citenamefont {Adhikari}
		\emph {et~al.}}]{aasi2013enhanced}%
	\BibitemOpen
	\bibfield  {author} {\bibinfo {author} {\bibfnamefont {J.}~\bibnamefont
			{Aasi}}, \bibinfo {author} {\bibfnamefont {J.}~\bibnamefont {Abadie}},
		\bibinfo {author} {\bibfnamefont {B.}~\bibnamefont {Abbott}}, \bibinfo
		{author} {\bibfnamefont {R.}~\bibnamefont {Abbott}}, \bibinfo {author}
		{\bibfnamefont {T.}~\bibnamefont {Abbott}}, \bibinfo {author} {\bibfnamefont
			{M.}~\bibnamefont {Abernathy}}, \bibinfo {author} {\bibfnamefont
			{C.}~\bibnamefont {Adams}}, \bibinfo {author} {\bibfnamefont
			{T.}~\bibnamefont {Adams}}, \bibinfo {author} {\bibfnamefont
			{P.}~\bibnamefont {Addesso}}, \bibinfo {author} {\bibfnamefont
			{R.}~\bibnamefont {Adhikari}}, \emph {et~al.},\ }\bibfield  {title} {\bibinfo
		{title} {Enhanced sensitivity of the ligo gravitational wave detector by
			using squeezed states of light},\ }\href
	{https://www.nature.com/articles/nphoton.2013.177} {\bibfield  {journal}
		{\bibinfo  {journal} {Nature Photonics}\ }\textbf {\bibinfo {volume} {7}},\
		\bibinfo {pages} {613} (\bibinfo {year} {2013})}\BibitemShut {NoStop}%
	\bibitem [{\citenamefont {Mason}\ \emph {et~al.}(2019)\citenamefont {Mason},
		\citenamefont {Chen}, \citenamefont {Rossi}, \citenamefont {Tsaturyan},\ and\
		\citenamefont {Schliesser}}]{mason2019continuous}%
	\BibitemOpen
	\bibfield  {author} {\bibinfo {author} {\bibfnamefont {D.}~\bibnamefont
			{Mason}}, \bibinfo {author} {\bibfnamefont {J.}~\bibnamefont {Chen}},
		\bibinfo {author} {\bibfnamefont {M.}~\bibnamefont {Rossi}}, \bibinfo
		{author} {\bibfnamefont {Y.}~\bibnamefont {Tsaturyan}},\ and\ \bibinfo
		{author} {\bibfnamefont {A.}~\bibnamefont {Schliesser}},\ }\bibfield  {title}
	{\bibinfo {title} {Continuous force and displacement measurement below the
			standard quantum limit},\ }\href
	{https://www.nature.com/articles/s41567-019-0533-5} {\bibfield  {journal}
		{\bibinfo  {journal} {Nature Physics}\ }\textbf {\bibinfo {volume} {15}},\
		\bibinfo {pages} {745} (\bibinfo {year} {2019})}\BibitemShut {NoStop}%
	\bibitem [{\citenamefont {Whittle}\ \emph {et~al.}(2021)\citenamefont
		{Whittle}, \citenamefont {Hall}, \citenamefont {Dwyer}, \citenamefont
		{Mavalvala}, \citenamefont {Sudhir}, \citenamefont {Abbott}, \citenamefont
		{Ananyeva}, \citenamefont {Austin}, \citenamefont {Barsotti}, \citenamefont
		{Betzwieser} \emph {et~al.}}]{whittle2021approaching}%
	\BibitemOpen
	\bibfield  {author} {\bibinfo {author} {\bibfnamefont {C.}~\bibnamefont
			{Whittle}}, \bibinfo {author} {\bibfnamefont {E.~D.}\ \bibnamefont {Hall}},
		\bibinfo {author} {\bibfnamefont {S.}~\bibnamefont {Dwyer}}, \bibinfo
		{author} {\bibfnamefont {N.}~\bibnamefont {Mavalvala}}, \bibinfo {author}
		{\bibfnamefont {V.}~\bibnamefont {Sudhir}}, \bibinfo {author} {\bibfnamefont
			{R.}~\bibnamefont {Abbott}}, \bibinfo {author} {\bibfnamefont
			{A.}~\bibnamefont {Ananyeva}}, \bibinfo {author} {\bibfnamefont
			{C.}~\bibnamefont {Austin}}, \bibinfo {author} {\bibfnamefont
			{L.}~\bibnamefont {Barsotti}}, \bibinfo {author} {\bibfnamefont
			{J.}~\bibnamefont {Betzwieser}}, \emph {et~al.},\ }\bibfield  {title}
	{\bibinfo {title} {Approaching the motional ground state of a 10-kg object},\
	}\href {https://www.nature.com/articles/nature20604} {\bibfield  {journal}
		{\bibinfo  {journal} {Science}\ }\textbf {\bibinfo {volume} {372}},\ \bibinfo
		{pages} {1333} (\bibinfo {year} {2021})}\BibitemShut {NoStop}%
	\bibitem [{\citenamefont {Pechal}\ \emph {et~al.}(2018)\citenamefont {Pechal},
		\citenamefont {Arrangoiz-Arriola},\ and\ \citenamefont
		{Safavi-Naeini}}]{pechal2018superconducting}%
	\BibitemOpen
	\bibfield  {author} {\bibinfo {author} {\bibfnamefont {M.}~\bibnamefont
			{Pechal}}, \bibinfo {author} {\bibfnamefont {P.}~\bibnamefont
			{Arrangoiz-Arriola}},\ and\ \bibinfo {author} {\bibfnamefont {A.~H.}\
			\bibnamefont {Safavi-Naeini}},\ }\bibfield  {title} {\bibinfo {title}
		{Superconducting circuit quantum computing with nanomechanical resonators as
			storage},\ }\href
	{https://iopscience.iop.org/article/10.1088/2058-9565/aadc6c/meta} {\bibfield
		{journal} {\bibinfo  {journal} {Quantum Science and Technology}\ }\textbf
		{\bibinfo {volume} {4}},\ \bibinfo {pages} {015006} (\bibinfo {year}
		{2018})}\BibitemShut {NoStop}%
	\bibitem [{\citenamefont {Wallucks}\ \emph {et~al.}(2020)\citenamefont
		{Wallucks}, \citenamefont {Marinkovi{\'c}}, \citenamefont {Hensen},
		\citenamefont {Stockill},\ and\ \citenamefont
		{Gr{\"o}blacher}}]{wallucks2020quantum}%
	\BibitemOpen
	\bibfield  {author} {\bibinfo {author} {\bibfnamefont {A.}~\bibnamefont
			{Wallucks}}, \bibinfo {author} {\bibfnamefont {I.}~\bibnamefont
			{Marinkovi{\'c}}}, \bibinfo {author} {\bibfnamefont {B.}~\bibnamefont
			{Hensen}}, \bibinfo {author} {\bibfnamefont {R.}~\bibnamefont {Stockill}},\
		and\ \bibinfo {author} {\bibfnamefont {S.}~\bibnamefont {Gr{\"o}blacher}},\
	}\bibfield  {title} {\bibinfo {title} {A quantum memory at telecom
			wavelengths},\ }\href {https://www.nature.com/articles/s41567-020-0891-z}
	{\bibfield  {journal} {\bibinfo  {journal} {Nature Physics}\ }\textbf
		{\bibinfo {volume} {16}},\ \bibinfo {pages} {772} (\bibinfo {year}
		{2020})}\BibitemShut {NoStop}%
	\bibitem [{\citenamefont {Fiaschi}\ \emph {et~al.}(2021)\citenamefont
		{Fiaschi}, \citenamefont {Hensen}, \citenamefont {Wallucks}, \citenamefont
		{Benevides}, \citenamefont {Li}, \citenamefont {Alegre},\ and\ \citenamefont
		{Gr{\"o}blacher}}]{fiaschi2021optomechanical}%
	\BibitemOpen
	\bibfield  {author} {\bibinfo {author} {\bibfnamefont {N.}~\bibnamefont
			{Fiaschi}}, \bibinfo {author} {\bibfnamefont {B.}~\bibnamefont {Hensen}},
		\bibinfo {author} {\bibfnamefont {A.}~\bibnamefont {Wallucks}}, \bibinfo
		{author} {\bibfnamefont {R.}~\bibnamefont {Benevides}}, \bibinfo {author}
		{\bibfnamefont {J.}~\bibnamefont {Li}}, \bibinfo {author} {\bibfnamefont
			{T.~P.~M.}\ \bibnamefont {Alegre}},\ and\ \bibinfo {author} {\bibfnamefont
			{S.}~\bibnamefont {Gr{\"o}blacher}},\ }\bibfield  {title} {\bibinfo {title}
		{Optomechanical quantum teleportation},\ }\href
	{https://www.nature.com/articles/s41566-021-00866-z} {\bibfield  {journal}
		{\bibinfo  {journal} {Nature Photonics}\ }\textbf {\bibinfo {volume} {15}},\
		\bibinfo {pages} {817} (\bibinfo {year} {2021})}\BibitemShut {NoStop}%
	\bibitem [{\citenamefont {Marinkovi{\'c}}\ \emph {et~al.}(2018)\citenamefont
		{Marinkovi{\'c}}, \citenamefont {Wallucks}, \citenamefont {Riedinger},
		\citenamefont {Hong}, \citenamefont {Aspelmeyer},\ and\ \citenamefont
		{Gr{\"o}blacher}}]{marinkovic2018optomechanical}%
	\BibitemOpen
	\bibfield  {author} {\bibinfo {author} {\bibfnamefont {I.}~\bibnamefont
			{Marinkovi{\'c}}}, \bibinfo {author} {\bibfnamefont {A.}~\bibnamefont
			{Wallucks}}, \bibinfo {author} {\bibfnamefont {R.}~\bibnamefont {Riedinger}},
		\bibinfo {author} {\bibfnamefont {S.}~\bibnamefont {Hong}}, \bibinfo {author}
		{\bibfnamefont {M.}~\bibnamefont {Aspelmeyer}},\ and\ \bibinfo {author}
		{\bibfnamefont {S.}~\bibnamefont {Gr{\"o}blacher}},\ }\bibfield  {title}
	{\bibinfo {title} {Optomechanical bell test},\ }\href
	{https://journals.aps.org/prl/abstract/10.1103/PhysRevLett.121.220404}
	{\bibfield  {journal} {\bibinfo  {journal} {Physical review letters}\
		}\textbf {\bibinfo {volume} {121}},\ \bibinfo {pages} {220404} (\bibinfo
		{year} {2018})}\BibitemShut {NoStop}%
	\bibitem [{\citenamefont {Carney}\ \emph {et~al.}(2021)\citenamefont {Carney},
		\citenamefont {Krnjaic}, \citenamefont {Moore}, \citenamefont {Regal},
		\citenamefont {Afek}, \citenamefont {Bhave}, \citenamefont {Brubaker},
		\citenamefont {Corbitt}, \citenamefont {Cripe}, \citenamefont {Crisosto}
		\emph {et~al.}}]{carney2021mechanical}%
	\BibitemOpen
	\bibfield  {author} {\bibinfo {author} {\bibfnamefont {D.}~\bibnamefont
			{Carney}}, \bibinfo {author} {\bibfnamefont {G.}~\bibnamefont {Krnjaic}},
		\bibinfo {author} {\bibfnamefont {D.~C.}\ \bibnamefont {Moore}}, \bibinfo
		{author} {\bibfnamefont {C.~A.}\ \bibnamefont {Regal}}, \bibinfo {author}
		{\bibfnamefont {G.}~\bibnamefont {Afek}}, \bibinfo {author} {\bibfnamefont
			{S.}~\bibnamefont {Bhave}}, \bibinfo {author} {\bibfnamefont
			{B.}~\bibnamefont {Brubaker}}, \bibinfo {author} {\bibfnamefont
			{T.}~\bibnamefont {Corbitt}}, \bibinfo {author} {\bibfnamefont
			{J.}~\bibnamefont {Cripe}}, \bibinfo {author} {\bibfnamefont
			{N.}~\bibnamefont {Crisosto}}, \emph {et~al.},\ }\bibfield  {title} {\bibinfo
		{title} {Mechanical quantum sensing in the search for dark matter},\ }\href
	{https://iopscience.iop.org/article/10.1088/2058-9565/abcfcd/meta} {\bibfield
		{journal} {\bibinfo  {journal} {Quantum Science and Technology}\ }\textbf
		{\bibinfo {volume} {6}},\ \bibinfo {pages} {024002} (\bibinfo {year}
		{2021})}\BibitemShut {NoStop}%
	\bibitem [{\citenamefont {Manley}\ \emph {et~al.}(2021)\citenamefont {Manley},
		\citenamefont {Chowdhury}, \citenamefont {Grin}, \citenamefont {Singh},\ and\
		\citenamefont {Wilson}}]{manley2021searching}%
	\BibitemOpen
	\bibfield  {author} {\bibinfo {author} {\bibfnamefont {J.}~\bibnamefont
			{Manley}}, \bibinfo {author} {\bibfnamefont {M.~D.}\ \bibnamefont
			{Chowdhury}}, \bibinfo {author} {\bibfnamefont {D.}~\bibnamefont {Grin}},
		\bibinfo {author} {\bibfnamefont {S.}~\bibnamefont {Singh}},\ and\ \bibinfo
		{author} {\bibfnamefont {D.~J.}\ \bibnamefont {Wilson}},\ }\bibfield  {title}
	{\bibinfo {title} {Searching for vector dark matter with an optomechanical
			accelerometer},\ }\href
	{https://journals.aps.org/prl/abstract/10.1103/PhysRevLett.126.061301}
	{\bibfield  {journal} {\bibinfo  {journal} {Physical review letters}\
		}\textbf {\bibinfo {volume} {126}},\ \bibinfo {pages} {061301} (\bibinfo
		{year} {2021})}\BibitemShut {NoStop}%
	\bibitem [{\citenamefont {Aspelmeyer}\ \emph {et~al.}(2014)\citenamefont
		{Aspelmeyer}, \citenamefont {Kippenberg},\ and\ \citenamefont
		{Marquardt}}]{RMP_optomechanics}%
	\BibitemOpen
	\bibfield  {author} {\bibinfo {author} {\bibfnamefont {M.}~\bibnamefont
			{Aspelmeyer}}, \bibinfo {author} {\bibfnamefont {T.~J.}\ \bibnamefont
			{Kippenberg}},\ and\ \bibinfo {author} {\bibfnamefont {F.}~\bibnamefont
			{Marquardt}},\ }\bibfield  {title} {\bibinfo {title} {Cavity optomechanics},\
	}\href {https://journals.aps.org/rmp/abstract/10.1103/RevModPhys.86.1391}
	{\bibfield  {journal} {\bibinfo  {journal} {Reviews of Modern Physics}\
		}\textbf {\bibinfo {volume} {86}},\ \bibinfo {pages} {1391} (\bibinfo {year}
		{2014})}\BibitemShut {NoStop}%
	\bibitem [{\citenamefont {Clerk}\ \emph {et~al.}(2020)\citenamefont {Clerk},
		\citenamefont {Lehnert}, \citenamefont {Bertet}, \citenamefont {Petta},\ and\
		\citenamefont {Nakamura}}]{clerk2020hybrid}%
	\BibitemOpen
	\bibfield  {author} {\bibinfo {author} {\bibfnamefont {A.}~\bibnamefont
			{Clerk}}, \bibinfo {author} {\bibfnamefont {K.}~\bibnamefont {Lehnert}},
		\bibinfo {author} {\bibfnamefont {P.}~\bibnamefont {Bertet}}, \bibinfo
		{author} {\bibfnamefont {J.}~\bibnamefont {Petta}},\ and\ \bibinfo {author}
		{\bibfnamefont {Y.}~\bibnamefont {Nakamura}},\ }\bibfield  {title} {\bibinfo
		{title} {Hybrid quantum systems with circuit quantum electrodynamics},\
	}\href {https://www.nature.com/articles/s41567-020-0797-9} {\bibfield
		{journal} {\bibinfo  {journal} {Nature Physics}\ }\textbf {\bibinfo {volume}
			{16}},\ \bibinfo {pages} {257} (\bibinfo {year} {2020})}\BibitemShut
	{NoStop}%
	\bibitem [{\citenamefont {Chu}\ and\ \citenamefont
		{Gr{\"o}blacher}(2020)}]{chu2020perspective}%
	\BibitemOpen
	\bibfield  {author} {\bibinfo {author} {\bibfnamefont {Y.}~\bibnamefont
			{Chu}}\ and\ \bibinfo {author} {\bibfnamefont {S.}~\bibnamefont
			{Gr{\"o}blacher}},\ }\bibfield  {title} {\bibinfo {title} {A perspective on
			hybrid quantum opto-and electromechanical systems},\ }\href
	{https://aip.scitation.org/doi/full/10.1063/5.0021088} {\bibfield  {journal}
		{\bibinfo  {journal} {Applied Physics Letters}\ }\textbf {\bibinfo {volume}
			{117}},\ \bibinfo {pages} {150503} (\bibinfo {year} {2020})}\BibitemShut
	{NoStop}%
	\bibitem [{\citenamefont {Wollman}\ \emph {et~al.}(2015)\citenamefont
		{Wollman}, \citenamefont {Lei}, \citenamefont {Weinstein}, \citenamefont
		{Suh}, \citenamefont {Kronwald}, \citenamefont {Marquardt}, \citenamefont
		{Clerk},\ and\ \citenamefont {Schwab}}]{wollman2015quantum}%
	\BibitemOpen
	\bibfield  {author} {\bibinfo {author} {\bibfnamefont {E.~E.}\ \bibnamefont
			{Wollman}}, \bibinfo {author} {\bibfnamefont {C.}~\bibnamefont {Lei}},
		\bibinfo {author} {\bibfnamefont {A.}~\bibnamefont {Weinstein}}, \bibinfo
		{author} {\bibfnamefont {J.}~\bibnamefont {Suh}}, \bibinfo {author}
		{\bibfnamefont {A.}~\bibnamefont {Kronwald}}, \bibinfo {author}
		{\bibfnamefont {F.}~\bibnamefont {Marquardt}}, \bibinfo {author}
		{\bibfnamefont {A.~A.}\ \bibnamefont {Clerk}},\ and\ \bibinfo {author}
		{\bibfnamefont {K.}~\bibnamefont {Schwab}},\ }\bibfield  {title} {\bibinfo
		{title} {Quantum squeezing of motion in a mechanical resonator},\ }\href
	{https://www.science.org/doi/10.1126/science.aac5138} {\bibfield  {journal}
		{\bibinfo  {journal} {Science}\ }\textbf {\bibinfo {volume} {349}},\ \bibinfo
		{pages} {952} (\bibinfo {year} {2015})}\BibitemShut {NoStop}%
	\bibitem [{\citenamefont {Pirkkalainen}\ \emph {et~al.}(2015)\citenamefont
		{Pirkkalainen}, \citenamefont {Damsk{\"a}gg}, \citenamefont {Brandt},
		\citenamefont {Massel},\ and\ \citenamefont
		{Sillanp{\"a}{\"a}}}]{pirkkalainen2015squeezing}%
	\BibitemOpen
	\bibfield  {author} {\bibinfo {author} {\bibfnamefont {J.-M.}\ \bibnamefont
			{Pirkkalainen}}, \bibinfo {author} {\bibfnamefont {E.}~\bibnamefont
			{Damsk{\"a}gg}}, \bibinfo {author} {\bibfnamefont {M.}~\bibnamefont
			{Brandt}}, \bibinfo {author} {\bibfnamefont {F.}~\bibnamefont {Massel}},\
		and\ \bibinfo {author} {\bibfnamefont {M.~A.}\ \bibnamefont
			{Sillanp{\"a}{\"a}}},\ }\bibfield  {title} {\bibinfo {title} {Squeezing of
			quantum noise of motion in a micromechanical resonator},\ }\href
	{https://journals.aps.org/prl/abstract/10.1103/PhysRevLett.115.243601}
	{\bibfield  {journal} {\bibinfo  {journal} {Physical Review Letters}\
		}\textbf {\bibinfo {volume} {115}},\ \bibinfo {pages} {243601} (\bibinfo
		{year} {2015})}\BibitemShut {NoStop}%
	\bibitem [{\citenamefont {Lecocq}\ \emph {et~al.}(2015)\citenamefont {Lecocq},
		\citenamefont {Clark}, \citenamefont {Simmonds}, \citenamefont {Aumentado},\
		and\ \citenamefont {Teufel}}]{lecocq2015quantum}%
	\BibitemOpen
	\bibfield  {author} {\bibinfo {author} {\bibfnamefont {F.}~\bibnamefont
			{Lecocq}}, \bibinfo {author} {\bibfnamefont {J.~B.}\ \bibnamefont {Clark}},
		\bibinfo {author} {\bibfnamefont {R.~W.}\ \bibnamefont {Simmonds}}, \bibinfo
		{author} {\bibfnamefont {J.}~\bibnamefont {Aumentado}},\ and\ \bibinfo
		{author} {\bibfnamefont {J.~D.}\ \bibnamefont {Teufel}},\ }\bibfield  {title}
	{\bibinfo {title} {Quantum nondemolition measurement of a nonclassical state
			of a massive object},\ }\href
	{https://journals.aps.org/prx/abstract/10.1103/PhysRevX.5.041037} {\bibfield
		{journal} {\bibinfo  {journal} {Physical Review X}\ }\textbf {\bibinfo
			{volume} {5}},\ \bibinfo {pages} {041037} (\bibinfo {year}
		{2015})}\BibitemShut {NoStop}%
	\bibitem [{\citenamefont {Reed}\ \emph {et~al.}(2017)\citenamefont {Reed},
		\citenamefont {Mayer}, \citenamefont {Teufel}, \citenamefont {Burkhart},
		\citenamefont {Pfaff}, \citenamefont {Reagor}, \citenamefont {Sletten},
		\citenamefont {Ma}, \citenamefont {Schoelkopf}, \citenamefont {Knill} \emph
		{et~al.}}]{reed2017faithful}%
	\BibitemOpen
	\bibfield  {author} {\bibinfo {author} {\bibfnamefont {A.}~\bibnamefont
			{Reed}}, \bibinfo {author} {\bibfnamefont {K.}~\bibnamefont {Mayer}},
		\bibinfo {author} {\bibfnamefont {J.}~\bibnamefont {Teufel}}, \bibinfo
		{author} {\bibfnamefont {L.}~\bibnamefont {Burkhart}}, \bibinfo {author}
		{\bibfnamefont {W.}~\bibnamefont {Pfaff}}, \bibinfo {author} {\bibfnamefont
			{M.}~\bibnamefont {Reagor}}, \bibinfo {author} {\bibfnamefont
			{L.}~\bibnamefont {Sletten}}, \bibinfo {author} {\bibfnamefont
			{X.}~\bibnamefont {Ma}}, \bibinfo {author} {\bibfnamefont {R.}~\bibnamefont
			{Schoelkopf}}, \bibinfo {author} {\bibfnamefont {E.}~\bibnamefont {Knill}},
		\emph {et~al.},\ }\bibfield  {title} {\bibinfo {title} {Faithful conversion
			of propagating quantum information to mechanical motion},\ }\href
	{https://www.nature.com/articles/nphys4251} {\bibfield  {journal} {\bibinfo
			{journal} {Nature Physics}\ }\textbf {\bibinfo {volume} {13}},\ \bibinfo
		{pages} {1163} (\bibinfo {year} {2017})}\BibitemShut {NoStop}%
	\bibitem [{\citenamefont {Chu}\ \emph {et~al.}(2018)\citenamefont {Chu},
		\citenamefont {Kharel}, \citenamefont {Yoon}, \citenamefont {Frunzio},
		\citenamefont {Rakich},\ and\ \citenamefont {Schoelkopf}}]{chu2018creation}%
	\BibitemOpen
	\bibfield  {author} {\bibinfo {author} {\bibfnamefont {Y.}~\bibnamefont
			{Chu}}, \bibinfo {author} {\bibfnamefont {P.}~\bibnamefont {Kharel}},
		\bibinfo {author} {\bibfnamefont {T.}~\bibnamefont {Yoon}}, \bibinfo {author}
		{\bibfnamefont {L.}~\bibnamefont {Frunzio}}, \bibinfo {author} {\bibfnamefont
			{P.~T.}\ \bibnamefont {Rakich}},\ and\ \bibinfo {author} {\bibfnamefont
			{R.~J.}\ \bibnamefont {Schoelkopf}},\ }\bibfield  {title} {\bibinfo {title}
		{Creation and control of multi-phonon fock states in a bulk acoustic-wave
			resonator},\ }\href {https://www.nature.com/articles/s41586-018-0717-7}
	{\bibfield  {journal} {\bibinfo  {journal} {Nature}\ }\textbf {\bibinfo
			{volume} {563}},\ \bibinfo {pages} {666} (\bibinfo {year}
		{2018})}\BibitemShut {NoStop}%
	\bibitem [{\citenamefont {Mirhosseini}\ \emph {et~al.}(2020)\citenamefont
		{Mirhosseini}, \citenamefont {Sipahigil}, \citenamefont {Kalaee},\ and\
		\citenamefont {Painter}}]{mirhosseini2020superconducting}%
	\BibitemOpen
	\bibfield  {author} {\bibinfo {author} {\bibfnamefont {M.}~\bibnamefont
			{Mirhosseini}}, \bibinfo {author} {\bibfnamefont {A.}~\bibnamefont
			{Sipahigil}}, \bibinfo {author} {\bibfnamefont {M.}~\bibnamefont {Kalaee}},\
		and\ \bibinfo {author} {\bibfnamefont {O.}~\bibnamefont {Painter}},\
	}\bibfield  {title} {\bibinfo {title} {Superconducting qubit to optical
			photon transduction},\ }\href
	{https://www.nature.com/articles/s41586-020-3038-6} {\bibfield  {journal}
		{\bibinfo  {journal} {Nature}\ }\textbf {\bibinfo {volume} {588}},\ \bibinfo
		{pages} {599} (\bibinfo {year} {2020})}\BibitemShut {NoStop}%
	\bibitem [{\citenamefont {Andrews}\ \emph {et~al.}(2014)\citenamefont
		{Andrews}, \citenamefont {Peterson}, \citenamefont {Purdy}, \citenamefont
		{Cicak}, \citenamefont {Simmonds} \emph {et~al.}}]{andrews2014bidirectional}%
	\BibitemOpen
	\bibfield  {author} {\bibinfo {author} {\bibfnamefont {R.~W.}\ \bibnamefont
			{Andrews}}, \bibinfo {author} {\bibfnamefont {R.~W.}\ \bibnamefont
			{Peterson}}, \bibinfo {author} {\bibfnamefont {T.~P.}\ \bibnamefont {Purdy}},
		\bibinfo {author} {\bibfnamefont {K.}~\bibnamefont {Cicak}}, \bibinfo
		{author} {\bibfnamefont {R.~W.}\ \bibnamefont {Simmonds}}, \emph {et~al.},\
	}\bibfield  {title} {\bibinfo {title} {Bidirectional and efficient conversion
			between microwave and optical light},\ }\href
	{https://www.nature.com/articles/nphys2911} {\bibfield  {journal} {\bibinfo
			{journal} {Nature Physics}\ }\textbf {\bibinfo {volume} {10}},\ \bibinfo
		{pages} {321} (\bibinfo {year} {2014})}\BibitemShut {NoStop}%
	\bibitem [{\citenamefont {MacCabe}\ \emph {et~al.}(2020)\citenamefont
		{MacCabe}, \citenamefont {Ren}, \citenamefont {Luo}, \citenamefont {Cohen},
		\citenamefont {Zhou}, \citenamefont {Sipahigil}, \citenamefont
		{Mirhosseini},\ and\ \citenamefont {Painter}}]{maccabe2020nano}%
	\BibitemOpen
	\bibfield  {author} {\bibinfo {author} {\bibfnamefont {G.~S.}\ \bibnamefont
			{MacCabe}}, \bibinfo {author} {\bibfnamefont {H.}~\bibnamefont {Ren}},
		\bibinfo {author} {\bibfnamefont {J.}~\bibnamefont {Luo}}, \bibinfo {author}
		{\bibfnamefont {J.~D.}\ \bibnamefont {Cohen}}, \bibinfo {author}
		{\bibfnamefont {H.}~\bibnamefont {Zhou}}, \bibinfo {author} {\bibfnamefont
			{A.}~\bibnamefont {Sipahigil}}, \bibinfo {author} {\bibfnamefont
			{M.}~\bibnamefont {Mirhosseini}},\ and\ \bibinfo {author} {\bibfnamefont
			{O.}~\bibnamefont {Painter}},\ }\bibfield  {title} {\bibinfo {title}
		{Nano-acoustic resonator with ultralong phonon lifetime},\ }\href
	{https://www.science.org/doi/full/10.1126/science.abc7312} {\bibfield
		{journal} {\bibinfo  {journal} {Science}\ }\textbf {\bibinfo {volume}
			{370}},\ \bibinfo {pages} {840} (\bibinfo {year} {2020})}\BibitemShut
	{NoStop}%
	\bibitem [{\citenamefont {Rossi}\ \emph {et~al.}(2018)\citenamefont {Rossi},
		\citenamefont {Mason}, \citenamefont {Chen}, \citenamefont {Tsaturyan},\ and\
		\citenamefont {Schliesser}}]{rossi2018measurement}%
	\BibitemOpen
	\bibfield  {author} {\bibinfo {author} {\bibfnamefont {M.}~\bibnamefont
			{Rossi}}, \bibinfo {author} {\bibfnamefont {D.}~\bibnamefont {Mason}},
		\bibinfo {author} {\bibfnamefont {J.}~\bibnamefont {Chen}}, \bibinfo {author}
		{\bibfnamefont {Y.}~\bibnamefont {Tsaturyan}},\ and\ \bibinfo {author}
		{\bibfnamefont {A.}~\bibnamefont {Schliesser}},\ }\bibfield  {title}
	{\bibinfo {title} {Measurement-based quantum control of mechanical motion},\
	}\href {https://doi.org/10.1038/s41586-018-0643-8} {\bibfield  {journal}
		{\bibinfo  {journal} {Nature}\ }\textbf {\bibinfo {volume} {563}},\ \bibinfo
		{pages} {53} (\bibinfo {year} {2018})}\BibitemShut {NoStop}%
	\bibitem [{\citenamefont {Palomaki}\ \emph
		{et~al.}(2013{\natexlab{a}})\citenamefont {Palomaki}, \citenamefont {Harlow},
		\citenamefont {Teufel}, \citenamefont {Simmonds},\ and\ \citenamefont
		{Lehnert}}]{palomaki2013coherent}%
	\BibitemOpen
	\bibfield  {author} {\bibinfo {author} {\bibfnamefont {T.}~\bibnamefont
			{Palomaki}}, \bibinfo {author} {\bibfnamefont {J.}~\bibnamefont {Harlow}},
		\bibinfo {author} {\bibfnamefont {J.}~\bibnamefont {Teufel}}, \bibinfo
		{author} {\bibfnamefont {R.}~\bibnamefont {Simmonds}},\ and\ \bibinfo
		{author} {\bibfnamefont {K.~W.}\ \bibnamefont {Lehnert}},\ }\bibfield
	{title} {\bibinfo {title} {Coherent state transfer between itinerant
			microwave fields and a mechanical oscillator},\ }\href
	{https://www.nature.com/articles/nature11915} {\bibfield  {journal} {\bibinfo
			{journal} {Nature}\ }\textbf {\bibinfo {volume} {495}},\ \bibinfo {pages}
		{210} (\bibinfo {year} {2013}{\natexlab{a}})}\BibitemShut {NoStop}%
	\bibitem [{\citenamefont {Magrini}\ \emph {et~al.}(2021)\citenamefont
		{Magrini}, \citenamefont {Rosenzweig}, \citenamefont {Bach}, \citenamefont
		{Deutschmann-Olek}, \citenamefont {Hofer}, \citenamefont {Hong},
		\citenamefont {Kiesel}, \citenamefont {Kugi},\ and\ \citenamefont
		{Aspelmeyer}}]{magrini2021real}%
	\BibitemOpen
	\bibfield  {author} {\bibinfo {author} {\bibfnamefont {L.}~\bibnamefont
			{Magrini}}, \bibinfo {author} {\bibfnamefont {P.}~\bibnamefont {Rosenzweig}},
		\bibinfo {author} {\bibfnamefont {C.}~\bibnamefont {Bach}}, \bibinfo {author}
		{\bibfnamefont {A.}~\bibnamefont {Deutschmann-Olek}}, \bibinfo {author}
		{\bibfnamefont {S.~G.}\ \bibnamefont {Hofer}}, \bibinfo {author}
		{\bibfnamefont {S.}~\bibnamefont {Hong}}, \bibinfo {author} {\bibfnamefont
			{N.}~\bibnamefont {Kiesel}}, \bibinfo {author} {\bibfnamefont
			{A.}~\bibnamefont {Kugi}},\ and\ \bibinfo {author} {\bibfnamefont
			{M.}~\bibnamefont {Aspelmeyer}},\ }\bibfield  {title} {\bibinfo {title}
		{Real-time optimal quantum control of mechanical motion at room
			temperature},\ }\href {https://www.nature.com/articles/s41586-021-03602-3}
	{\bibfield  {journal} {\bibinfo  {journal} {Nature}\ }\textbf {\bibinfo
			{volume} {595}},\ \bibinfo {pages} {373} (\bibinfo {year}
		{2021})}\BibitemShut {NoStop}%
	\bibitem [{\citenamefont {Wollack}\ \emph {et~al.}(2022)\citenamefont
		{Wollack}, \citenamefont {Cleland}, \citenamefont {Gruenke}, \citenamefont
		{Wang}, \citenamefont {Arrangoiz-Arriola},\ and\ \citenamefont
		{Safavi-Naeini}}]{wollack2022quantum}%
	\BibitemOpen
	\bibfield  {author} {\bibinfo {author} {\bibfnamefont {E.~A.}\ \bibnamefont
			{Wollack}}, \bibinfo {author} {\bibfnamefont {A.~Y.}\ \bibnamefont
			{Cleland}}, \bibinfo {author} {\bibfnamefont {R.~G.}\ \bibnamefont
			{Gruenke}}, \bibinfo {author} {\bibfnamefont {Z.}~\bibnamefont {Wang}},
		\bibinfo {author} {\bibfnamefont {P.}~\bibnamefont {Arrangoiz-Arriola}},\
		and\ \bibinfo {author} {\bibfnamefont {A.~H.}\ \bibnamefont
			{Safavi-Naeini}},\ }\bibfield  {title} {\bibinfo {title} {Quantum state
			preparation and tomography of entangled mechanical resonators},\ }\href
	{https://www.nature.com/articles/s41586-022-04500-y} {\bibfield  {journal}
		{\bibinfo  {journal} {Nature}\ }\textbf {\bibinfo {volume} {604}},\ \bibinfo
		{pages} {463} (\bibinfo {year} {2022})}\BibitemShut {NoStop}%
	\bibitem [{\citenamefont {Satzinger}\ \emph {et~al.}(2018)\citenamefont
		{Satzinger}, \citenamefont {Zhong}, \citenamefont {Chang}, \citenamefont
		{Peairs}, \citenamefont {Bienfait}, \citenamefont {Chou}, \citenamefont
		{Cleland}, \citenamefont {Conner}, \citenamefont {Dumur}, \citenamefont
		{Grebel} \emph {et~al.}}]{satzinger2018quantum}%
	\BibitemOpen
	\bibfield  {author} {\bibinfo {author} {\bibfnamefont {K.~J.}\ \bibnamefont
			{Satzinger}}, \bibinfo {author} {\bibfnamefont {Y.}~\bibnamefont {Zhong}},
		\bibinfo {author} {\bibfnamefont {H.-S.}\ \bibnamefont {Chang}}, \bibinfo
		{author} {\bibfnamefont {G.~A.}\ \bibnamefont {Peairs}}, \bibinfo {author}
		{\bibfnamefont {A.}~\bibnamefont {Bienfait}}, \bibinfo {author}
		{\bibfnamefont {M.-H.}\ \bibnamefont {Chou}}, \bibinfo {author}
		{\bibfnamefont {A.}~\bibnamefont {Cleland}}, \bibinfo {author} {\bibfnamefont
			{C.~R.}\ \bibnamefont {Conner}}, \bibinfo {author} {\bibfnamefont
			{{\'E}.}~\bibnamefont {Dumur}}, \bibinfo {author} {\bibfnamefont
			{J.}~\bibnamefont {Grebel}}, \emph {et~al.},\ }\bibfield  {title} {\bibinfo
		{title} {Quantum control of surface acoustic-wave phonons},\ }\href
	{https://www.nature.com/articles/s41586-018-0719-5} {\bibfield  {journal}
		{\bibinfo  {journal} {Nature}\ }\textbf {\bibinfo {volume} {563}},\ \bibinfo
		{pages} {661} (\bibinfo {year} {2018})}\BibitemShut {NoStop}%
	\bibitem [{\citenamefont {Gaebler}\ \emph {et~al.}(2016)\citenamefont
		{Gaebler}, \citenamefont {Tan}, \citenamefont {Lin}, \citenamefont {Wan},
		\citenamefont {Bowler}, \citenamefont {Keith}, \citenamefont {Glancy},
		\citenamefont {Coakley}, \citenamefont {Knill}, \citenamefont {Leibfried}
		\emph {et~al.}}]{gaebler2016high}%
	\BibitemOpen
	\bibfield  {author} {\bibinfo {author} {\bibfnamefont {J.~P.}\ \bibnamefont
			{Gaebler}}, \bibinfo {author} {\bibfnamefont {T.~R.}\ \bibnamefont {Tan}},
		\bibinfo {author} {\bibfnamefont {Y.}~\bibnamefont {Lin}}, \bibinfo {author}
		{\bibfnamefont {Y.}~\bibnamefont {Wan}}, \bibinfo {author} {\bibfnamefont
			{R.}~\bibnamefont {Bowler}}, \bibinfo {author} {\bibfnamefont {A.~C.}\
			\bibnamefont {Keith}}, \bibinfo {author} {\bibfnamefont {S.}~\bibnamefont
			{Glancy}}, \bibinfo {author} {\bibfnamefont {K.}~\bibnamefont {Coakley}},
		\bibinfo {author} {\bibfnamefont {E.}~\bibnamefont {Knill}}, \bibinfo
		{author} {\bibfnamefont {D.}~\bibnamefont {Leibfried}}, \emph {et~al.},\
	}\bibfield  {title} {\bibinfo {title} {High-fidelity universal gate set for
			be 9+ ion qubits},\ }\href
	{https://journals.aps.org/prl/abstract/10.1103/PhysRevLett.117.060505}
	{\bibfield  {journal} {\bibinfo  {journal} {Physical review letters}\
		}\textbf {\bibinfo {volume} {117}},\ \bibinfo {pages} {060505} (\bibinfo
		{year} {2016})}\BibitemShut {NoStop}%
	\bibitem [{\citenamefont {Leibfried}\ \emph {et~al.}(2003)\citenamefont
		{Leibfried}, \citenamefont {Blatt}, \citenamefont {Monroe},\ and\
		\citenamefont {Wineland}}]{leibfried2003quantum}%
	\BibitemOpen
	\bibfield  {author} {\bibinfo {author} {\bibfnamefont {D.}~\bibnamefont
			{Leibfried}}, \bibinfo {author} {\bibfnamefont {R.}~\bibnamefont {Blatt}},
		\bibinfo {author} {\bibfnamefont {C.}~\bibnamefont {Monroe}},\ and\ \bibinfo
		{author} {\bibfnamefont {D.}~\bibnamefont {Wineland}},\ }\bibfield  {title}
	{\bibinfo {title} {Quantum dynamics of single trapped ions},\ }\href
	{https://journals.aps.org/rmp/abstract/10.1103/RevModPhys.75.281} {\bibfield
		{journal} {\bibinfo  {journal} {Reviews of Modern Physics}\ }\textbf
		{\bibinfo {volume} {75}},\ \bibinfo {pages} {281} (\bibinfo {year}
		{2003})}\BibitemShut {NoStop}%
	\bibitem [{\citenamefont {Gely}\ and\ \citenamefont
		{Steele}(2021{\natexlab{a}})}]{gely2021phonon}%
	\BibitemOpen
	\bibfield  {author} {\bibinfo {author} {\bibfnamefont {M.~F.}\ \bibnamefont
			{Gely}}\ and\ \bibinfo {author} {\bibfnamefont {G.~A.}\ \bibnamefont
			{Steele}},\ }\bibfield  {title} {\bibinfo {title} {Phonon-number resolution
			of voltage-biased mechanical oscillators with weakly anharmonic
			superconducting circuits},\ }\href
	{https://journals.aps.org/pra/abstract/10.1103/PhysRevA.104.053509}
	{\bibfield  {journal} {\bibinfo  {journal} {Physical Review A}\ }\textbf
		{\bibinfo {volume} {104}},\ \bibinfo {pages} {053509} (\bibinfo {year}
		{2021}{\natexlab{a}})}\BibitemShut {NoStop}%
	\bibitem [{\citenamefont {Gely}\ and\ \citenamefont
		{Steele}(2021{\natexlab{b}})}]{gely2021superconducting}%
	\BibitemOpen
	\bibfield  {author} {\bibinfo {author} {\bibfnamefont {M.~F.}\ \bibnamefont
			{Gely}}\ and\ \bibinfo {author} {\bibfnamefont {G.~A.}\ \bibnamefont
			{Steele}},\ }\bibfield  {title} {\bibinfo {title} {Superconducting
			electro-mechanics to test {D}i{\'o}si--{P}enrose effects of general
			relativity in massive superpositions},\ }\href
	{https://avs.scitation.org/doi/full/10.1116/5.0050988} {\bibfield  {journal}
		{\bibinfo  {journal} {AVS Quantum Science}\ }\textbf {\bibinfo {volume}
			{3}},\ \bibinfo {pages} {035601} (\bibinfo {year}
		{2021}{\natexlab{b}})}\BibitemShut {NoStop}%
	\bibitem [{\citenamefont {Liu}\ \emph {et~al.}(2021{\natexlab{a}})\citenamefont
		{Liu}, \citenamefont {Mummery}, \citenamefont {Zhou},\ and\ \citenamefont
		{Sillanp{\"a}{\"a}}}]{liu2021gravitational}%
	\BibitemOpen
	\bibfield  {author} {\bibinfo {author} {\bibfnamefont {Y.}~\bibnamefont
			{Liu}}, \bibinfo {author} {\bibfnamefont {J.}~\bibnamefont {Mummery}},
		\bibinfo {author} {\bibfnamefont {J.}~\bibnamefont {Zhou}},\ and\ \bibinfo
		{author} {\bibfnamefont {M.~A.}\ \bibnamefont {Sillanp{\"a}{\"a}}},\
	}\bibfield  {title} {\bibinfo {title} {Gravitational forces between
			nonclassical mechanical oscillators},\ }\href
	{https://journals.aps.org/prapplied/abstract/10.1103/PhysRevApplied.15.034004}
	{\bibfield  {journal} {\bibinfo  {journal} {Physical Review Applied}\
		}\textbf {\bibinfo {volume} {15}},\ \bibinfo {pages} {034004} (\bibinfo
		{year} {2021}{\natexlab{a}})}\BibitemShut {NoStop}%
	\bibitem [{\citenamefont {Kotler}\ \emph {et~al.}(2021)\citenamefont {Kotler},
		\citenamefont {Peterson}, \citenamefont {Shojaee}, \citenamefont {Lecocq},
		\citenamefont {Cicak}, \citenamefont {Kwiatkowski}, \citenamefont {Geller},
		\citenamefont {Glancy}, \citenamefont {Knill}, \citenamefont {Simmonds} \emph
		{et~al.}}]{kotler2021direct}%
	\BibitemOpen
	\bibfield  {author} {\bibinfo {author} {\bibfnamefont {S.}~\bibnamefont
			{Kotler}}, \bibinfo {author} {\bibfnamefont {G.~A.}\ \bibnamefont
			{Peterson}}, \bibinfo {author} {\bibfnamefont {E.}~\bibnamefont {Shojaee}},
		\bibinfo {author} {\bibfnamefont {F.}~\bibnamefont {Lecocq}}, \bibinfo
		{author} {\bibfnamefont {K.}~\bibnamefont {Cicak}}, \bibinfo {author}
		{\bibfnamefont {A.}~\bibnamefont {Kwiatkowski}}, \bibinfo {author}
		{\bibfnamefont {S.}~\bibnamefont {Geller}}, \bibinfo {author} {\bibfnamefont
			{S.}~\bibnamefont {Glancy}}, \bibinfo {author} {\bibfnamefont
			{E.}~\bibnamefont {Knill}}, \bibinfo {author} {\bibfnamefont {R.~W.}\
			\bibnamefont {Simmonds}}, \emph {et~al.},\ }\bibfield  {title} {\bibinfo
		{title} {Direct observation of deterministic macroscopic entanglement},\
	}\href {https://science.sciencemag.org/content/372/6542/622} {\bibfield
		{journal} {\bibinfo  {journal} {Science}\ }\textbf {\bibinfo {volume}
			{372}},\ \bibinfo {pages} {622} (\bibinfo {year} {2021})}\BibitemShut
	{NoStop}%
	\bibitem [{\citenamefont {Delaney}\ \emph {et~al.}(2019)\citenamefont
		{Delaney}, \citenamefont {Reed}, \citenamefont {Andrews},\ and\ \citenamefont
		{Lehnert}}]{delaney2019measurement}%
	\BibitemOpen
	\bibfield  {author} {\bibinfo {author} {\bibfnamefont {R.~D.}\ \bibnamefont
			{Delaney}}, \bibinfo {author} {\bibfnamefont {A.~P.}\ \bibnamefont {Reed}},
		\bibinfo {author} {\bibfnamefont {R.~W.}\ \bibnamefont {Andrews}},\ and\
		\bibinfo {author} {\bibfnamefont {K.~W.}\ \bibnamefont {Lehnert}},\
	}\bibfield  {title} {\bibinfo {title} {Measurement of motion beyond the
			quantum limit by transient amplification},\ }\href
	{https://journals.aps.org/prl/abstract/10.1103/PhysRevLett.123.183603}
	{\bibfield  {journal} {\bibinfo  {journal} {Physical review letters}\
		}\textbf {\bibinfo {volume} {123}},\ \bibinfo {pages} {183603} (\bibinfo
		{year} {2019})}\BibitemShut {NoStop}%
	\bibitem [{\citenamefont {Gardiner}\ \emph {et~al.}(2004)\citenamefont
		{Gardiner}, \citenamefont {Zoller},\ and\ \citenamefont
		{Zoller}}]{gardiner2004quantum}%
	\BibitemOpen
	\bibfield  {author} {\bibinfo {author} {\bibfnamefont {C.}~\bibnamefont
			{Gardiner}}, \bibinfo {author} {\bibfnamefont {P.}~\bibnamefont {Zoller}},\
		and\ \bibinfo {author} {\bibfnamefont {P.}~\bibnamefont {Zoller}},\
	}\href@noop {} {\emph {\bibinfo {title} {Quantum noise: a handbook of
				Markovian and non-Markovian quantum stochastic methods with applications to
				quantum optics}}}\ (\bibinfo  {publisher} {Springer Science \& Business
		Media},\ \bibinfo {year} {2004})\BibitemShut {NoStop}%
	\bibitem [{\citenamefont {Tebbenjohanns}\ \emph {et~al.}(2021)\citenamefont
		{Tebbenjohanns}, \citenamefont {Mattana}, \citenamefont {Rossi},
		\citenamefont {Frimmer},\ and\ \citenamefont
		{Novotny}}]{tebbenjohanns2021quantum}%
	\BibitemOpen
	\bibfield  {author} {\bibinfo {author} {\bibfnamefont {F.}~\bibnamefont
			{Tebbenjohanns}}, \bibinfo {author} {\bibfnamefont {M.~L.}\ \bibnamefont
			{Mattana}}, \bibinfo {author} {\bibfnamefont {M.}~\bibnamefont {Rossi}},
		\bibinfo {author} {\bibfnamefont {M.}~\bibnamefont {Frimmer}},\ and\ \bibinfo
		{author} {\bibfnamefont {L.}~\bibnamefont {Novotny}},\ }\bibfield  {title}
	{\bibinfo {title} {Quantum control of a nanoparticle optically levitated in
			cryogenic free space},\ }\href
	{https://www.nature.com/articles/s41586-021-03617-w} {\bibfield  {journal}
		{\bibinfo  {journal} {Nature}\ }\textbf {\bibinfo {volume} {595}},\ \bibinfo
		{pages} {378} (\bibinfo {year} {2021})}\BibitemShut {NoStop}%
	\bibitem [{\citenamefont {Deli{\'c}}\ \emph {et~al.}(2020)\citenamefont
		{Deli{\'c}}, \citenamefont {Reisenbauer}, \citenamefont {Dare}, \citenamefont
		{Grass}, \citenamefont {Vuleti{\'c}}, \citenamefont {Kiesel},\ and\
		\citenamefont {Aspelmeyer}}]{delic2020cooling}%
	\BibitemOpen
	\bibfield  {author} {\bibinfo {author} {\bibfnamefont {U.}~\bibnamefont
			{Deli{\'c}}}, \bibinfo {author} {\bibfnamefont {M.}~\bibnamefont
			{Reisenbauer}}, \bibinfo {author} {\bibfnamefont {K.}~\bibnamefont {Dare}},
		\bibinfo {author} {\bibfnamefont {D.}~\bibnamefont {Grass}}, \bibinfo
		{author} {\bibfnamefont {V.}~\bibnamefont {Vuleti{\'c}}}, \bibinfo {author}
		{\bibfnamefont {N.}~\bibnamefont {Kiesel}},\ and\ \bibinfo {author}
		{\bibfnamefont {M.}~\bibnamefont {Aspelmeyer}},\ }\bibfield  {title}
	{\bibinfo {title} {Cooling of a levitated nanoparticle to the motional
			quantum ground state},\ }\href {https://doi.org/10.1126/science.aba3993}
	{\bibfield  {journal} {\bibinfo  {journal} {Science}\ }\textbf {\bibinfo
			{volume} {367}},\ \bibinfo {pages} {892} (\bibinfo {year}
		{2020})}\BibitemShut {NoStop}%
	\bibitem [{\citenamefont {Piotrowski}\ \emph {et~al.}(2023)\citenamefont
		{Piotrowski}, \citenamefont {Windey}, \citenamefont {Vijayan}, \citenamefont
		{Gonzalez-Ballestero}, \citenamefont {de~los R{\'\i}os~Sommer}, \citenamefont
		{Meyer}, \citenamefont {Quidant}, \citenamefont {Romero-Isart}, \citenamefont
		{Reimann},\ and\ \citenamefont {Novotny}}]{piotrowski2022simultaneous}%
	\BibitemOpen
	\bibfield  {author} {\bibinfo {author} {\bibfnamefont {J.}~\bibnamefont
			{Piotrowski}}, \bibinfo {author} {\bibfnamefont {D.}~\bibnamefont {Windey}},
		\bibinfo {author} {\bibfnamefont {J.}~\bibnamefont {Vijayan}}, \bibinfo
		{author} {\bibfnamefont {C.}~\bibnamefont {Gonzalez-Ballestero}}, \bibinfo
		{author} {\bibfnamefont {A.}~\bibnamefont {de~los R{\'\i}os~Sommer}},
		\bibinfo {author} {\bibfnamefont {N.}~\bibnamefont {Meyer}}, \bibinfo
		{author} {\bibfnamefont {R.}~\bibnamefont {Quidant}}, \bibinfo {author}
		{\bibfnamefont {O.}~\bibnamefont {Romero-Isart}}, \bibinfo {author}
		{\bibfnamefont {R.}~\bibnamefont {Reimann}},\ and\ \bibinfo {author}
		{\bibfnamefont {L.}~\bibnamefont {Novotny}},\ }\bibfield  {title} {\bibinfo
		{title} {Simultaneous ground-state cooling of two mechanical modes of a
			levitated nanoparticle},\ }\href {https://doi.org/10.1038/s41567-023-01956-1}
	{\bibfield  {journal} {\bibinfo  {journal} {Nature Physics}\ ,\ \bibinfo
			{pages} {1}} (\bibinfo {year} {2023})}\BibitemShut {NoStop}%
	\bibitem [{\citenamefont {Teufel}\ \emph {et~al.}(2011)\citenamefont {Teufel},
		\citenamefont {Donner}, \citenamefont {Li}, \citenamefont {Harlow},
		\citenamefont {Allman}, \citenamefont {Cicak}, \citenamefont {Sirois},
		\citenamefont {Whittaker}, \citenamefont {Lehnert},\ and\ \citenamefont
		{Simmonds}}]{teufel2011sideband}%
	\BibitemOpen
	\bibfield  {author} {\bibinfo {author} {\bibfnamefont {J.~D.}\ \bibnamefont
			{Teufel}}, \bibinfo {author} {\bibfnamefont {T.}~\bibnamefont {Donner}},
		\bibinfo {author} {\bibfnamefont {D.}~\bibnamefont {Li}}, \bibinfo {author}
		{\bibfnamefont {J.~W.}\ \bibnamefont {Harlow}}, \bibinfo {author}
		{\bibfnamefont {M.}~\bibnamefont {Allman}}, \bibinfo {author} {\bibfnamefont
			{K.}~\bibnamefont {Cicak}}, \bibinfo {author} {\bibfnamefont {A.~J.}\
			\bibnamefont {Sirois}}, \bibinfo {author} {\bibfnamefont {J.~D.}\
			\bibnamefont {Whittaker}}, \bibinfo {author} {\bibfnamefont {K.~W.}\
			\bibnamefont {Lehnert}},\ and\ \bibinfo {author} {\bibfnamefont {R.~W.}\
			\bibnamefont {Simmonds}},\ }\bibfield  {title} {\bibinfo {title} {Sideband
			cooling of micromechanical motion to the quantum ground state},\ }\href
	{https://www.nature.com/articles/nature10261} {\bibfield  {journal} {\bibinfo
			{journal} {Nature}\ }\textbf {\bibinfo {volume} {475}},\ \bibinfo {pages}
		{359} (\bibinfo {year} {2011})}\BibitemShut {NoStop}%
	\bibitem [{\citenamefont {Ockeloen-Korppi}\ \emph {et~al.}(2018)\citenamefont
		{Ockeloen-Korppi}, \citenamefont {Damsk{\"a}gg}, \citenamefont
		{Pirkkalainen}, \citenamefont {Asjad}, \citenamefont {Clerk} \emph
		{et~al.}}]{ockeloen2018stabilized}%
	\BibitemOpen
	\bibfield  {author} {\bibinfo {author} {\bibfnamefont {C.}~\bibnamefont
			{Ockeloen-Korppi}}, \bibinfo {author} {\bibfnamefont {E.}~\bibnamefont
			{Damsk{\"a}gg}}, \bibinfo {author} {\bibfnamefont {J.-M.}\ \bibnamefont
			{Pirkkalainen}}, \bibinfo {author} {\bibfnamefont {M.}~\bibnamefont {Asjad}},
		\bibinfo {author} {\bibfnamefont {A.}~\bibnamefont {Clerk}}, \emph {et~al.},\
	}\bibfield  {title} {\bibinfo {title} {Stabilized entanglement of massive
			mechanical oscillators},\ }\href
	{https://www.nature.com/articles/s41586-018-0038-x} {\bibfield  {journal}
		{\bibinfo  {journal} {Nature}\ }\textbf {\bibinfo {volume} {556}},\ \bibinfo
		{pages} {478} (\bibinfo {year} {2018})}\BibitemShut {NoStop}%
	\bibitem [{\citenamefont {Palomaki}\ \emph
		{et~al.}(2013{\natexlab{b}})\citenamefont {Palomaki}, \citenamefont {Teufel},
		\citenamefont {Simmonds},\ and\ \citenamefont
		{Lehnert}}]{palomaki2013entangling}%
	\BibitemOpen
	\bibfield  {author} {\bibinfo {author} {\bibfnamefont {T.}~\bibnamefont
			{Palomaki}}, \bibinfo {author} {\bibfnamefont {J.}~\bibnamefont {Teufel}},
		\bibinfo {author} {\bibfnamefont {R.}~\bibnamefont {Simmonds}},\ and\
		\bibinfo {author} {\bibfnamefont {K.~W.}\ \bibnamefont {Lehnert}},\
	}\bibfield  {title} {\bibinfo {title} {Entangling mechanical motion with
			microwave fields},\ }\href
	{https://science.sciencemag.org/content/342/6159/710} {\bibfield  {journal}
		{\bibinfo  {journal} {Science}\ }\textbf {\bibinfo {volume} {342}},\ \bibinfo
		{pages} {710} (\bibinfo {year} {2013}{\natexlab{b}})}\BibitemShut {NoStop}%
	\bibitem [{\citenamefont {Bernier}\ \emph {et~al.}(2017)\citenamefont
		{Bernier}, \citenamefont {Tóth}, \citenamefont {Koottandavida},
		\citenamefont {Ioannou}, \citenamefont {Malz} \emph {et~al.}}]{Bernier2017}%
	\BibitemOpen
	\bibfield  {author} {\bibinfo {author} {\bibfnamefont {N.~R.}\ \bibnamefont
			{Bernier}}, \bibinfo {author} {\bibfnamefont {L.~D.}\ \bibnamefont {Tóth}},
		\bibinfo {author} {\bibfnamefont {A.}~\bibnamefont {Koottandavida}}, \bibinfo
		{author} {\bibfnamefont {M.~A.}\ \bibnamefont {Ioannou}}, \bibinfo {author}
		{\bibfnamefont {D.}~\bibnamefont {Malz}}, \emph {et~al.},\ }\bibfield
	{title} {\bibinfo {title} {Nonreciprocal reconfigurable microwave
			optomechanical circuit},\ }\bibfield  {journal} {\bibinfo  {journal} {Nature
			Communications}\ }\textbf {\bibinfo {volume} {8}},\ \href
	{https://doi.org/10.1038/s41467-017-00447-1} {10.1038/s41467-017-00447-1}
	(\bibinfo {year} {2017})\BibitemShut {NoStop}%
	\bibitem [{\citenamefont {Seis}\ \emph {et~al.}(2022)\citenamefont {Seis},
		\citenamefont {Capelle}, \citenamefont {Langman}, \citenamefont {Saarinen},
		\citenamefont {Planz},\ and\ \citenamefont {Schliesser}}]{seis2022ground}%
	\BibitemOpen
	\bibfield  {author} {\bibinfo {author} {\bibfnamefont {Y.}~\bibnamefont
			{Seis}}, \bibinfo {author} {\bibfnamefont {T.}~\bibnamefont {Capelle}},
		\bibinfo {author} {\bibfnamefont {E.}~\bibnamefont {Langman}}, \bibinfo
		{author} {\bibfnamefont {S.}~\bibnamefont {Saarinen}}, \bibinfo {author}
		{\bibfnamefont {E.}~\bibnamefont {Planz}},\ and\ \bibinfo {author}
		{\bibfnamefont {A.}~\bibnamefont {Schliesser}},\ }\bibfield  {title}
	{\bibinfo {title} {Ground state cooling of an ultracoherent electromechanical
			system},\ }\href {https://www.nature.com/articles/s41467-022-29115-9}
	{\bibfield  {journal} {\bibinfo  {journal} {Nature communications}\ }\textbf
		{\bibinfo {volume} {13}},\ \bibinfo {pages} {1} (\bibinfo {year}
		{2022})}\BibitemShut {NoStop}%
	\bibitem [{\citenamefont {Liu}\ \emph {et~al.}(2021{\natexlab{b}})\citenamefont
		{Liu}, \citenamefont {Liu}, \citenamefont {Wang}, \citenamefont {Chen},
		\citenamefont {Sillanp{\"a}{\"a}},\ and\ \citenamefont
		{Li}}]{liu2021optomechanical}%
	\BibitemOpen
	\bibfield  {author} {\bibinfo {author} {\bibfnamefont {Y.}~\bibnamefont
			{Liu}}, \bibinfo {author} {\bibfnamefont {Q.}~\bibnamefont {Liu}}, \bibinfo
		{author} {\bibfnamefont {S.}~\bibnamefont {Wang}}, \bibinfo {author}
		{\bibfnamefont {Z.}~\bibnamefont {Chen}}, \bibinfo {author} {\bibfnamefont
			{M.~A.}\ \bibnamefont {Sillanp{\"a}{\"a}}},\ and\ \bibinfo {author}
		{\bibfnamefont {T.}~\bibnamefont {Li}},\ }\bibfield  {title} {\bibinfo
		{title} {Optomechanical anti-lasing with infinite group delay at a phase
			singularity},\ }\href
	{https://journals.aps.org/prl/abstract/10.1103/PhysRevLett.127.273603}
	{\bibfield  {journal} {\bibinfo  {journal} {Physical Review Letters}\
		}\textbf {\bibinfo {volume} {127}},\ \bibinfo {pages} {273603} (\bibinfo
		{year} {2021}{\natexlab{b}})}\BibitemShut {NoStop}%
	\bibitem [{\citenamefont {Tsaturyan}\ \emph {et~al.}(2017)\citenamefont
		{Tsaturyan}, \citenamefont {Barg}, \citenamefont {Polzik},\ and\
		\citenamefont {Schliesser}}]{tsaturyan2017ultracoherent}%
	\BibitemOpen
	\bibfield  {author} {\bibinfo {author} {\bibfnamefont {Y.}~\bibnamefont
			{Tsaturyan}}, \bibinfo {author} {\bibfnamefont {A.}~\bibnamefont {Barg}},
		\bibinfo {author} {\bibfnamefont {E.~S.}\ \bibnamefont {Polzik}},\ and\
		\bibinfo {author} {\bibfnamefont {A.}~\bibnamefont {Schliesser}},\ }\bibfield
	{title} {\bibinfo {title} {Ultracoherent nanomechanical resonators via soft
			clamping and dissipation dilution},\ }\href
	{https://www.nature.com/articles/nnano.2017.101} {\bibfield  {journal}
		{\bibinfo  {journal} {Nature nanotechnology}\ }\textbf {\bibinfo {volume}
			{12}},\ \bibinfo {pages} {776} (\bibinfo {year} {2017})}\BibitemShut
	{NoStop}%
	\bibitem [{\citenamefont {Schmid}\ \emph {et~al.}(2011)\citenamefont {Schmid},
		\citenamefont {Jensen}, \citenamefont {Nielsen},\ and\ \citenamefont
		{Boisen}}]{schmid2011damping}%
	\BibitemOpen
	\bibfield  {author} {\bibinfo {author} {\bibfnamefont {S.}~\bibnamefont
			{Schmid}}, \bibinfo {author} {\bibfnamefont {K.}~\bibnamefont {Jensen}},
		\bibinfo {author} {\bibfnamefont {K.}~\bibnamefont {Nielsen}},\ and\ \bibinfo
		{author} {\bibfnamefont {A.}~\bibnamefont {Boisen}},\ }\bibfield  {title}
	{\bibinfo {title} {Damping mechanisms in high-q micro and nanomechanical
			string resonators},\ }\href
	{https://journals.aps.org/prb/abstract/10.1103/PhysRevB.84.165307} {\bibfield
		{journal} {\bibinfo  {journal} {Physical Review B}\ }\textbf {\bibinfo
			{volume} {84}},\ \bibinfo {pages} {165307} (\bibinfo {year}
		{2011})}\BibitemShut {NoStop}%
	\bibitem [{\citenamefont {Weinstein}\ \emph {et~al.}(2014)\citenamefont
		{Weinstein}, \citenamefont {Lei}, \citenamefont {Wollman}, \citenamefont
		{Suh}, \citenamefont {Metelmann}, \citenamefont {Clerk},\ and\ \citenamefont
		{Schwab}}]{weinstein2014observation}%
	\BibitemOpen
	\bibfield  {author} {\bibinfo {author} {\bibfnamefont {A.}~\bibnamefont
			{Weinstein}}, \bibinfo {author} {\bibfnamefont {C.}~\bibnamefont {Lei}},
		\bibinfo {author} {\bibfnamefont {E.}~\bibnamefont {Wollman}}, \bibinfo
		{author} {\bibfnamefont {J.}~\bibnamefont {Suh}}, \bibinfo {author}
		{\bibfnamefont {A.}~\bibnamefont {Metelmann}}, \bibinfo {author}
		{\bibfnamefont {A.}~\bibnamefont {Clerk}},\ and\ \bibinfo {author}
		{\bibfnamefont {K.}~\bibnamefont {Schwab}},\ }\bibfield  {title} {\bibinfo
		{title} {Observation and interpretation of motional sideband asymmetry in a
			quantum electromechanical device},\ }\href
	{https://journals.aps.org/prx/abstract/10.1103/PhysRevX.4.041003} {\bibfield
		{journal} {\bibinfo  {journal} {Physical Review X}\ }\textbf {\bibinfo
			{volume} {4}},\ \bibinfo {pages} {041003} (\bibinfo {year}
		{2014})}\BibitemShut {NoStop}%
	\bibitem [{\citenamefont {Macklin}\ \emph {et~al.}(2015)\citenamefont
		{Macklin}, \citenamefont {O’Brien}, \citenamefont {Hover}, \citenamefont
		{Schwartz}, \citenamefont {Bolkhovsky} \emph {et~al.}}]{macklin2015near}%
	\BibitemOpen
	\bibfield  {author} {\bibinfo {author} {\bibfnamefont {C.}~\bibnamefont
			{Macklin}}, \bibinfo {author} {\bibfnamefont {K.}~\bibnamefont {O’Brien}},
		\bibinfo {author} {\bibfnamefont {D.}~\bibnamefont {Hover}}, \bibinfo
		{author} {\bibfnamefont {M.}~\bibnamefont {Schwartz}}, \bibinfo {author}
		{\bibfnamefont {V.}~\bibnamefont {Bolkhovsky}}, \emph {et~al.},\ }\bibfield
	{title} {\bibinfo {title} {A near--quantum-limited josephson traveling-wave
			parametric amplifier},\ }\href
	{https://science.sciencemag.org/content/350/6258/307} {\bibfield  {journal}
		{\bibinfo  {journal} {Science}\ }\textbf {\bibinfo {volume} {350}},\ \bibinfo
		{pages} {307} (\bibinfo {year} {2015})}\BibitemShut {NoStop}%
	\bibitem [{\citenamefont {Kronwald}\ \emph {et~al.}(2013)\citenamefont
		{Kronwald}, \citenamefont {Marquardt},\ and\ \citenamefont
		{Clerk}}]{kronwald2013arbitrarily}%
	\BibitemOpen
	\bibfield  {author} {\bibinfo {author} {\bibfnamefont {A.}~\bibnamefont
			{Kronwald}}, \bibinfo {author} {\bibfnamefont {F.}~\bibnamefont
			{Marquardt}},\ and\ \bibinfo {author} {\bibfnamefont {A.~A.}\ \bibnamefont
			{Clerk}},\ }\bibfield  {title} {\bibinfo {title} {Arbitrarily large
			steady-state bosonic squeezing via dissipation},\ }\href
	{https://journals.aps.org/pra/abstract/10.1103/PhysRevA.88.063833} {\bibfield
		{journal} {\bibinfo  {journal} {Physical Review A}\ }\textbf {\bibinfo
			{volume} {88}},\ \bibinfo {pages} {063833} (\bibinfo {year}
		{2013})}\BibitemShut {NoStop}%
\end{thebibliography}

\begin{thebibliography}{25}%
	\makeatletter
	\providecommand \@ifxundefined [1]{%
		\@ifx{#1\undefined}
	}%
	\providecommand \@ifnum [1]{%
		\ifnum #1\expandafter \@firstoftwo
		\else \expandafter \@secondoftwo
		\fi
	}%
	\providecommand \@ifx [1]{%
		\ifx #1\expandafter \@firstoftwo
		\else \expandafter \@secondoftwo
		\fi
	}%
	\providecommand \natexlab [1]{#1}%
	\providecommand \enquote  [1]{``#1''}%
	\providecommand \bibnamefont  [1]{#1}%
	\providecommand \bibfnamefont [1]{#1}%
	\providecommand \citenamefont [1]{#1}%
	\providecommand \href@noop [0]{\@secondoftwo}%
	\providecommand \href [0]{\begingroup \@sanitize@url \@href}%
	\providecommand \@href[1]{\@@startlink{#1}\@@href}%
	\providecommand \@@href[1]{\endgroup#1\@@endlink}%
	\providecommand \@sanitize@url [0]{\catcode `\\12\catcode `\$12\catcode
		`\&12\catcode `\#12\catcode `\^12\catcode `\_12\catcode `\%12\relax}%
	\providecommand \@@startlink[1]{}%
	\providecommand \@@endlink[0]{}%
	\providecommand \url  [0]{\begingroup\@sanitize@url \@url }%
	\providecommand \@url [1]{\endgroup\@href {#1}{\urlprefix }}%
	\providecommand \urlprefix  [0]{URL }%
	\providecommand \Eprint [0]{\href }%
	\providecommand \doibase [0]{https://doi.org/}%
	\providecommand \selectlanguage [0]{\@gobble}%
	\providecommand \bibinfo  [0]{\@secondoftwo}%
	\providecommand \bibfield  [0]{\@secondoftwo}%
	\providecommand \translation [1]{[#1]}%
	\providecommand \BibitemOpen [0]{}%
	\providecommand \bibitemStop [0]{}%
	\providecommand \bibitemNoStop [0]{.\EOS\space}%
	\providecommand \EOS [0]{\spacefactor3000\relax}%
	\providecommand \BibitemShut  [1]{\csname bibitem#1\endcsname}%
	\let\auto@bib@innerbib\@empty
	\bibitem [{\citenamefont {Reed}\ \emph {et~al.}(2017)\citenamefont {Reed},
		\citenamefont {Mayer}, \citenamefont {Teufel}, \citenamefont {Burkhart},
		\citenamefont {Pfaff}, \citenamefont {Reagor}, \citenamefont {Sletten},
		\citenamefont {Ma}, \citenamefont {Schoelkopf}, \citenamefont {Knill} \emph
		{et~al.}}]{SI_reed2017faithful}%
	\BibitemOpen
	\bibfield  {author} {\bibinfo {author} {\bibfnamefont {A.}~\bibnamefont
			{Reed}}, \bibinfo {author} {\bibfnamefont {K.}~\bibnamefont {Mayer}},
		\bibinfo {author} {\bibfnamefont {J.}~\bibnamefont {Teufel}}, \bibinfo
		{author} {\bibfnamefont {L.}~\bibnamefont {Burkhart}}, \bibinfo {author}
		{\bibfnamefont {W.}~\bibnamefont {Pfaff}}, \bibinfo {author} {\bibfnamefont
			{M.}~\bibnamefont {Reagor}}, \bibinfo {author} {\bibfnamefont
			{L.}~\bibnamefont {Sletten}}, \bibinfo {author} {\bibfnamefont
			{X.}~\bibnamefont {Ma}}, \bibinfo {author} {\bibfnamefont {R.}~\bibnamefont
			{Schoelkopf}}, \bibinfo {author} {\bibfnamefont {E.}~\bibnamefont {Knill}},
		\emph {et~al.},\ }\bibfield  {title} {\bibinfo {title} {Faithful conversion
			of propagating quantum information to mechanical motion},\ }\href
	{https://www.nature.com/articles/nphys4251} {\bibfield  {journal} {\bibinfo
			{journal} {Nature Physics}\ }\textbf {\bibinfo {volume} {13}},\ \bibinfo
		{pages} {1163} (\bibinfo {year} {2017})}\BibitemShut {NoStop}%
	\bibitem [{\citenamefont {Delaney}\ \emph {et~al.}(2019)\citenamefont
		{Delaney}, \citenamefont {Reed}, \citenamefont {Andrews},\ and\ \citenamefont
		{Lehnert}}]{SI_delaney2019measurement}%
	\BibitemOpen
	\bibfield  {author} {\bibinfo {author} {\bibfnamefont {R.~D.}\ \bibnamefont
			{Delaney}}, \bibinfo {author} {\bibfnamefont {A.~P.}\ \bibnamefont {Reed}},
		\bibinfo {author} {\bibfnamefont {R.~W.}\ \bibnamefont {Andrews}},\ and\
		\bibinfo {author} {\bibfnamefont {K.~W.}\ \bibnamefont {Lehnert}},\
	}\bibfield  {title} {\bibinfo {title} {Measurement of motion beyond the
			quantum limit by transient amplification},\ }\href
	{https://journals.aps.org/prl/abstract/10.1103/PhysRevLett.123.183603}
	{\bibfield  {journal} {\bibinfo  {journal} {Physical review letters}\
		}\textbf {\bibinfo {volume} {123}},\ \bibinfo {pages} {183603} (\bibinfo
		{year} {2019})}\BibitemShut {NoStop}%
	\bibitem [{\citenamefont {Fedorov}(2020)}]{SI_fedorov2020mechanical}%
	\BibitemOpen
	\bibfield  {author} {\bibinfo {author} {\bibfnamefont {S.}~\bibnamefont
			{Fedorov}},\ }\href {http://dx.doi.org/10.5075/epfl-thesis-10421} {\emph
		{\bibinfo {title} {Mechanical resonators with high dissipation dilution in
				precision and quantum measurements}}},\ \bibinfo {type} {Tech. Rep.}\
	(\bibinfo  {institution} {EPFL},\ \bibinfo {year} {2020})\BibitemShut
	{NoStop}%
	\bibitem [{\citenamefont {Cole}\ \emph {et~al.}(2011)\citenamefont {Cole},
		\citenamefont {Wilson-Rae}, \citenamefont {Werbach}, \citenamefont {Vanner},\
		and\ \citenamefont {Aspelmeyer}}]{SI_cole2011phonon}%
	\BibitemOpen
	\bibfield  {author} {\bibinfo {author} {\bibfnamefont {G.~D.}\ \bibnamefont
			{Cole}}, \bibinfo {author} {\bibfnamefont {I.}~\bibnamefont {Wilson-Rae}},
		\bibinfo {author} {\bibfnamefont {K.}~\bibnamefont {Werbach}}, \bibinfo
		{author} {\bibfnamefont {M.~R.}\ \bibnamefont {Vanner}},\ and\ \bibinfo
		{author} {\bibfnamefont {M.}~\bibnamefont {Aspelmeyer}},\ }\bibfield  {title}
	{\bibinfo {title} {Phonon-tunnelling dissipation in mechanical resonators},\
	}\href {https://www.nature.com/articles/ncomms1212} {\bibfield  {journal}
		{\bibinfo  {journal} {Nature communications}\ }\textbf {\bibinfo {volume}
			{2}},\ \bibinfo {pages} {1} (\bibinfo {year} {2011})}\BibitemShut {NoStop}%
	\bibitem [{\citenamefont {Wilson-Rae}\ \emph {et~al.}(2011)\citenamefont
		{Wilson-Rae}, \citenamefont {Barton}, \citenamefont {Verbridge},
		\citenamefont {Southworth}, \citenamefont {Ilic}, \citenamefont {Craighead},\
		and\ \citenamefont {Parpia}}]{SI_wilson2011high}%
	\BibitemOpen
	\bibfield  {author} {\bibinfo {author} {\bibfnamefont {I.}~\bibnamefont
			{Wilson-Rae}}, \bibinfo {author} {\bibfnamefont {R.}~\bibnamefont {Barton}},
		\bibinfo {author} {\bibfnamefont {S.}~\bibnamefont {Verbridge}}, \bibinfo
		{author} {\bibfnamefont {D.}~\bibnamefont {Southworth}}, \bibinfo {author}
		{\bibfnamefont {B.}~\bibnamefont {Ilic}}, \bibinfo {author} {\bibfnamefont
			{H.~G.}\ \bibnamefont {Craighead}},\ and\ \bibinfo {author} {\bibfnamefont
			{J.}~\bibnamefont {Parpia}},\ }\bibfield  {title} {\bibinfo {title} {High-q
			nanomechanics via destructive interference of elastic waves},\ }\href
	{https://journals.aps.org/prl/abstract/10.1103/PhysRevLett.106.047205}
	{\bibfield  {journal} {\bibinfo  {journal} {Physical review letters}\
		}\textbf {\bibinfo {volume} {106}},\ \bibinfo {pages} {047205} (\bibinfo
		{year} {2011})}\BibitemShut {NoStop}%
	\bibitem [{\citenamefont {Cattiaux}\ \emph {et~al.}(2021)\citenamefont
		{Cattiaux}, \citenamefont {Golokolenov}, \citenamefont {Kumar}, \citenamefont
		{Sillanp{\"a}{\"a}}, \citenamefont {Mercier~de L{\'e}pinay}, \citenamefont
		{Gazizulin}, \citenamefont {Zhou}, \citenamefont {Armour}, \citenamefont
		{Bourgeois}, \citenamefont {Fefferman} \emph
		{et~al.}}]{SI_cattiaux2021macroscopic}%
	\BibitemOpen
	\bibfield  {author} {\bibinfo {author} {\bibfnamefont {D.}~\bibnamefont
			{Cattiaux}}, \bibinfo {author} {\bibfnamefont {I.}~\bibnamefont
			{Golokolenov}}, \bibinfo {author} {\bibfnamefont {S.}~\bibnamefont {Kumar}},
		\bibinfo {author} {\bibfnamefont {M.}~\bibnamefont {Sillanp{\"a}{\"a}}},
		\bibinfo {author} {\bibfnamefont {L.}~\bibnamefont {Mercier~de L{\'e}pinay}},
		\bibinfo {author} {\bibfnamefont {R.}~\bibnamefont {Gazizulin}}, \bibinfo
		{author} {\bibfnamefont {X.}~\bibnamefont {Zhou}}, \bibinfo {author}
		{\bibfnamefont {A.}~\bibnamefont {Armour}}, \bibinfo {author} {\bibfnamefont
			{O.}~\bibnamefont {Bourgeois}}, \bibinfo {author} {\bibfnamefont
			{A.}~\bibnamefont {Fefferman}}, \emph {et~al.},\ }\bibfield  {title}
	{\bibinfo {title} {A macroscopic object passively cooled into its quantum
			ground state of motion beyond single-mode cooling},\ }\href
	{https://www.nature.com/articles/s41467-021-26457-8} {\bibfield  {journal}
		{\bibinfo  {journal} {Nature communications}\ }\textbf {\bibinfo {volume}
			{12}},\ \bibinfo {pages} {1} (\bibinfo {year} {2021})}\BibitemShut {NoStop}%
	\bibitem [{\citenamefont {Fedorov}\ \emph {et~al.}(2019)\citenamefont
		{Fedorov}, \citenamefont {Engelsen}, \citenamefont {Ghadimi}, \citenamefont
		{Bereyhi}, \citenamefont {Schilling}, \citenamefont {Wilson},\ and\
		\citenamefont {Kippenberg}}]{SI_fedorov2019generalized}%
	\BibitemOpen
	\bibfield  {author} {\bibinfo {author} {\bibfnamefont {S.~A.}\ \bibnamefont
			{Fedorov}}, \bibinfo {author} {\bibfnamefont {N.~J.}\ \bibnamefont
			{Engelsen}}, \bibinfo {author} {\bibfnamefont {A.~H.}\ \bibnamefont
			{Ghadimi}}, \bibinfo {author} {\bibfnamefont {M.~J.}\ \bibnamefont
			{Bereyhi}}, \bibinfo {author} {\bibfnamefont {R.}~\bibnamefont {Schilling}},
		\bibinfo {author} {\bibfnamefont {D.~J.}\ \bibnamefont {Wilson}},\ and\
		\bibinfo {author} {\bibfnamefont {T.~J.}\ \bibnamefont {Kippenberg}},\
	}\bibfield  {title} {\bibinfo {title} {Generalized dissipation dilution in
			strained mechanical resonators},\ }\href
	{https://journals.aps.org/prb/abstract/10.1103/PhysRevB.99.054107} {\bibfield
		{journal} {\bibinfo  {journal} {Physical Review B}\ }\textbf {\bibinfo
			{volume} {99}},\ \bibinfo {pages} {054107} (\bibinfo {year}
		{2019})}\BibitemShut {NoStop}%
	\bibitem [{\citenamefont {Ekin}(2006)}]{SI_ekin2006experimental}%
	\BibitemOpen
	\bibfield  {author} {\bibinfo {author} {\bibfnamefont {J.}~\bibnamefont
			{Ekin}},\ }\href {https://doi.org/10.1093/acprof:oso/9780198570547.001.0001}
	{\emph {\bibinfo {title} {Experimental techniques for low-temperature
				measurements: cryostat design, material properties and superconductor
				critical-current testing}}}\ (\bibinfo  {publisher} {Oxford university
		press},\ \bibinfo {year} {2006})\BibitemShut {NoStop}%
	\bibitem [{\citenamefont {Cicak}\ \emph {et~al.}(2010)\citenamefont {Cicak},
		\citenamefont {Li}, \citenamefont {Strong}, \citenamefont {Allman},
		\citenamefont {Altomare}, \citenamefont {Sirois}, \citenamefont {Whittaker},
		\citenamefont {Teufel},\ and\ \citenamefont {Simmonds}}]{SI_cicak2010low}%
	\BibitemOpen
	\bibfield  {author} {\bibinfo {author} {\bibfnamefont {K.}~\bibnamefont
			{Cicak}}, \bibinfo {author} {\bibfnamefont {D.}~\bibnamefont {Li}}, \bibinfo
		{author} {\bibfnamefont {J.~A.}\ \bibnamefont {Strong}}, \bibinfo {author}
		{\bibfnamefont {M.~S.}\ \bibnamefont {Allman}}, \bibinfo {author}
		{\bibfnamefont {F.}~\bibnamefont {Altomare}}, \bibinfo {author}
		{\bibfnamefont {A.~J.}\ \bibnamefont {Sirois}}, \bibinfo {author}
		{\bibfnamefont {J.~D.}\ \bibnamefont {Whittaker}}, \bibinfo {author}
		{\bibfnamefont {J.~D.}\ \bibnamefont {Teufel}},\ and\ \bibinfo {author}
		{\bibfnamefont {R.~W.}\ \bibnamefont {Simmonds}},\ }\bibfield  {title}
	{\bibinfo {title} {Low-loss superconducting resonant circuits using
			vacuum-gap-based microwave components},\ }\href@noop {} {\bibfield  {journal}
		{\bibinfo  {journal} {Applied Physics Letters}\ }\textbf {\bibinfo {volume}
			{96}},\ \bibinfo {pages} {093502} (\bibinfo {year} {2010})}\BibitemShut
	{NoStop}%
	\bibitem [{\citenamefont {Teufel}\ \emph
		{et~al.}(2011{\natexlab{a}})\citenamefont {Teufel}, \citenamefont {Li},
		\citenamefont {Allman}, \citenamefont {Cicak}, \citenamefont {Sirois},
		\citenamefont {Whittaker},\ and\ \citenamefont
		{Simmonds}}]{SI_teufel2011circuit}%
	\BibitemOpen
	\bibfield  {author} {\bibinfo {author} {\bibfnamefont {J.~D.}\ \bibnamefont
			{Teufel}}, \bibinfo {author} {\bibfnamefont {D.}~\bibnamefont {Li}}, \bibinfo
		{author} {\bibfnamefont {M.}~\bibnamefont {Allman}}, \bibinfo {author}
		{\bibfnamefont {K.}~\bibnamefont {Cicak}}, \bibinfo {author} {\bibfnamefont
			{A.}~\bibnamefont {Sirois}}, \bibinfo {author} {\bibfnamefont
			{J.}~\bibnamefont {Whittaker}},\ and\ \bibinfo {author} {\bibfnamefont
			{R.}~\bibnamefont {Simmonds}},\ }\bibfield  {title} {\bibinfo {title}
		{Circuit cavity electromechanics in the strong-coupling regime},\ }\href
	{https://www.nature.com/articles/nature09898} {\bibfield  {journal} {\bibinfo
			{journal} {Nature}\ }\textbf {\bibinfo {volume} {471}},\ \bibinfo {pages}
		{204} (\bibinfo {year} {2011}{\natexlab{a}})}\BibitemShut {NoStop}%
	\bibitem [{\citenamefont {Suh}\ \emph {et~al.}(2014)\citenamefont {Suh},
		\citenamefont {Weinstein}, \citenamefont {Lei}, \citenamefont {Wollman},
		\citenamefont {Steinke}, \citenamefont {Meystre}, \citenamefont {Clerk},\
		and\ \citenamefont {Schwab}}]{SI_suh2014mechanically}%
	\BibitemOpen
	\bibfield  {author} {\bibinfo {author} {\bibfnamefont {J.}~\bibnamefont
			{Suh}}, \bibinfo {author} {\bibfnamefont {A.}~\bibnamefont {Weinstein}},
		\bibinfo {author} {\bibfnamefont {C.}~\bibnamefont {Lei}}, \bibinfo {author}
		{\bibfnamefont {E.}~\bibnamefont {Wollman}}, \bibinfo {author} {\bibfnamefont
			{S.}~\bibnamefont {Steinke}}, \bibinfo {author} {\bibfnamefont
			{P.}~\bibnamefont {Meystre}}, \bibinfo {author} {\bibfnamefont {A.~A.}\
			\bibnamefont {Clerk}},\ and\ \bibinfo {author} {\bibfnamefont
			{K.}~\bibnamefont {Schwab}},\ }\bibfield  {title} {\bibinfo {title}
		{Mechanically detecting and avoiding the quantum fluctuations of a microwave
			field},\ }\href {https://www.science.org/doi/full/10.1126/science.1253258}
	{\bibfield  {journal} {\bibinfo  {journal} {Science}\ }\textbf {\bibinfo
			{volume} {344}},\ \bibinfo {pages} {1262} (\bibinfo {year}
		{2014})}\BibitemShut {NoStop}%
	\bibitem [{\citenamefont {Pirkkalainen}\ \emph {et~al.}(2015)\citenamefont
		{Pirkkalainen}, \citenamefont {Damsk{\"a}gg}, \citenamefont {Brandt},
		\citenamefont {Massel},\ and\ \citenamefont
		{Sillanp{\"a}{\"a}}}]{SI_pirkkalainen2015squeezing}%
	\BibitemOpen
	\bibfield  {author} {\bibinfo {author} {\bibfnamefont {J.-M.}\ \bibnamefont
			{Pirkkalainen}}, \bibinfo {author} {\bibfnamefont {E.}~\bibnamefont
			{Damsk{\"a}gg}}, \bibinfo {author} {\bibfnamefont {M.}~\bibnamefont
			{Brandt}}, \bibinfo {author} {\bibfnamefont {F.}~\bibnamefont {Massel}},\
		and\ \bibinfo {author} {\bibfnamefont {M.~A.}\ \bibnamefont
			{Sillanp{\"a}{\"a}}},\ }\bibfield  {title} {\bibinfo {title} {Squeezing of
			quantum noise of motion in a micromechanical resonator},\ }\href
	{https://journals.aps.org/prl/abstract/10.1103/PhysRevLett.115.243601}
	{\bibfield  {journal} {\bibinfo  {journal} {Physical Review Letters}\
		}\textbf {\bibinfo {volume} {115}},\ \bibinfo {pages} {243601} (\bibinfo
		{year} {2015})}\BibitemShut {NoStop}%
	\bibitem [{\citenamefont {Toth}\ \emph {et~al.}(2017)\citenamefont {Toth},
		\citenamefont {Bernier}, \citenamefont {Nunnenkamp}, \citenamefont
		{Feofanov},\ and\ \citenamefont {Kippenberg}}]{SI_toth2017dissipative}%
	\BibitemOpen
	\bibfield  {author} {\bibinfo {author} {\bibfnamefont {L.~D.}\ \bibnamefont
			{Toth}}, \bibinfo {author} {\bibfnamefont {N.~R.}\ \bibnamefont {Bernier}},
		\bibinfo {author} {\bibfnamefont {A.}~\bibnamefont {Nunnenkamp}}, \bibinfo
		{author} {\bibfnamefont {A.}~\bibnamefont {Feofanov}},\ and\ \bibinfo
		{author} {\bibfnamefont {T.}~\bibnamefont {Kippenberg}},\ }\bibfield  {title}
	{\bibinfo {title} {A dissipative quantum reservoir for microwave light using
			a mechanical oscillator},\ }\href {https://www.nature.com/articles/nphys4121}
	{\bibfield  {journal} {\bibinfo  {journal} {Nature Physics}\ }\textbf
		{\bibinfo {volume} {13}},\ \bibinfo {pages} {787} (\bibinfo {year}
		{2017})}\BibitemShut {NoStop}%
	\bibitem [{\citenamefont {Youssefi}\ \emph {et~al.}(2021)\citenamefont
		{Youssefi}, \citenamefont {Bancora}, \citenamefont {Kono}, \citenamefont
		{Chegnizadeh}, \citenamefont {Vovk}, \citenamefont {Pan},\ and\ \citenamefont
		{Kippenberg}}]{SI_youssefi2021superconducting}%
	\BibitemOpen
	\bibfield  {author} {\bibinfo {author} {\bibfnamefont {A.}~\bibnamefont
			{Youssefi}}, \bibinfo {author} {\bibfnamefont {A.}~\bibnamefont {Bancora}},
		\bibinfo {author} {\bibfnamefont {S.}~\bibnamefont {Kono}}, \bibinfo {author}
		{\bibfnamefont {M.}~\bibnamefont {Chegnizadeh}}, \bibinfo {author}
		{\bibfnamefont {T.}~\bibnamefont {Vovk}}, \bibinfo {author} {\bibfnamefont
			{J.}~\bibnamefont {Pan}},\ and\ \bibinfo {author} {\bibfnamefont {T.~J.}\
			\bibnamefont {Kippenberg}},\ }\bibfield  {title} {\bibinfo {title}
		{Topological lattices realized in superconducting circuit optomechanics},\
	}\href {https://arxiv.org/abs/2111.09133} {\bibfield  {journal} {\bibinfo
			{journal} {arXiv preprint arXiv:2111.09133}\ } (\bibinfo {year}
		{2021})}\BibitemShut {NoStop}%
	\bibitem [{\citenamefont {Macklin}\ \emph {et~al.}(2015)\citenamefont
		{Macklin}, \citenamefont {O’brien}, \citenamefont {Hover}, \citenamefont
		{Schwartz}, \citenamefont {Bolkhovsky}, \citenamefont {Zhang}, \citenamefont
		{Oliver},\ and\ \citenamefont {Siddiqi}}]{SI_macklin2015near}%
	\BibitemOpen
	\bibfield  {author} {\bibinfo {author} {\bibfnamefont {C.}~\bibnamefont
			{Macklin}}, \bibinfo {author} {\bibfnamefont {K.}~\bibnamefont {O’brien}},
		\bibinfo {author} {\bibfnamefont {D.}~\bibnamefont {Hover}}, \bibinfo
		{author} {\bibfnamefont {M.}~\bibnamefont {Schwartz}}, \bibinfo {author}
		{\bibfnamefont {V.}~\bibnamefont {Bolkhovsky}}, \bibinfo {author}
		{\bibfnamefont {X.}~\bibnamefont {Zhang}}, \bibinfo {author} {\bibfnamefont
			{W.}~\bibnamefont {Oliver}},\ and\ \bibinfo {author} {\bibfnamefont
			{I.}~\bibnamefont {Siddiqi}},\ }\bibfield  {title} {\bibinfo {title} {A
			near--quantum-limited josephson traveling-wave parametric amplifier},\
	}\href@noop {} {\bibfield  {journal} {\bibinfo  {journal} {Science}\ }\textbf
		{\bibinfo {volume} {350}},\ \bibinfo {pages} {307} (\bibinfo {year}
		{2015})}\BibitemShut {NoStop}%
	\bibitem [{\citenamefont {Aspelmeyer}\ \emph {et~al.}(2014)\citenamefont
		{Aspelmeyer}, \citenamefont {Kippenberg},\ and\ \citenamefont
		{Marquardt}}]{SI_aspelmeyer2014cavity}%
	\BibitemOpen
	\bibfield  {author} {\bibinfo {author} {\bibfnamefont {M.}~\bibnamefont
			{Aspelmeyer}}, \bibinfo {author} {\bibfnamefont {T.~J.}\ \bibnamefont
			{Kippenberg}},\ and\ \bibinfo {author} {\bibfnamefont {F.}~\bibnamefont
			{Marquardt}},\ }\bibfield  {title} {\bibinfo {title} {Cavity optomechanics},\
	}\href {https://journals.aps.org/rmp/abstract/10.1103/RevModPhys.86.1391}
	{\bibfield  {journal} {\bibinfo  {journal} {Reviews of Modern Physics}\
		}\textbf {\bibinfo {volume} {86}},\ \bibinfo {pages} {1391} (\bibinfo {year}
		{2014})}\BibitemShut {NoStop}%
	\bibitem [{\citenamefont {Rabl}\ \emph {et~al.}(2009)\citenamefont {Rabl},
		\citenamefont {Genes}, \citenamefont {Hammerer},\ and\ \citenamefont
		{Aspelmeyer}}]{SI_rabl2009phase}%
	\BibitemOpen
	\bibfield  {author} {\bibinfo {author} {\bibfnamefont {P.}~\bibnamefont
			{Rabl}}, \bibinfo {author} {\bibfnamefont {C.}~\bibnamefont {Genes}},
		\bibinfo {author} {\bibfnamefont {K.}~\bibnamefont {Hammerer}},\ and\
		\bibinfo {author} {\bibfnamefont {M.}~\bibnamefont {Aspelmeyer}},\ }\bibfield
	{title} {\bibinfo {title} {Phase-noise induced limitations on cooling and
			coherent evolution in optomechanical systems},\ }\href@noop {} {\bibfield
		{journal} {\bibinfo  {journal} {Physical Review A}\ }\textbf {\bibinfo
			{volume} {80}},\ \bibinfo {pages} {063819} (\bibinfo {year}
		{2009})}\BibitemShut {NoStop}%
	\bibitem [{\citenamefont {Joshi}\ \emph {et~al.}(2021)\citenamefont {Joshi},
		\citenamefont {Sauerwein}, \citenamefont {Youssefi}, \citenamefont {Uhrich},\
		and\ \citenamefont {Kippenberg}}]{SI_joshi2021automated}%
	\BibitemOpen
	\bibfield  {author} {\bibinfo {author} {\bibfnamefont {Y.~J.}\ \bibnamefont
			{Joshi}}, \bibinfo {author} {\bibfnamefont {N.}~\bibnamefont {Sauerwein}},
		\bibinfo {author} {\bibfnamefont {A.}~\bibnamefont {Youssefi}}, \bibinfo
		{author} {\bibfnamefont {P.}~\bibnamefont {Uhrich}},\ and\ \bibinfo {author}
		{\bibfnamefont {T.~J.}\ \bibnamefont {Kippenberg}},\ }\bibfield  {title}
	{\bibinfo {title} {Automated wide-ranged finely tunable microwave cavity for
			narrowband phase noise filtering},\ }\href
	{https://doi.org/10.1063/5.0034696} {\bibfield  {journal} {\bibinfo
			{journal} {Review of Scientific Instruments}\ }\textbf {\bibinfo {volume}
			{92}},\ \bibinfo {pages} {034710} (\bibinfo {year} {2021})}\BibitemShut
	{NoStop}%
	\bibitem [{\citenamefont {Bernier}(2019)}]{SI_bernier2019multimode}%
	\BibitemOpen
	\bibfield  {author} {\bibinfo {author} {\bibfnamefont {N.~R.}\ \bibnamefont
			{Bernier}},\ }\href@noop {} {\emph {\bibinfo {title} {Multimode microwave
				circuit optomechanics as a platform to study coupled quantum harmonic
				oscillators}}},\ \bibinfo {type} {Tech. Rep.}\ (\bibinfo  {institution}
	{EPFL},\ \bibinfo {year} {2019})\BibitemShut {NoStop}%
	\bibitem [{\citenamefont {Bowen}\ and\ \citenamefont
		{Milburn}(2015)}]{SI_bowen2015quantum}%
	\BibitemOpen
	\bibfield  {author} {\bibinfo {author} {\bibfnamefont {W.~P.}\ \bibnamefont
			{Bowen}}\ and\ \bibinfo {author} {\bibfnamefont {G.~J.}\ \bibnamefont
			{Milburn}},\ }\href@noop {} {\emph {\bibinfo {title} {Quantum
				optomechanics}}}\ (\bibinfo  {publisher} {CRC press},\ \bibinfo {year}
	{2015})\BibitemShut {NoStop}%
	\bibitem [{\citenamefont {Teufel}\ \emph
		{et~al.}(2011{\natexlab{b}})\citenamefont {Teufel}, \citenamefont {Donner},
		\citenamefont {Li}, \citenamefont {Harlow}, \citenamefont {Allman},
		\citenamefont {Cicak}, \citenamefont {Sirois}, \citenamefont {Whittaker},
		\citenamefont {Lehnert},\ and\ \citenamefont
		{Simmonds}}]{SI_teufel2011sideband}%
	\BibitemOpen
	\bibfield  {author} {\bibinfo {author} {\bibfnamefont {J.~D.}\ \bibnamefont
			{Teufel}}, \bibinfo {author} {\bibfnamefont {T.}~\bibnamefont {Donner}},
		\bibinfo {author} {\bibfnamefont {D.}~\bibnamefont {Li}}, \bibinfo {author}
		{\bibfnamefont {J.~W.}\ \bibnamefont {Harlow}}, \bibinfo {author}
		{\bibfnamefont {M.}~\bibnamefont {Allman}}, \bibinfo {author} {\bibfnamefont
			{K.}~\bibnamefont {Cicak}}, \bibinfo {author} {\bibfnamefont {A.~J.}\
			\bibnamefont {Sirois}}, \bibinfo {author} {\bibfnamefont {J.~D.}\
			\bibnamefont {Whittaker}}, \bibinfo {author} {\bibfnamefont {K.~W.}\
			\bibnamefont {Lehnert}},\ and\ \bibinfo {author} {\bibfnamefont {R.~W.}\
			\bibnamefont {Simmonds}},\ }\bibfield  {title} {\bibinfo {title} {Sideband
			cooling of micromechanical motion to the quantum ground state},\ }\href
	{https://www.nature.com/articles/nature10261} {\bibfield  {journal} {\bibinfo
			{journal} {Nature}\ }\textbf {\bibinfo {volume} {475}},\ \bibinfo {pages}
		{359} (\bibinfo {year} {2011}{\natexlab{b}})}\BibitemShut {NoStop}%
	\bibitem [{\citenamefont {Weinstein}\ \emph {et~al.}(2014)\citenamefont
		{Weinstein}, \citenamefont {Lei}, \citenamefont {Wollman}, \citenamefont
		{Suh}, \citenamefont {Metelmann}, \citenamefont {Clerk},\ and\ \citenamefont
		{Schwab}}]{SI_weinstein2014observation}%
	\BibitemOpen
	\bibfield  {author} {\bibinfo {author} {\bibfnamefont {A.}~\bibnamefont
			{Weinstein}}, \bibinfo {author} {\bibfnamefont {C.}~\bibnamefont {Lei}},
		\bibinfo {author} {\bibfnamefont {E.}~\bibnamefont {Wollman}}, \bibinfo
		{author} {\bibfnamefont {J.}~\bibnamefont {Suh}}, \bibinfo {author}
		{\bibfnamefont {A.}~\bibnamefont {Metelmann}}, \bibinfo {author}
		{\bibfnamefont {A.}~\bibnamefont {Clerk}},\ and\ \bibinfo {author}
		{\bibfnamefont {K.}~\bibnamefont {Schwab}},\ }\bibfield  {title} {\bibinfo
		{title} {Observation and interpretation of motional sideband asymmetry in a
			quantum electromechanical device},\ }\href@noop {} {\bibfield  {journal}
		{\bibinfo  {journal} {Physical Review X}\ }\textbf {\bibinfo {volume} {4}},\
		\bibinfo {pages} {041003} (\bibinfo {year} {2014})}\BibitemShut {NoStop}%
	\bibitem [{\citenamefont {Kronwald}\ \emph {et~al.}(2013)\citenamefont
		{Kronwald}, \citenamefont {Marquardt},\ and\ \citenamefont
		{Clerk}}]{SI_kronwald2013arbitrarily}%
	\BibitemOpen
	\bibfield  {author} {\bibinfo {author} {\bibfnamefont {A.}~\bibnamefont
			{Kronwald}}, \bibinfo {author} {\bibfnamefont {F.}~\bibnamefont
			{Marquardt}},\ and\ \bibinfo {author} {\bibfnamefont {A.~A.}\ \bibnamefont
			{Clerk}},\ }\bibfield  {title} {\bibinfo {title} {Arbitrarily large
			steady-state bosonic squeezing via dissipation},\ }\href
	{https://journals.aps.org/pra/abstract/10.1103/PhysRevA.88.063833} {\bibfield
		{journal} {\bibinfo  {journal} {Physical Review A}\ }\textbf {\bibinfo
			{volume} {88}},\ \bibinfo {pages} {063833} (\bibinfo {year}
		{2013})}\BibitemShut {NoStop}%
	\bibitem [{\citenamefont {Wollman}\ \emph {et~al.}(2015)\citenamefont
		{Wollman}, \citenamefont {Lei}, \citenamefont {Weinstein}, \citenamefont
		{Suh}, \citenamefont {Kronwald}, \citenamefont {Marquardt}, \citenamefont
		{Clerk},\ and\ \citenamefont {Schwab}}]{SI_wollman2015quantum}%
	\BibitemOpen
	\bibfield  {author} {\bibinfo {author} {\bibfnamefont {E.~E.}\ \bibnamefont
			{Wollman}}, \bibinfo {author} {\bibfnamefont {C.}~\bibnamefont {Lei}},
		\bibinfo {author} {\bibfnamefont {A.}~\bibnamefont {Weinstein}}, \bibinfo
		{author} {\bibfnamefont {J.}~\bibnamefont {Suh}}, \bibinfo {author}
		{\bibfnamefont {A.}~\bibnamefont {Kronwald}}, \bibinfo {author}
		{\bibfnamefont {F.}~\bibnamefont {Marquardt}}, \bibinfo {author}
		{\bibfnamefont {A.~A.}\ \bibnamefont {Clerk}},\ and\ \bibinfo {author}
		{\bibfnamefont {K.}~\bibnamefont {Schwab}},\ }\bibfield  {title} {\bibinfo
		{title} {Quantum squeezing of motion in a mechanical resonator},\ }\href
	{https://www.science.org/doi/10.1126/science.aac5138} {\bibfield  {journal}
		{\bibinfo  {journal} {Science}\ }\textbf {\bibinfo {volume} {349}},\ \bibinfo
		{pages} {952} (\bibinfo {year} {2015})}\BibitemShut {NoStop}%
	\bibitem [{\citenamefont {Shomroni}\ \emph {et~al.}(2019)\citenamefont
		{Shomroni}, \citenamefont {Youssefi}, \citenamefont {Sauerwein},
		\citenamefont {Qiu}, \citenamefont {Seidler}, \citenamefont {Malz},
		\citenamefont {Nunnenkamp},\ and\ \citenamefont
		{Kippenberg}}]{SI_shomroni2019two}%
	\BibitemOpen
	\bibfield  {author} {\bibinfo {author} {\bibfnamefont {I.}~\bibnamefont
			{Shomroni}}, \bibinfo {author} {\bibfnamefont {A.}~\bibnamefont {Youssefi}},
		\bibinfo {author} {\bibfnamefont {N.}~\bibnamefont {Sauerwein}}, \bibinfo
		{author} {\bibfnamefont {L.}~\bibnamefont {Qiu}}, \bibinfo {author}
		{\bibfnamefont {P.}~\bibnamefont {Seidler}}, \bibinfo {author} {\bibfnamefont
			{D.}~\bibnamefont {Malz}}, \bibinfo {author} {\bibfnamefont {A.}~\bibnamefont
			{Nunnenkamp}},\ and\ \bibinfo {author} {\bibfnamefont {T.~J.}\ \bibnamefont
			{Kippenberg}},\ }\bibfield  {title} {\bibinfo {title} {Two-tone
			optomechanical instability and its fundamental implications for
			backaction-evading measurements},\ }\href
	{https://journals.aps.org/prx/abstract/10.1103/PhysRevX.9.041022} {\bibfield
		{journal} {\bibinfo  {journal} {Physical Review X}\ }\textbf {\bibinfo
			{volume} {9}},\ \bibinfo {pages} {041022} (\bibinfo {year}
		{2019})}\BibitemShut {NoStop}%
\end{thebibliography}
\end{document}